\documentclass[12pt,preprint]{aastex}
\usepackage{amsmath}

\newcommand{\bdv}[1]{\mbox{\boldmath$#1$}}

\def\au{{\rm au}} 
 
\def\kms{{\rm km}\,{\rm s}^{-1}}
\def\masyr{{\rm mas}\,{\rm yr}^{-1}}
\def\kpc{{\rm kpc}}
\def\mas{{\rm mas}}
\def\anom{{\rm anom}}

\def\muas{\mu{\rm as}}

\def\min{{\rm min}}
\def\rel{{\rm rel}}

\def\hel{{\rm hel}}
\def\geo{{\rm geo}}
\def\e{{\rm E}}
\def\bpi{{\bdv\pi}}

\def\bmu{{\bdv\mu}}

\def\bgamma{{\bdv\gamma}}

\def\bv{{\bf v}}

\begin{document}
\title{Mass Production of 2023 KMTNet Microlensing Planets I: Low Mass Ratio}

\author{\textsc{
Yoon-Hyun Ryu$^{1}$,
Andrzej Udalski$^{2}$,
Hongjing Yang$^{3,4}$, 
Kyu-Ha Hwang$^{1}$, 
Weicheng Zang$^{5}$,
Yang Huang$^{6}$,
Andrew Gould$^{7}$\\
and\\
Michael D. Albrow$^{8}$, 
Ping Chen$^{9,10}$,
Sun-Ju Chung$^{1}$,
Subo Dong$^{11,12,13}$,
Cheongho Han$^{14}$, 
Youn Kil Jung$^{1,15}$, 
In-Gu Shin$^{3}$, 
Yossi Shvartzvald$^{16}$, 
Jennifer C. Yee$^{5}$, 
Sang-Mok Cha$^{1,17}$, 
Dong-Jin Kim$^{1}$,
Seung-Lee Kim$^{1}$, 
Chung-Uk Lee$^{1}$, 
Dong-Joo Lee$^{1}$,
Yongseok Lee$^{1,17}$, 
Byeong-Gon Park$^{1}$, 
Richard W. Pogge$^{7,18}$\\
(The KMTNet Collaboration)\\
Przemek Mr{\'o}z$^{2}$,
Rados{\l}aw Poleski$^{2}$,
Jan Skowron$^{2}$,
Micha{\l} K. Szyma{\'n}ski$^{2}$,
Igor Soszy\'nski $^{2}$,
Pawe{\l} Pietrukowicz$^{2}$,
Szymon Koz{\l}owski$^{2}$,
Krzysztof Ulaczyk$^{19}$,
Krzysztof A. Rybicki$^{2}$,
Patryk  Iwanek$^{2}$,
Marcin Wrona$^{2,20}$,
Mariusz Gromadzki$^{2}$,
Mateusz J. Mr{\'o}z$^{2}$\\
(The OGLE Collaboration)\\
} }

\affil{$^{1}$Korea Astronomy and Space Science Institute, Daejon
34055, Republic of Korea}

\affil{$^{2}$Astronomical Observatory, University of Warsaw,
Al.~Ujazdowskie~4, 00-478~Warszawa, Poland}

\affil{$^{3}$School of Science, Westlake University, Hangzhou, Zhejiang 310030, China}

\affil{$^{4}$ Department of Astronomy, 
Tsinghua University, Beijing 100084, China}

\affil{$^{5}$ Center for Astrophysics $|$ Harvard \& Smithsonian, 60 Garden
St., Cambridge, MA 02138, USA}

\affil{$^{6}$School of Astronomy and Space Sciences, University of Chinese Academy of Sciences, 100049, Beijing, China}

\affil{$^{7}$Department of Astronomy, Ohio State University, 140 W.
18th Ave., Columbus, OH 43210, USA}

\affil{$^{8}$University of Canterbury, Department of Physical and Chemical 
Sciences, Private Bag 4800, Christchurch 8020, New Zealand}

\affil{$^{9}$Institute for Advanced Study in Physics, Zhejiang University, Hangzhou 310027, China}

\affil{$^{10}$Institute for Astronomy, School of Physics, Zhejiang University, Hangzhou 310027, China}

\affil{$^{11}$Department of Astronomy, School of Physics, Peking University,
      5 Yiheyuan Road, Haidian District, Beijing 100871, People's
Republic of China}

\affil{$^{12}$Kavli Institute of Astronomy and Astrophysics, Peking University,
      5 Yiheyuan Road, Haidian District, Beijing 100871, People's
Republic of China}

\affil{$^{13}$National Astronomical Observatories, Chinese Academy of Science,
20A Datun Road, Chaoyang District, Beijing 100101, China}

\affil{$^{14}$Department of Physics, Chungbuk National University,
Cheongju 28644, Republic of Korea}

\affil{$^{15}$National University of Science and Technology (UST), Daejeon 34113, Republic of Korea}

\affil{$^{16}$Department of Particle Physics and Astrophysics, 
Weizmann Institute of Science, Rehovot 7610001, Israel}

\affil{$^{17}$School of Space Research, Kyung Hee University,
Yongin, Kyeonggi 17104, Republic of Korea}

\affil{$^{18}$Center for Cosmology and AstroParticle Physics, Ohio State University, 191 West Woodruff Ave., Columbus, OH 43210, USA}

\affil{$^{19}$Department of Physics, University of Warwick, Gibbet Hill
Road, Coventry, CV4~7AL,~UK}

\affil{$^{20}$Villanova University, Department of Astrophysics and Planetary
Sciences, 800 Lancaster Ave., Villanova, PA 19085, USA}






\begin{abstract}

  We initiate the systematic search for planets in the 2023 data of the
  Korea Microlensing Telescope Network (KMTNet), focusing on those
  planets found by the KMTNet AnomalyFinder with low preliminary
  estimates of the mass-ratio, $q<2\times 10^{-4}$.
  The 2023 season is the first for which the photometry of all events was
  re-reduced prior to the AnomalyFinder search, potentially increasing its
  sensitivity to planets.  We find three strong low-$q$ planet candidates,
  KMT-2023-BLG-0164 ($q\sim 1.3\times 10^{-4}$),
  KMT-2023-BLG-1286 ($q\sim 1.9\times 10^{-4}$), and
  KMT-2023-BLG-1746 ($q\sim 8\times 10^{-5}$).  
  KMT-2023-BLG-0164 is notable in that the source is projected on a very
  bright ($I=16.0$) foreground star, which is either the planet's host or
  (more likely) a companion to the host.  We obtain a spectrum, finding
  that its mass and distance are $M\sim 1.0\,M_\odot$ and $D\sim 1.5\,\kpc$,
  the latter being the distance of the lens ($D_L$) regardless of whether
  the spectroscopic target is the host or its companion.  We also analyze two
  other candidates, KMT-2023-BLG-0614 and KMT-2023-BLG-1593, which are
  unlikely to enter the statistical sample due to their ambiguous 
  interpretations as possible non-planetary events.
  
\end{abstract}

\keywords{gravitational lensing: micro}

\section{{Introduction}
\label{sec:intro}}

\subsection{From Mass-ratio Function to Planet Mass Function}
\label{sec:massfunction}

The Korea Microlensing Telescope Network (KMTNet) is engaged in
a decade-plus effort to assemble a statistical sample of bound
planets discovered via microlensing observations toward the Galactic
Bulge.
The immediate objective is to measure the
mass-ratio function, that is, the number of planets as a function
of planet/host mass ratio, $q$.  A longer-term objective
is to measure the planet mass function.

The initial results for the mass-ratio function, based on 2016-2019
data, suggest that there are two populations of planets: one with mass
ratios of order those of the Solar System's gas giants (that is,
Jupiter and Saturn), and the other with mass ratios of the Solar
System's solid-dominated planets (that is, Venus, Earth, Uranus, and
Neptune) \citep{massrat}.  This study was based on just 63 (out of the final 2016-2019 sample of 101) planets, so
it is important to confirm (or contradict) its results using a larger
sample.



At present, the longer-term objective of measuring the planet mass
function is less well articulated compared to measuring the planet-host
mass-ratio function,
although \citet{gould22} has initiated such an articulation.
To turn a planet-host mass-ratio
measurement (which are directly inferred from the microlensing light
curves) into a planet-mass measurement, it is of course necessary to
measure the host mass.  In principle, this can be extracted from the
light curve by measuring the Einstein radius $\theta_\e$ and the
microlens parallax, $\pi_\e$, via $M=\theta_\e/\kappa\pi_\e$ where
$\kappa=4G/c^2\au$ \citep{gould92}.  In practice, while $\theta_\e$ is
measured in roughly 2/3 of all planetary events \citep{gould22},
measurements of $\pi_\e$ are made only occasionally.  Although measuring
the mass function does not require knowledge of the mass of every
planet within a statistical sample, determining a substantial majority of
planet masses is essential.

\subsection{Most Planet Masses Will Come From Late-time Imaging}
\label{sec:imaging}

Fortunately, there is a proven method
that will eventually work for most microlensing planets, which was first applied
to OGLE-2005-BLG-169 \citep{ob05169bat,ob05169ben}: simply wait for
the host star to be sufficiently separated from the source star that
it had previously microlensed and take a high-resolution image.  From
the host-star flux (and an assumed mass-luminosity relation), one
infers a relation between the mass of and the distance to the host.
And from the host-source separation $\Delta\theta$ (and the length of
time since the event $\Delta t$), one can infer their relative proper
motion, $\mu_\rel = \Delta\theta/\Delta t$, and hence the Einstein
radius, $\theta_\e = \mu_\rel t_\e$, where $t_\e$ is the Einstein timescale,
which has already been derived from the light curve.  Then, because the
source distance $D_S$ is approximately known, one obtains a second relation
between the lens mass and distance, $\theta_\e^2 = \kappa
M[\au(D_L^{-1}-D_S^{-1})]$.  Combining the two relations yields separate
determinations of $M$ and $D_L$. See Figure~2 of \citet{gould22}.

There are numerous potential
pitfalls to this approach, but \citet{gould22} shows how these can be
systematically resolved.  The main difficulty today is that, even with
the largest existing telescope (Keck), the FWHM in $K$-band is 55 mas.
Thus, it has only been possible to separately resolve the two stars by
waiting for an exceptionally long time, or by looking at systems for
which the host and source have roughly equal brightness.  As a result,
by 2025, only 11 host mass measurements have been made using this
method, the most recent being an event that occurred in
2014.\footnote{These 11 are: OGLE-2005-BLG-169 \citep{ob05169bat,ob05169ben},
OGLE-2012-BLG-0950 \citep{ob120950b}, OGLE-2005-BLG-071
\citep{ob05071c}, MOA-2013-BLG-220 \citep{mb13220b}, MOA-2009-BLG-319
\citep{mb09319b}, MOA-2009-BLG-400 \citep{mb09400b},
OGLE-2013-BLG-0132 \citep{ob130132b}, MOA-2008-BLG-379
\citep{mb08379b}, MOA-2007-BLG-192 \citep{mb07192b},
OGLE-2012-BLG-0563 \citep{ob120563b}, OGLE-2014-BLG-1760
\citep{ob141760b}}.  In particular, there are no such
mass measurements for events from 2016 and later, that is, the period of
the KMT project.

One possible solution would be to continue waiting,
a strategy that we can assess by examining Figure~\ref{fig:event-hist},
  which shows all 24 mass measurements out of the 52 microlens planets
  that were observed during 2005-2014.  These include 20 measurements
  using the two methods just described, that is,
  (1) Late-time high-resolution imaging \citep{ob05169bat,ob05169ben,gould22}
    (``IMAGE'', red, 11); and
    (2) Combining light curve measurements of the microlens parallax ($\pi_\e$)
    and Einstein radius ($\theta_\e$) to yield $M=\theta_\e/\kappa\pi_\e$
    \citep{gould92} (``PAR'', green, 9); as well as a total of four
    measurements made using two other methods,
    (3) Combining light curve measurements of $\pi_\e$ with excess light,
    relative to the source flux derived from the light curve \citep{ob06109b},
    (``PAR+EXC-LIGHT'', cyan, 2); and
    (4) Mass derived from excess-light alone \citep{mb11293b}
    (``EXC-LIGHT'', blue, 2).

\subsection{EELT Will Dramatically Shorten Wait Times}
\label{sec:wait}

From Figure~\ref{fig:event-hist}, one sees that by waiting about
20 years after the KMTNet mass-ratio sample is collected, it
will be possible to apply 
the imaging method
to close to half the sample.
And perhaps after another decade or two, the great majority of the sample
will become accessible.  However, long before this wait time
expires ($\sim 2050$ for the planets from the first full KMTNet
  observing season),
a much larger telescope is expected to come on line: the 39m
European Extremely Large Telescope (EELT),
which completely changes the calculus.

 Among the other three methods, the most promising would appear
  from Figure~\ref{fig:event-hist}
  to be ``PAR'', i.e., $M=\theta_\e/\kappa\pi_\e$,
  with 9/52=17\% mass measurements.  However, due to the very
  different methods of identifying planet candidates in that earlier
  era (real-time by-eye recognition of promising events
  and -- most often -- real-time
  decision for intensive follow-up observations), the conditions for
  making such light-curve-based mass measurements were much more
  favorable than is the case for the current KMTNet sample.  Thus, in
  the future, planet mass measurements will be dominated by the
  imaging method.

\subsection{2016-2019 Sample: Best for EELT First-light Mass-function Measurement}
\label{sec:eeltbest}

KMTNet has already assembled and fully analyzed a planet-host mass-ratio
  sample for 2016-2019, which contains 101 planets.  This sample will be far 
  more suitable for the first-light effort to systematically measure masses
  with EELT than planets from subsequent
years, for example, the next four-year sample from 2021-2024.  In 2030,
hosts of planetary events from 2024 will have separated from the source
star for 6 years, compared to 11 years for 2019 events.  Because EELT
is 3.9 times larger than Keck, its images will have a 3.9 times smaller FWHM
than Keck, or 14 mas.  Thus waiting for 6 years on EELT is equivalent
to waiting 23 years on Keck.  This is only slightly longer than the time span
shown in Figure~\ref{fig:event-hist}.  Of course, the situation will be
somewhat better for 2021-2023.  But for the existing 2016-2019 sample, the
EELT wait intervals will be equivalent to waiting 43 to 55 years on Keck.
Hence, we may expect that the completeness of the mass measurements will be
dramatically better than shown in Figure~\ref{fig:event-hist}.

In this regard, there is a subtle effect that is not conveyed by
Figure~\ref{fig:event-hist}, 
These pre-2015 planets were overwhelmingly recognized
``by eye'', whereas \citet{2019subprime} showed (their Table 17) that
$\sim 40\%$ of the KMT planet sample was found by the machine-based
AnomalyFinder \citep{ob191053,af2}.
And their Figure 20 shows that $\sim 65\%$ of non-caustic
crossing planets (the main source of events without $\theta_\e$ (and $\mu_\rel$)
measurements) come from this $40\%$ of the sample.  These non-$\mu_\rel$ planets
pose a special challenge for the high-resolution follow-up program:
if $\mu_\rel$ is known, then one knows the exact annulus in which to search
for the host, and a non-detection therefore directly yields an upper limit
on the host flux that can be detected in this annulus.  One can then decide
whether this limit is adequate to the goal of measuring the mass function, or
whether additional late-time imaging is warranted.  However, if there is
no $\mu_\rel$ measurement from the microlensing event, a non-detection
could be due to an abnormally low proper motion, 
potentially requiring additional observations. 
The only way to ameliorate
this problem is to minimize the number of non-detections by having the first
observation as late as possible.  This again favors the 2016-2019 sample
over later-year planets.

A third important advantage of the 2016-2019 events is that 10 of the
hosts have mass measurements from {\it Spitzer} satellite-parallax observations,
of which four 
(OGLE-2017-BLG-0406,  
OGLE-2017-BLG-1140,  
OGLE-2018-BLG-0596, and  
OGLE-2018-BLG-0932)  
have giant-star sources \citep{gould22}. Due to the high ($\sim 100$)
to extremely high ($\sim 10^4$) source-host contrast ratios of these
giant-source events, host-source separations of 5 or even more FWHM
are required for host detection.  Because {\it Spitzer} ceased
operation after 2019, it does not provide supplementary information on
the difficult class of giant-source events for any subsequent years.
See Figure~6 from \citet{gould22} for how microlens-parallax information
  can greatly improve a planet-mass measurement relative to imaging-only.

\subsection{Can the 2016-2019 Sample Be Improved by Universal TLC Reductions?}
\label{sec:tlc}

In brief, there are compelling reasons to focus on the 2016-2019 sample
when EELT first becomes available, and this raises the question of whether
anything can be done to increase the size of 
this sample?

The answer is ``plausibly yes'', but doing so would require long
lead times
and substantial effort, so that a cost-benefit analysis of this effort
should be undertaken as soon as possible.

The KMT planets from 2016-2022 were found either by eye or with the
AnomalyFinder by searching the event light curves from
KMT's standard pySIS \citep{albrow09} reductions.  These pipeline reductions
are usually adequate to recognize the planetary anomaly, after which the
data are subjected to tender loving care (TLC) re-reductions, and it is these
that are presented in publications.  However, it is likely that some
planetary anomalies are not recognizable from the pipeline reductions,
in which case the data would not be re-reduced.  With this in mind,
\citet{yang24} developed an end-of-season pipeline by optimizing pySIS.
This does not change the real-time photometry (and so does not impact
the functioning of KMT alerts), but it does enable higher-quality end-of-season
light curves and thereby potentially enables more end-of-season planets.
In principle, one could extract the roughly 3000/year microlensing-event
light curves from the KMT archives, apply the \citet{yang24} TLC pysis,
rerun the AnomalyFinder on the result, and update the 2016-2019 sample.
The human and machine effort would be considerable and would inevitably 
delay processing of subsequent seasons of planets (which would be 100\% new
discoveries).




The general goal of the ``mass-production'' and ``complete-sample''
papers, which contribute to assembling the underlying samples for such
analyses, is to identify and analyze all events that have at least
one ``planetary'' solution, defined as having a mass ratio $q\leq 0.03$.
To this end, we would carry out an initial analysis of all candidates
identified by AnomalyFinder and rereduce the data for all candidates
that have at least one $q<0.06$ solution. However, rereduction is
unnecessary for 2023 data because (for the first time),
the entire season's
light curves were rereduced prior to carrying out the search.
For events with $0.03<q<0.05$, we would report
the result, but not present a full analysis.

Here we begin the first assessment of the likely benefit
of rereducing and reanalyzing the $\sim 12,000$
  light curves from 2016-2019.
The 2023 data have been re-reduced by the \citet{yang24} pipeline, and
the AnomalyFinder has been run both on the old and new reductions.  In
the end, we must evaluate the whole sample of planets identified by both
reductions, which requires the detailed analysis of more than 50 candidates.
In this ``mass production'' paper, we begin this process by systematically
investigating the 11 candidates with preliminary mass-ratio estimates
of $\log q < -3.7$.

The 11 candidates\footnote{
Excluding the five planetary events that have already been published,
KMT-2023-BLG-(0416,1866,1896,2209) \citep{kb230416,kb231866,kb231896,kb232209} 
or are works in progress, KMT-2023-BLG-0382 (Krzysztof et al. in prep).}
that we examined are \hfil\break
KMT-2023-BLG-(0161,0164,0614,0705,0780,1286,1467,1593,1746,1790,1799).
Of these 11, six were rejected because an examination of the images showed
that the candidate planetary anomalies were not of astrophysical origin.
These are KMT-2023-BLG-(0161,0705,0780,1467,1790,1799).  This
leaves five candidates, KMT-2023-BLG-(0164,0614,1286,1593,1746).

It is not our purpose to choose a final sample to measure the mass-ratio
function or mass function, but rather to provide all the information
necessary to do so.  The main (though not the only) reason for excluding
planetary candidates from such samples is that there are alternative
non-planetary solutions, either binary-stellar solutions or
binary-source (1L2S) solutions.  As we will show, two of the five planetary
candidates, KMT-2023-BLG-0614 and KMT-2023-BLG-1593, have such 1L2S solutions.
In our view, these two are unlikely to be included in subsequent,
statistical analyses, but we nevertheless present complete analyses of
these events.

In Section~\ref{sec:method}, we present our general methodology.
In Section~\ref{sec:anal}, we present the analysis of five planetary candidates,
three of which are secure planetary events, while the remaining two
are ambiguous.
We present statistical estimates of the physical parameters of the
three secure planetary systems in Section~\ref{sec:phys}, and
in Section~\ref{sec:conclude}, we briefly summarize our results.

\section{{Methodology}
\label{sec:method}}

\subsection{{Observations}
\label{sec:obs}}

All of the planetary candidates in this paper were identified in AnomalyFinder
searches of KMT events that were announced by the KMT AlertFinder
\citep{alertfinder} as the 2023 season progressed.
KMTNet observes
from three 1.6m telescopes \citep{kmtnet} that are equipped with
$(2^\circ\times 2^\circ)$ cameras at CTIO in Chile (KMTC),
SAAO in South Africa (KMTS), and SSO in Australia (KMTA), mainly
in the $I$ band, with 60 second exposures, but with 9\% of the 
observations in the $V$ band.
The data were initially reduced using pySIS \citep{albrow09}, a form of
difference image analysis (DIA, \citealt{tomaney96,alard98}).
As described in Section~\ref{sec:intro}, all light curves were reduced
using the TLC pySIS pipeline of \citet{yang24} prior to applying
AnomalyFinder, and
we manually examined the images during the anomaly
to rule out image artifacts as a potential explanation for the
light-curve deviations.

Two of the five events reported here were also alerted by other
surveys: KMT-2023-BLG-1286 (OGLE-2023-BLG-0803) was alerted by OGLE, and
KMT-2023-BLG-0164 (OGLE-2023-BLG-0116, MOA-2023-BLG-107)
was alerted by both OGLE and MOA.  We include the
OGLE data in the analysis of both of these events.  However, we
do not include the MOA data because the scatter is too large to help
constrain either the anomaly itself or the parallax measurement of
KMT-2023-BLG-0164.

The OGLE data were obtained using a 1.4 ${\rm deg}^2$ camera on the
1.3m Warsaw telescope at Las Campanas Observatory in Chile and were
primarily 100 second exposures in standard Cousins $I$-band,
with a small minority of $V$-band observations.

 The photometric error bars were renormalized according to the prescription
  of Section 2.1 of \citet{mb11293} on an observatory-by-observatory basis.
  Briefly, the original errors, $\sigma_i$, are expressed (or re-expressed)
  in magnitudes, and the renormalization is expressed in terms of two parameters
  $(k,e_\min)$ as $\sigma^\prime=k\sqrt{\sigma^2 + e_\min^2}$. First an initial
  model is found.  Then, $k=1$ is held constant while $e_\min$ is adjusted until
  the cumulative distribution of $\chi^2$ (sorted by magnification $A_i$)
  is approximately a straight line, and the total $\chi^2_{\rm tot}(e_\min)$
  is noted for this choice of $e_\min$.  Finally, $k$ is set for each data by
  $k=\sqrt{N_{\rm data}/\chi^2_{\rm tot}(e_\min)}$, where $N_{\rm data}$ is the number
  of data points for that observatory, so that $\chi^2$ per degree of freedom
  is approximately unity for each observatory separately and for all
  observatories together.

As far as we are aware, there were no follow-up observations, either
by follow-up groups or the surveys themselves.

Table~\ref{tab:names} gives the event names, 
observational cadences $\Gamma$, discovery dates and sky locations.
Because each of the five events was first discovered by KMT, we use
the KMT name for all five, following the standard naming convention.

{\subsection{{Light-curve Modeling}
\label{sec:anal-preamble}}

Our approach to analyzing events is identical to that described
in several earlier mass-production papers that report collections
of AnomalyFinder planets and planet candidates.  Here we repeat,
essentially verbatim, the ``preamble'' to the light-curve
analysis of one these papers, \citet{kb210712}, in order to permit
the reader direct access to the relevant formulae and descriptions.

All of the events in this paper can be analyzed to a first approximation
as 1L1S events, which are characterized by
three \citet{pac86} parameters,
$(t_0,u_0,t_\e)$, that is, the time of lens-source closest approach, the impact
parameter (normalized to the Einstein radius, $\theta_\e$), and the
Einstein-radius crossing time
\begin{equation}
t_\e = {\theta_\e\over\mu_\rel};
\qquad
\theta_\e =\sqrt{\kappa M \pi_\rel},
\qquad
\kappa\equiv {4G\over c^2\,\au}\simeq 8.14{\mas\over M_\odot}.
\label{eqn:thetae}
\end{equation}
Here, $M$ is the mass of the lens, $(\pi_\rel,\bmu_\rel)$ are the lens-source
relative parallax and proper motion, $\mu_\rel\equiv |\bmu_\rel|$,
and $n$L$m$S means ``$n$ lenses and $m$ sources''.

A 2L1S model always requires at least three additional parameters
$(s,q,\alpha)$, that is, the separation (normalized to $\theta_\e$)
and mass ratio of the two lens components, as well as the angle between
the line connecting these and the direction of $\bmu_\rel$.  
If there are finite-source effects due to the
source approaching or crossing caustic structures that are
generated by the lens, then
one must also specify $\rho\equiv \theta_*/\theta_\e$, where $\theta_*$
is the angular radius of the source.

For 1L2S models, which can generate featureless bumps that can be mistaken
for 2L1S ``planets'' \citep{gaudi98},  the minimal number of parameters is 6, 
including $(t_{0,1},t_{0,2})$ and $(u_{0,1},u_{0,2})$
for the two times of closest approach and impact parameters, respectively,
$t_\e$ for the Einstein timescale, and $q_F$, that is, the flux ratio
of the two sources in the $I$-band.  In many cases, one or both of the two 
normalized source radii must be specified, $\rho_1=\theta_{*,1}/\theta_\e$ and
$\rho_2=\theta_{*,2}/\theta_\e$.  More complex models involving orbital
motion of the binary-source system may also be needed.

If the microlens parallax effect can be detected (or constrained),
then one should include the microlens parallax vector 
\citep{gould92,gould00,gould04},
\begin{equation}
\bpi_\e = {\pi_\rel\over\theta_\e}\,{\bmu_\rel\over\mu_\rel},
\label{eqn:bpie}
\end{equation}
which is normally expressed in equatorial coordinates
$\bpi_\e=(\pi_{\e,N},\pi_{\e,E})$.  In these cases, one usually must also 
fit, at least initially, for the first derivatives in time of the lens 
angular position, $\bgamma=[(ds/dt)/s,d\alpha/dt]$, because
$\bpi_\e$ and $\bgamma$ can be correlated or even degenerate.
In these cases, we restrict such fits to $\beta<0.8$, where
\citep{eb2k5,ob05071b},
\begin{equation}
\beta\equiv {\kappa M_\odot{\rm yr}^2\over 8\pi^2}{\pi_\e\over\theta_\e}
\gamma^2\biggl({s\over \pi_\e +\pi_s/\theta_\e}\biggr)^3,
\label{eqn:beta}
\end{equation}
and where $\pi_S$ is the source parallax.

In our initial heuristic analyses, we often predict $s^\dagger_\pm$
and $\alpha$ from the morphology of the light curve
\citep{kb190253,kb211391},
\begin{equation}
s^\dagger_\pm = {\sqrt{4 +u_\anom^2}\pm u_\anom\over 2};
\qquad \tan\alpha = {u_0\over \tau_\anom},
\label{eqn:sdagger}
\end{equation}
under the assumption that the anomaly occurs when the source crosses
the binary axis.  Here, $u_\anom = \sqrt{\tau_\anom^2 + u_0^2}$,
$\tau_\anom = (t_\anom - t_0)/t_\e$, $t_\anom$ is the midpoint of the anomaly,
and the ``$\pm$'' refers to major/minor image perturbations.  If there
are two solutions, with normalized separation values $s_\pm$, 
as often occurs (see \citealt{zhang22} for a theoretical discussion of such 
degeneracies), we expect that the empirical quantity $s^\dagger=\sqrt{s_+ s_-}$
(without subscript)
will be approximately equal to the subscripted quantity from 
Equation~(\ref{eqn:sdagger}).

We also note that \citet{kb190253} showed that for minor-image perturbations,
the mass ratio can be estimated by
\begin{equation}
q = \biggl({\Delta t_{\rm dip}\over 4\, t_\e}\biggr)^2
{s\sin^2\alpha\over u_{\rm anom}} 
= \biggl({\Delta t_{\rm dip}\over 4\, t_\e}\biggr)^2
{s\over |u_0|}|\sin^3\alpha| ,
\label{eqn:qeval}
\end{equation}
where $\Delta t_{\rm dip}$ is the full width of the depression in the
light curve relative to the 1L1S model.

Finally we sometimes report the ``source self crossing time'',
$t_* \equiv \rho t_\e$.  We note that this is a derived quantity and is
not fit independently.

\subsection{{Source Properties}
\label{sec:cmd}}

Our evaluation of the source properties exactly follows the goals
and procedures of previous ``mass production'' papers.
In this introduction, we therefore repeat, essentially verbatim, the
corresponding introduction from Section 4.1 of \citet{kb210712}.  
In particular (as in Section~\ref{sec:anal-preamble}), we seek to
document all procedures and notation without referring the reader to
other papers.

We analyze the color-magnitude diagram (CMD) of each
event, primarily to measure $\theta_*$ and so to determine
\begin{equation}
\theta_\e = {\theta_*\over\rho};\
\qquad
\mu_\rel = {\theta_\e\over t_\e}.
\label{eqn:thetae_murel}
\end{equation}
We follow the method of \citet{ob03262}.  We first find the offset
of the source from the red clump
\begin{equation}
\Delta[(V-I),I] = [(V-I),I]_S - [(V-I),I]_{\rm cl}.
\label{eqn:deltacmd}
\end{equation}
We adopt  $(V-I)_{\rm cl,0}=1.06$ from \citet{bensby13} and evaluate
$I_{\rm cl,0}$ from Table~1 of \citet{nataf13}, based on the Galactic
longitude of the event, which yields the  dereddened color and magnitude
of the source,
\begin{equation}
[(V-I),I]_{S,0} = [(V-I),I]_{\rm cl,0} + \Delta[(V-I),I].
\label{eqn:cmd0}
\end{equation}
Next, we transform from $V/I$ to $V/K$ using the $VIK$ color-color 
relations of \citet{bb88}, and we apply the color/surface-brightness relations
of \citet{kervella04a,kervella04b} to obtain $\theta_*$.
After propagating the
measurement errors, we add 5\% to the error in quadrature to take account
of systematic errors due to the method as a whole.

To obtain $[(V-I),I]_S$, we always begin with pyDIA reductions \citep{pydia},
which put the light curve and field-star photometry on the same system.
With two exceptions (see below), we determine $(V-I)_S$ by regression of the 
$V$-band data 
on the $I$-band data, and we determine $I_S$ by regression of the $I$-band
data on the best-fit model.  For three of the five events analyzed in this
paper, there is calibrated OGLE-III field-star photometry \citep{oiiicat}.
For these three cases, we transform $[(V-I),I]_S$ to the OGLE-III
system.  For the two remaining cases, we work in the instrumental 
KMT pyDIA system.

For KMT-2023-BLG-0614 and KMT-2023-BLG-1593, the sources are too faint in
the $V$ band to measure the source color from the light curve.
We therefore employ a different technique, as described in
Sections~\ref{sec:cmd-kb230614} and \ref{sec:cmd-kb231593}.

The CMDs are shown in Figure~\ref{fig:allcmd}.

The elements of these calculations are summarized in
Table~\ref{tab:cmd}.   
In all cases, the source
flux is that of the best solution.  Under the assumption of fixed
source color, $\theta_*$ scales as $10^{-\Delta I_S/5}$ for the other solutions,
where $\Delta I_S$ is the difference in source magnitudes, as given
in the Tables of Section~\ref{sec:anal}.
The inferred values (or limits
upon) $\theta_\e$, and $\mu_\rel$ are given in the 
individual events subsections below, where we also discuss other
issues, when relevant.

\subsection{{Bayesian Analysis}
\label{sec:bayes}}

None of the three planets and two candidate planets have sufficient
information to precisely specify the host mass and distance.
For any given solution, we can incorporate
Galactic-model priors into standard Bayesian techniques to obtain estimates
of the host mass $M_{\rm host}$ and distance $D_L$, 
as well as the planet mass $M_{\rm planet}$ 
and planet-host projected separation $a_\perp$.
See \citet{ob180567} for a description of the Galactic model and Bayesian
techniques.  However, in most cases we still have to decide how to combine these separate estimates into a single ``quotable result''.

 We show the posterior distributions for the host mass and distance in
the individual subsections below, and

we tabulate the posterior estimates and uncertainties for $M_{\rm host}$,
$M_{\rm planet}$, $D_{L}$, and $a_\bot$ in Section~\ref{sec:phys}.


\section{{Analysis of Selected Events from the KMTNet 2023 Season}
\label{sec:anal}}

{\subsection{{KMT-2023-BLG-0164}\label{sec:anal-kb230164}}}

Figure~\ref{fig:0164lc} shows an otherwise standard 1L1S light curve
with \citet{pac86} parameters $(t_0,u_0,t_\e)=(51.23,0.072,75\,{\rm day})$,
superposed by a $\Delta t_{\rm dip}=3.0$ day dip, centered at $t_\anom=63.0$.

{\subsubsection{{Heuristic Analysis}\label{sec:heuristic-kb230164}}}

These parameters imply $\tau_\anom=0.157$, $u_\anom=0.173$, and thus
\begin{equation}
\alpha = 204.6^\circ;
\qquad
s^\dagger_- = 0.917;
\qquad
q = 9\times 10^{-5}.
\label{eqn:kb23_0164_heur}
\end{equation}

{\subsubsection{{Static Analysis}\label{sec:static-kb230164}}}

The grid search on the $(s,q)$ plane returns two solutions,
whose refinements with all parameters set free are 
shown in
Table~\ref{tab:0164parms}. 
We see that $\alpha=(204.8^\circ,204.7^\circ)$
in the two solutions, 
while $s^\dagger\equiv \sqrt{s_+ s_-} = 0.913$, which are both
extremely close to the heuristic predictions.  The two values of $q$
are also in qualitative agreement with Equation~(\ref{eqn:kb23_0164_heur}).
Because the anomaly takes the form of shallow dip, we expect only
upper limits on the normalized source flux $\rho$.
Qualitatively, we expect $\rho t_\e \la (\Delta t_{\rm dip}/2)|\sin\alpha|$,
or $\rho\la 8\times 10^{-3}$.
As seen from Table~\ref{tab:0164parms},
this expectation
proves qualitatively correct, but
we defer quantitative discussion of the limit on $\rho$ until
Section~\ref{sec:parallax-kb230164}.

Within the context of static models, the degeneracy between the inner
and outer solutions is often broken when the source trajectory
is close to planet-host axis, as for example, $|\sin\alpha| \simeq 0.4$
in the present case.  This is because the source trajectory is
substantially closer (or farther) from the vertically symmetric
caustic at the beginning of the anomaly than at the end, so that the
anomaly is not time-symmetric about its center.  See the
caustic-geometry panels in Figure~\ref{fig:0164lc}.
Such would appear to be true for KMT-2023-BLG-0164
for which the inner model is favored by $\Delta\chi^2=50$.
However, because the event is relatively long, $t_\e\sim 75\,$day,
both parallax and orbital motion can play a significant role, and
these additional parameters may absorb some or all of the distinguishing
features of the two light-curve models.  Therefore, we investigate
both the inner and outer solutions
in the parallax analysis, to which we now turn.

{\subsubsection{{Parallax Analysis}\label{sec:parallax-kb230164}}}

As is almost always the case (except for some extremely long
events), for each standard solution,
there are two parallax solutions, which differ by the
sign of $u_0$.  See \citet{smp03} for an explanation and
Figure~4 of \citet{gould04} for the sign convention.  Hence, there
are a total of four solutions, which
are summarized in
Table~\ref{tab:0164par_parms} and illustrated in
Figure~\ref{fig:0164lc}.  
As described in Section~\ref{sec:anal-preamble}, we simultaneously fit for the
first derivatives of the planet position due to its orbital motion
$\bgamma=[(ds/dt)/s,d\alpha/dt]$.  Table~\ref{tab:0164par_parms} shows
that in each of the four solutions, $\bpi_\e$ is very well localized,
and this is further illustrated by Figure~\ref{fig:0164parallax}.
In both $u_0>0$ solutions,
$\bpi_{\e,+} = (\pi_{\e,N},\pi_{\e,E})\simeq (0.52,0.16)$, while
in both $u_0<0$ solutions,
$\bpi_{\e,-} \simeq (0.71,0.15)$.
That is, the magnitude and direction (north through east), $\psi_\pi$, are
$(\pi_\e,\psi_\pi)_+ = (0.54,18^\circ)$ and
$(\pi_\e,\psi_\pi)_- = (0.73,12^\circ)$, respectively.
These angles are close to the Galactic plane which is inclined by
$30^\circ$ relative to Equatorial north.

The large value of $\pi_\e$ implies that the lens almost certainly
lies in foreground disk unless it is an extreme substellar object.  That is,
$\pi_\rel=\pi_\e^2\kappa M= 0.3\,\mas(\pi_\e/0.6)^2(M/0.1\,M_\odot)$.
In fact, we will show in Section~\ref{sec:cmd-kb230164}}
that the source is blended with a bright foreground star that is
almost certainly either the lens itself or a stellar companion to it.
The Gaia proper motion for this object (and so, in either case, the
proper motion of the lens) is
$\bmu_l = (\mu_{l,N},\mu_{l,E}) = (+0.2,-1.9)\,\masyr$.  Thus, if the
source has a typical bulge proper motion, that is, similar to
SGR A*, $\bmu_{\rm sgrA*} = (-5.6,-2.7)\,\masyr$, then the
lens-source proper motion would be
$\bmu_\rel\sim(5.8,0.8)\,\masyr$, implying
$\mu_\rel \sim 5.9\,\masyr$ and $\psi \sim 8^\circ$.
Thus, the resulting direction is very similar
to the two values of $\psi_{\pi,\pm}$ derived above.

With these broad expectations in mind, we now turn to the constraints
on $\rho$. Figure~\ref{fig:0164rho} shows a scatter plot
of $\Delta\chi^2$ versus $\rho$ from the Monte Carlo fit.  As also
indicated in Table~\ref{tab:0164par_parms}, for the outer solutions
$\rho<2.3\times 10^{-4}$ at the $3\,\sigma$ level.  Such low values would
be in tension with the previous arguments because
\begin{equation}
  \rho \equiv {\theta_*\over\theta_\e} = {\theta_*\over \mu_\rel t_\e}=
  7.4\times 10^{-4}\biggl({\mu_\rel\over 6\,\masyr}\biggr)^{-1},
  \label{eqn:0164rho}
\end{equation}

  where we have incorporated the measurement, $\theta_*=0.92\,\muas$, from
  Table~\ref{tab:cmd}.   

That is, given the characteristic
expected proper motion derived above, $\rho$ would be 3.2 times larger
than its $3\,\sigma$ limit.

This tension could be resolved by one of several paths.  First, the
source could be moving at $13\,\masyr$ relative to SGR A* (approximately
in the direction of negative Galactic rotation),
which is extremely unlikely ($p\sim 10^{-5}$).
Second, the $\chi^2$ deviation could be
greater than $3\,\sigma$, but according to Figure~\ref{fig:0164rho}, it
would have to be much greater.  Third, one of the two inner solutions could be
correct.

The last alternative seems obviously correct because
comparison of Equation~(\ref{eqn:0164rho}) and
Table~\ref{tab:0164par_parms} shows that they are compatible at the
$3\,\sigma$ level.  Moreover, they are only mildly in tension at
$2\,\sigma$ (or even $1\,\sigma$),
which could easily be accommodated by normal motion
of the source within the bulge.

Thus, the most likely interpretation is that one of the two inner
solutions is correct, and that $\mu_\rel\sim 6\,\masyr$, which
would imply
\begin{equation}
   M = 0.25\,M_\odot
  \biggl({\mu_\rel\over 6\,\masyr}\biggr)
  \biggl({\pi_\e\over 0.6}\biggr)^{-1};\qquad
  \pi_\rel = 0.74\,\mas
  \biggl({\mu_\rel\over 6\,\masyr}\biggr)
  \biggl({\pi_\e\over 0.6}\biggr).
  \label{eqn:md0164}
\end{equation}
That is, at the fiducial values of Equation~(\ref{eqn:md0164}), the
lens would be a middle M dwarf at $D_L\sim 1.2\,\kpc$.

Unfortunately, such a lens would be far too faint
($I\sim 20$) to account for the blended light ($I\sim 16$), which
we have identified as being either the lens itself or its companion.
One obvious solution is that blended light is the companion to the lens.
This would still be in some tension with the Gaia measurement
because the lens parallax would still be $\pi_l = 0.8\,\mas$,
whereas Gaia reports $\pi_{\rm Gaia} = 0.28\pm 0.17\,\mas$.  However,
Gaia also reports a large ``RUWE'' value, 2.25, which parameterizes
the excess scatter of the data about the 5-parameter astrometric
solution.  Such large RUWE values imply substantially less
reliability for results reported by Gaia (even though Gaia
renormalizes its error bars by the RUWE), as studied by
\citet{2018subprime}.  Note, in particular, their discussion of the
four events listed in their Table~14 that have large RUWE values,
$1.75\leq {\rm RUWE}\leq 2.15$.  At RUWE=2.25, the excess astrometric
scatter of KMT-2023-BLG-0164 is bigger than any of the four documented
problem cases.

In fact the large RUWE may tend to support the conjecture that
the bright Gaia star is a companion to the lens rather than the
lens itself.  The main cause of high RUWE is the presence of an
unresolved star near or somewhat beyond the Gaia point-spread function (PSF)
of the target star, which is asymmetric and
approximately $(135\times 375)\,\mas$ at $I$ band.  When the short axis
of the mirror is aligned with the position angle of the interloper, its light
maximally contributes to the mean astrometric position of the measurement,
whereas when the long axis of the mirror
is so aligned, the interloper contributes much less.
Thus, a binary companion at, for example, a separation of 150 mas, would
induce excess RUWE scatter through this mechanism.  This would represent
a projected separation of 180 AU in the current example.

This is a plausible scenario, but it is also possible that the proper motion
is substantially faster than assumed in Equation~(\ref{eqn:md0164}),
for example, $\mu_\rel=12\,\masyr$.  While this would require a mildly
improbable source proper motion ($p\sim 5\%$), the resulting mass,
$M\sim 0.5\,M_\odot$, and distance, $D_L\sim 0.6\,\kpc$, would yield
a lens apparent magnitude of $I\sim 16$, which is approximately what
is observed.  Of course, the larger lens parallax $\pi_l\sim 1.6\,\mas$
would be in even greater conflict with the Gaia measurement, but again,
we cannot place too much trust in such a high-RUWE measurement.

{\subsubsection{{Robustness Tests}\label{sec:robust-kb230164}}}

We note that the inferences just derived rest critically on two
broad physical pieces of information.  First, the parallax is
relatively large, $\pi_\e \sim 0.6$, and second, the lens proper motion
$\mu_l$ is small compared to the difference between the mean motion
of disk and bulge stars, that is $\mu_l\ll |\bmu_{\rm sgrA*}|\sim 6\,\masyr$,
which implies, roughly speaking, $\mu_\rel \sim \mu_{\rm sgrA*}$.

We now examine whether these two pieces of information are really robust.
Regarding $\pi_\e$, the main questions are: could the putative $\pi_\e$
be of non-astrophysical origin (that is, the result of systematic
errors in the measurement), or could it be due to astrophysical
effects that are unrelated to the lens distance?

To address the first of these, we examine separately data from the four
observatories, $n=1,2,3,4 \rightarrow$ (KMTC, KMTS, KMTA, OGLE).  We start
by excluding the part of the light curve that is significantly affected
by the anomaly ($60< {\rm HJD}^\prime <70$) and fitting the remaining light
curve to a five parameter 1L1S model ($t_0,u_0,t_\e,\pi_{\e N},\pi_{\e E}$),
thus deriving the best fit $(a_1,a_2)^n = \bpi_\e$ and the parallax covariance
matrix, $c^n_{ij}$, for each of the four
observatories\footnote{Of course, a proper measurement of $\bpi_\e$
requires inclusion of the planetary parameters, and hence of the anomalous
data.  However, here we are searching for evidence of systematic errors via
discrepancies between independent data sets, not $\bpi_\e$ itself.  And
this comparison can only be made for the 1L1S light curve because the
OGLE data do not have sufficient coverage, by themselves, to derive the
planetary parameters.}.  Here
HJD$^\prime = {\rm HJD}-2460000$.  For each of
the six resulting $(n,m)$ pairs we calculate
$\chi^2_{nm} = \sum_{ij} (a^n-a^m)_i(a^n-a^m)_j(c^n+c^m)^{-1}_{ij}$.  We find
that the three pairs of (KMTS, KMTA, OGLE) are all perfectly consistent
$(\chi^2\la 1)$, whereas the other three pairs (each containing KMTC)
each show indications of tension.  We therefore consider two complementary
data sets, namely KMTC-only and Complement(KMTC) = (KMTS+KMTA+OGLE).
And, for completeness, we also consider each of the other three data
sets and its complement.  To form the complement, we calculate
$b^{{\rm comp},n} = \sum_{m\not=n} b^m$, $d^{{\rm comp},n} = \sum_{m\not=n} d^m$,
$b^m=(c^m)^{-1}$, $d_i^m=\sum_j b_{ij}^m a_j^m$,
$c^{{\rm comp},n} = (b^{{\rm comp},n})^{-1}$,
$a^{{\rm comp},n}_i = \sum_j c^{{\rm comp},n}_{ij}d^{{\rm comp},n}_j$.
The results are shown in Table~\ref{tab:0164partest}.  The main
point is that while the two solutions (KMTC and Complement(KMTC))
have different best-fit
values of $\bpi_\e$, these values are not qualitatively different.  Hence,
even if one were to conclude that, for example, KMTC should excluded
as ``corrupted'' by systematic errors, the arguments given above
would not be qualitatively affected.  Moreover, we find that
the same test of difference in data sets given above yields
$\chi^2_{\rm KMTC,Complement}=5.7$ . This would appear to have a (two-dimensional)
false-alarm probability of $p=\exp(-\chi^2/2)=6\%$, which would itself
be marginal evidence for a systematic disagreement.  But we must
also consider that we have effectively considered four ``trials'', that is,
four configurations of
the data (excluding each of the four observatories). Hence, the
significance of this ``conflict'' is $<1\,\sigma$.  Therefore, in
our further analysis we do not exclude any of the data.

The only known type of astrophysical ``contamination'' of the
$\bpi_\e$ measurement would be ``xallarap'', that is, distortions
of the light curve due to orbital motion of the source about an unseen
companion \citep{griest-hu,han97}.  In principle, xallarap can
perfectly mimic any real parallax effect provided that the 7 Kepler
parameters of the binary-source orbit match those of Earth, including,
in particular that the orientation and phase of the orbit mimic the
parallax ellipse induced by Earth's motion about the Sun.  However,
the more precisely that xallarap can be measured, and the tighter
the resulting ``agreement'', the less probable is the xallarap interpretation
\citep{poindexter05}.  As usual, we do not fit for all seven Kepler
parameters, but assume a circular orbit.  

Figure~\ref{fig:0164xalper} shows $\chi^2$ as a function of assumed
xallarap period $P$, after marginalizing over all possible orbit-orientation
parameters, $(\alpha,\delta)$.  As expected if the actual cause of
the light-curve distortions is parallax (that is, Earth's motion about
the Sun), the best-fit xallarap period should be $P=1\,$yr.
Relative to the best
parallax fit (Table~\ref{tab:0164par_parms}), there is an improvement
of $\Delta\chi^2\sim 6$.  Under the assumption of Gaussian statistics
(which is optimistic for microlensing), this can occur with probability
$p=\exp(-\Delta\chi^2/2)=5\%$ for two degrees of freedom (dof), which
reflects only mild tension.

Nevertheless, we conduct a further test: we allow the xallarap orientation,
$(\alpha,\delta)$, to vary freely, while holding the xallarap period fixed at
$P=1\,$yr.
In an ideal world, the best fit would be at the ecliptic coordinates of the
event
$(\alpha,\delta)=(\lambda,\beta)\rightarrow(269.8^\circ,-8.3^\circ)$.
We find a best fit of $(\alpha,\delta)=(262^\circ,-5^\circ)$, which is
good agreement.  As a matter of due diligence, we repeat the 1-year xallarap
fit with the eccentricity and time of periastron fixed at the terrestrial
values, but (as one might expect), this does not materially affect the result.


Finally, we examine whether the apparently low Gaia $\mu_l=1.9\pm 0.2\,\masyr$
could be an artifact of whatever measurement problems
caused the high excess-scatter parameter, RUWE $=2.25$.  Recall that we
regarded the Gaia parallax measurement, $\pi_l=0.28\pm 0.17\,\mas$,
as not providing a strong constraint due to this same high RUWE.
And, of course, the Gaia parallax and proper-motion derive from the same
5-parameter fit.

However, there are three reasons that the same argument does not apply to
$\bmu_l$.  First, we expect that while the excess scatter can reflect
systematic errors that are larger than the reported statistical errors
(even after renormalization by the RUWE factor that is already included
in the Gaia result), these errors are still most likely caused by
a semi-random process.  That is, they are due to excess light entering
the aperture in semi-random amounts due to semi-random orientations
of the telescope.  Therefore, it is plausible to represent these errors
as a further inflation of the reported statistical errors, say, to
be conservative, by a factor five.  However, whereas the effective errors
of the parallax measurement would then be $\pi_l=0.28\pm 0.85\,\mas$
(thus, for example, allowing any $D_l \ga 1\,\kpc$), the same inflation would
lead to $\mu_l = 1.9\pm 1.0\,\masyr$, which would not qualitatively
change the argument given in Section~\ref{sec:parallax-kb230164} at all.

The second reason that the same argument does not apply is that
the parallax is a scalar while the proper motion is a vector.  That is,
as an exercise, let us assume that the true proper motion is much larger
but has been corrupted by a random systematic error of arbitrary size
(even though it is difficult to imagine the physical process that would then
yield the low observed scatter of the fit).  This systematic error would
be in a random direction relative to the true proper motion of the object.
However, there is only a relatively small range of angles for which
this would lead to a substantial reduction of the reported $\mu_l$
relative to the ``true'' (that is, putative) one.

The third reason is that we have an independent check on the
  Gaia proper motion from OGLE.  Specifically, in Gaia DR3 \citep{gaia},
  the blend (or technically, the blend augmented by $\sim 4\%$ of its light
from the source) is listed as having
\begin{equation}
  \pi_b = 0.277\pm 0.169,\quad
  \bmu_b=(\mu_{b,E},\mu_{b,N}) = (-1.91,+0.18)\pm (0.19,0.13)\,\masyr,
  \label{eqn:gaia}
\end{equation}
with a correlation coefficient of 0.356, and a RUWE=2.25 parameter.
We have already discussed the implications of the high RUWE parameter
above 
and, and we do not repeat that discussion here.
On the other hand, based on its own internal measurement, OGLE reports
the blend proper motion (but not the parallax) as 
\begin{equation}
  \bmu_b=(\mu_{b,E},\mu_{b,N}) = (-2.05,-0.20)\pm (0.07,0.05)\,\masyr,
  \qquad (OGLE).
  \label{eqn:ogle}
\end{equation}
Unlike Gaia, OGLE is ground-based rather than space-based, but it does
have two major advantages relative to Gaia: a far larger number of data points
and two-dimensional imaging, which is far less susceptible to contamination
by bright neighboring stars.  As a result, the formal errors are
roughly three times smaller and the poorly understood systematics signaled
by the large Gaia RUWE number are less of a concern.  While the
$0.4\,\masyr$ difference between Equations~(\ref{eqn:gaia}) and (\ref{eqn:ogle})
is formally significant (albeit marginally), it is very small compared
to the characteristic, $\sim 3\,\masyr$, dispersion of bulge
sources, which dominates the dispersions in the lens-source relative
proper motion.

Thus, the possibility that $\mu_l$ is substantially larger
than the value reported by Gaia, is effectively ruled out.


{\subsubsection{{CMD Analysis}
    \label{sec:cmd-kb230164}}}

The main issue that requires special care for this event is the blended light.
As can be seen in Figure~\ref{fig:allcmd}, the blend is 3.5 mag
brighter than the source and lies on the ``foreground'' main sequence.
That is, it suffers substantially less extinction and reddening due
to a smaller dust column, compared to the majority of stars in the
CMD, which lie in the bulge.  These characteristics imply that the blend 
dominates the ``catalog star'' that was monitored in order to find the
microlensing event, but is unrelated to the ``source star'' whose light was
impacted by the event.  In particular, the blend is not a companion to the
source, so it must be either the lens itself, a companion to the lens, or
an ambient field star that happens to be projected close to the
event.  The measured offset between the event and the baseline object
(which is overwhelmingly dominated by the blend) is
$\Delta\theta=70\,\mas$ as measured by KMT and
$\Delta\theta=44\,\mas$ as measured by OGLE.  Given
that the surface density of foreground stars brighter than the blend
is $n=3.0\,{\rm arcmin}^{-2}$, the probability of such a chance
projection, even using the larger of these two values
is $p=n\pi\Delta\theta^2=1.3\times 10^{-5}$.  Thus, the blend
is almost certainly either the lens itself or a companion to the lens.


In preparation for Section~\ref{sec:phys-kb230164},
we note that the proper motion in Equation~(\ref{eqn:gaia})
can be expressed in Galactic coordinates as
\begin{equation}
  \bmu_b=(\mu_{b,l},\mu_{b,b}) = (-0.89,+1.75)\pm (0.17,0.16)\,\masyr,
  \label{eqn:gaiagal}
\end{equation}
with a correlation coefficient of $-0.48$.  
In that section
we will make use of the full inverse covariance matrix
\begin{equation}
  b_{\mu,\rm gaia}=
  \begin{pmatrix}
    43.37 & 23.02\\
    23.02 & 53.77
  \end{pmatrix}
  \,(\masyr)^{-2}.
  \label{eqn:bmugaia}
\end{equation}

{\subsubsection{{Bayesian Analysis\label{sec:phys-kb230164}}}

The Bayesian analysis of KMT-2023-BLG-0164 is subject to four
separate constraints, which differ for each of the four solutions
shown in Table~\ref{tab:0164par_parms}.  Two of constraints
come directly from that table: on $t_\e$ and $\bpi_\e$.  For both of these
parameters, the central values are those listed in the table, 
but for $\bpi_\e$, we express the error ellipse as an inverse
covariance matrix, $b$, and evaluate $\chi^2=\sum_{ij} a_i b_{ij} a_j$
where $a_i$ is the difference between the measured parallax and
the random realization of the Galactic model.

The third constraint is the limit on $\rho$, which we characterize
as an envelope function based on Figure~\ref{fig:0164rho}.
That is, for each event realization, we evaluate $\theta_\e$
based on the mass of the host and the distances to the source
and lens.  We then evaluate $\rho=\theta_*/\theta_\e$ using the
central value of $\theta_*$ given in Table~\ref{tab:cmd}, and finally
we evaluate $\chi^2$ based on the envelope function derived from
Figure~\ref{fig:0164rho}.  We then repeat this for 17 values of
$\theta_*$, weighting each according to the error bar, $\sigma(\theta_*)$,
given in Table~\ref{tab:cmd}, that is,
$\theta_{*,k} = \theta_{*,{\rm best}} + (k/2)\sigma(\theta_*),\ (k=-8,\ldots,+8)$.

The fourth constraint is on the lens proper motion.  As discussed in
Sections~\ref{sec:robust-kb230164} and \ref{sec:cmd-kb230164},
the Gaia star is almost certainly either the lens itself or a companion
to it, and in either case its proper motion is the same as that of the lens.
However, due to the high RUWE value, the Gaia measurement error is almost
certainly underestimated.  To take account of this, we adopt
Equation~(\ref{eqn:gaiagal}) for the central value of the lens proper
motion (in Galactic coordinates), but we degrade the error ellipse of
this determination relative to Equation~(\ref{eqn:bmugaia}) by
setting $b_{\mu,l}=b_{\mu,\rm gaia}/10$.  We note that the
OGLE measurement that was reported in Equation~(\ref{eqn:ogle}),
which is much more precise, is then consistent with this inflated error
at the $1\,\sigma$ level.

Finally, each random realization  (including the 17 separate realizations
of $\theta_*$) is weighted by $\exp(-\chi^2/2)$, where
$\chi^2 = \chi^2_{t_\e} + \chi^2_{\bpi_\e} + \chi^2_{\rho}+ \chi^2_{\bmu,l}$.

The posterior distributions of the lens mass and distance
are shown in Figure~\ref{fig:0164bayes}.
As we tabulate in Section~\ref{sec:phys},
the two ``outer'' models are strongly inconsistent with
the Galactic-model priors, that is, they have $p=0.000$ weight in the
penultimate column.  The physical reasons for this were anticipated
in detail in Section~\ref{sec:parallax-kb230164}.  Both of the remaining
(that is, the ``inner'') solutions have M-dwarf hosts that lie a few kpc
from the Sun.  
See Figure~\ref{fig:0164bayes} and Section~\ref{sec:phys}.

Finally, we predict the two-dimensional (color-magnitude) posterior of
the lens light, assuming a main-sequence lens and accounting for the
increasing dust column as a function of distance.  This is shown for the
two inner solutions in Figure~\ref{fig:0164bayescmd}.  The color and magnitude
of the blend (which did not enter the Bayesian analysis in any way) are shown
as an ``X''.  Note that while, in both cases, the ``most likely'' lenses
are much fainter and redder than the blend, the blend is consistent with
the overall distribution at the 1--2 sigma level.

\subsubsection{{Spectrum of KMT-2023-BLG-0164}
  \label{sec:spectrum}}

Based on the analysis of Section~\ref{sec:parallax-kb230164}, as confirmed
by Section~\ref{sec:phys-kb230164}, we recognized that the nature of
KMT-2023-BLG-0164 could be significantly clarified by a
high-resolution spectrum, which was feasible with a relatively large
telescope due to the brightness of the blend, $(V,I)_b =(17.55,16.00)$,
as derived from OGLE-III calibrated photometry.

On UT 2025-07-18, P.C.\ obtained spectroscopic data consisting of two
$450\,$s exposures with a spectral resolution $R\sim 1000$ using the
Inamori Magellan Areal Camera and Spectrograph (IMACS;
\citealt{Dressler11}) mounted on the 6.5-m Magellan-Baade
telescope under excellent conditions ($\sim 0.5^{\prime\prime}$
seeing).  The spectrum (Figure~\ref{fig:results_spmatch}) directly
resolves one major issue, the lens distance, and it also opens the path
to a more complete understanding of the planetary system.

The light from the target is dominated by a star with ($T_{\rm eff}/{\rm K}$ ,
$\log g$, [Fe/H]) = ($5703 \pm 153$, $4.46 \pm 0.17$, $0.10 \pm
0.15$), which is similar to the Sun (5770, 4.44, 0). We infer a blend
mass, absolute magnitude and color of ($M_{\rm bl}/M_\odot$, $M_I$ , $(V -
I)_0$) = $(0.98 \pm 0.07, 4.17 \pm 0.33, 0.78 \pm 0.05)$.


The stellar atmospheric parameters are estimated by template matching
using a $\chi^2$ minimization technique, with templates drawn from the
MILES spectral library \citep{2006MNRAS.371..703S}.  We restrict the
fits to the spectral range from 4600 to 6800\,\AA. The
blue part of the spectrum is excluded due to its low signal-to-noise
ratio, while the red part is excluded because of the limited
wavelength coverage of MILES and the scarcity of absorption features
in that region.  In addition, the Na\,\textsc{i}\,D doublet region and
the diffuse interstellar band (DIB) features at $\lambda\lambda$\,5780
and 6283\,\AA\ are masked in the fitting, as they are affected by
interstellar absorption.  All MILES templates are convolved to a
resolution of $R \sim 1000$ to match the resolution of the observed
spectra.  The final atmospheric parameters are computed as the average
values from the five best-fitting templates with the lowest $\chi^2$
values. The associated uncertainties are taken as the standard
deviations among these five solutions.


Using the derived stellar atmospheric parameters, we estimated the
stellar mass, intrinsic color $(V-I)_0$, and $I$-band absolute
magnitude of the target by adopting a Bayesian method similar to that
of \citet{2022ApJ...925..164H}.  For the theoretical models, we
employed the PARSEC isochrones \citep{2012MNRAS.427..127B,
  2017ApJ...835...77M}.  This approach yields the posterior
probability distribution function (PDF) for each parameter.  The final
adopted value of each parameter is taken as the median of the
corresponding posterior PDF, and the associated uncertainty is defined
as half of the difference between the 84th and 16th percentiles.

Assuming for the moment that all the light from the observed blend is due to
this spectroscopic star (we have already subtracted the small amount of source
light), then its observed $(V-I)_b=1.55$ requires $E(V-I)=1.55-0.78=0.77$ of
reddening from dust along the line of sight.  From
Table~\ref{tab:cmd}, the ratio of total-to-selective extinction toward
the source is $(A_I/E(V-I))_s = (1.68/1.44)=1.17$.  In principle, the
dust behind the target could have different properties than the nearby
dust.  However, because the event is at relatively high latitude,
$b=-4.0$, most of the dust column is relatively near the Sun, implying that
this effect is likely to be small.  In any case we ignore it and
conclude that the extinction is $A_{I,b} = 0.77\times 1.17 = 0.90$,
which implies a distance of $D_L=10^{(I-M_I-A_I)/5 - 2}=1.53\,\kpc$ and
therefore a lens-source relative parallax of
\begin{equation}
\pi_\rel = {\au\over D_L}-{\au\over D_S}\simeq 0.53\,\mas.
\label{eqn:pirel}
\end{equation}
Notice that this estimate of $\pi_\rel$ does not depend on the spectroscopic
target actually being the host of the planet, it only depends on this
object being at the same distance as the host, that is, either the host
itself or a stellar companion of the host.  We already argued in
Section~\ref{sec:anal-kb230164} against the only other two
logical possibilities, that is, a companion to the source or a random field
star projected along the line of sight.  We note that if, for example, the
target were not the host and the actual host accounted for 20\% of the blend
light,
then the $I$-band flux
would be 20\% less bright than we have inferred and so $D_L$
would be 10\% bigger.  However, even this possibility would have only a modest
impact on $\pi_\rel$.

There are various errors associated with this estimate, but for the
moment we ignore these and just ask whether this spectroscopic object
is broadly consistent with being the lens, given the other information
that is available.

Regardless of whether the spectroscopic target is the lens or its companion,
Equation~(\ref{eqn:pirel}) implies that
\begin{equation}
  \pi_\e
  = \sqrt{\pi_\rel\over\kappa M} = 0.26\biggl({M\over 0.98\,M_\odot}\biggr)^{-1/2};
  \qquad \theta_\e
  =\sqrt{\kappa M\pi_\rel}=2.1\,\mas\biggl({M\over 0.98\,M_\odot}\biggr)^{1/2}.
\label{eqn:piethetae}
\end{equation}
The strongest constraint
comes from the microlens parallax measurement (Figure~\ref{fig:0164parallax}).
We have already argued in
Section~\ref{sec:parallax-kb230164}
that the two
``outer'' solutions are ruled out, which was confirmed from the formal
Bayesian analysis by
by their $p=0.000$ posterior probabilities.
The ``inner(+)''
solution has a smaller parallax, but as can been seen from
Figure~\ref{fig:0164parallax},
it is still in $4.5\,\sigma$ conflict with a value of $\pi_\e=0.26$.
From the parallax formula, there are only two ways to ``increase'' $\pi_\e$:
either increase $\pi_\rel$ or decrease $M$.  There is hardly any room for
the latter, given the inferred temperature.  Increasing $\pi_\rel$
(decreasing $D_L$) would require increasing $M_I$ or $A_I$ or decreasing $I$.
Small changes in any of these are possible, but just to move from $4.5\,\sigma$
to $3\,\sigma$ would require a combined change of 0.9 mag in these three
quantities.  Thus, if reconciliation is possible, it would be primarily
by ``blaming'' the light-curve measurement of $\bpi_\e$, not the interpretation
of the spectrum.  Such problems in microlensing parallaxes (and microlensing
light curves more generally) are hardly unknown, but before settling on
this explanation, we explore other evidence and other possibilities.

Next, we examine the problem of the implied source proper motion, $\bmu_s$.
We have
\begin{equation}
  \mu_{\rel,\geo} = {\theta_\e\over t_\e}=
     {2.1\,\mas(M/0.98 M_\odot)^{1/2}\over 70\,{\rm d}}
       =11.0\,\masyr\biggl({M\over 0.98\,M_\odot}\biggr)^{1/2},
\label{eqn:murel}
 \end{equation}
where we have inserted the scaled value of $\theta_\e$
that was just derived.  Note
that we have made explicit that this is the geocentric proper motion because
our information about the source and lens proper motions is, by contrast,
in the heliocentric frame.  The two are related by,
\begin{equation}
  \Delta \bmu_\rel = \bmu_{\rel,\hel} - \bmu_{\rel,\geo} =
      {\pi_\rel\over\au}\bv_{\perp,\oplus};\qquad
      \bv_{\perp,\oplus}(N,E)=(+3.88,+12.71)\,\kms,
      \label{eqn:vperp}
\end{equation}
where $\bv_{\perp,\oplus}$ is the projected velocity of Earth at $t_0$.
Inserting the above value of $\pi_\rel=0.53\,\mas$ yields
$\Delta\bmu_\rel(N,E) = (0.43,1.42)\,\masyr$.  

We now seek to express $\bmu_s$ in the bulge frame (in which the underlying
distribution is approximately symmetric about (0,0).  We
write $(\bmu_l - \bmu_s)_\hel = \bmu_{\rel,\geo} + \Delta\bmu_\rel$, implying
$  (\bmu_{s,\hel}-\bmu_{\rm SgrA*}) =  -\bmu_{\rel,\geo}  - \Delta\bmu_\rel
  -\bmu_{\rm SgrA*} +\bmu_{l,\hel}$, or
\begin{equation}
  (\bmu_{s,\hel}-\bmu_{\rm SgrA*}) 
  = \biggl[-11.0\biggl({M\over 0.98 M_\odot}\biggr)^{1/2}{\bf\hat n}
      + (+4.54,+2.98)_{l,b}\biggr]\masyr,
    \label{eqn:musource}
\end{equation}
where ${\bf\hat n}$ is the direction of the parallax vector $\bpi_\e$.
From the inner(+) solution as represented in Table~\ref{tab:0164par_parms},
the best fit for this direction is $17^\circ$ (north through east), which
is $13^\circ$ north of the Galactic plane.  Inserting this value into
Equation~(\ref{eqn:musource}), and adopting its fiducial mass, yields
$(\bmu_{s,\hel}-\bmu_{\rm SgrA*}) = (-6.18,+0.51)\,\masyr$.  As the one-dimensional dispersion
of bulge stars is slightly less than $\sigma=3\,\masyr$, such a value would
be in slightly more than $2\,\sigma$ tension with the expectation for bulge
sources, which is (other things being equal) acceptable.

Finally, we note that, given the value $\theta_*=0.92\,\muas$ from
Table~\ref{tab:cmd}, the inferred value of $\rho=\theta_*/\theta_\e$ as a
function of lens mass is $\rho = 4.4\times 10^{-4}(M/0.98\,M_\odot)^{-1/2}$.
From Table~\ref{tab:0164par_parms} (see Figure~\ref{fig:0164rho} for
more detail), this is acceptable for the fiducial mass, although it
degrades slowly toward lower masses.

Thus, to briefly summarize, the value of $\bpi_\e$ inferred under the
assumption that the spectroscopic target is the host is in $4.5\,\sigma$
tension with the light-curve analysis, which could only be resolved by
suggesting that there was a major problem with the light-curve
data.  This is possible,
but we have already shown in
Section~\ref{sec:robust-kb230164}
that the various
individual data sets are basically consistent.  By contrast, the inferred
values of $\bmu_s$ and $\rho$ are only in mild tension.

On the other hand, we should note that for Bayesian range
tabulated in Section~\ref{sec:phys}
for the inner(+) solution,
$M/M_\odot = 0.44\pm 0.22$, the inferred $\bpi_\e$ is perfectly
consistent, as are the inferred $\bmu_s$ and $\rho$.  Moreover,
even though no direct information about $D_L$ was injected into
the Bayesian analysis, the posterior range $D_L=1.16^{+0.64}_{-0.33}\,\kpc$,
is in excellent agreement with the spectroscopic determination of the
lens distance.

We conclude that the balance of evidence favors that the spectroscopic target
is a binary companion to the host, rather than the host itself.

\subsubsection{{Host Mass From Future Adaptive Optics (AO) Observation}
  \label{sec:ao}}

It will be possible measure the host mass of KMT-2023-BLG-0164 and
to distinguish whether it is the spectroscopic target or its companion
using AO on extremely large telescopes (ELTs), most plausibly the
EELT.  The method is slightly different from the one
pioneered by \citet{ob05169bat} and successfully applied to a dozen other
cases, as summarized in
Section~\ref{sec:intro}
and analyzed in depth
by \citet{gould22}.  The method is closer in spirit to the one first
advocated by \citet{pac95}, in which he pointed out that the
lens mass could be derived from (in modern notation) the measurements of
$\pi_\rel$ and $\mu_\rel$.  That is,
\begin{equation}
  M = {(\mu_\rel t_\e)^2\over \kappa \pi_\rel}.
  \label{eqn:refsdal}
\end{equation}
For Paczy\'nski, both $\mu_\rel$ and $\pi_\rel$ would be obtained astrometrically,
but in our case, $\pi_\rel$ is obtained spectroscopically, not necessarily from
the lens itself but most likely from a more luminous stellar companion.

It is far from clear that late-time AO observations will even detect the
lens.  For example, it would likely be several mag fainter than the
spectroscopic target and therefore, if the binary orbit is too tight
(for example, $\sim 20\,\au$ or $\sim 13\,\mas$), it would remain hidden
in the glare of the spectroscopic target, indefinitely.  Nevertheless,
because $\mu_\rel$ for this companion is essentially the same as for the
host, it can be derived from two late-time AO observations that separately
resolve the spectroscopic target and the source.  The spectroscopic target
is $\Delta I=3.5$ mag brighter than the source, and has a similar spectral
type, so similar colors.  Therefore, the contrast ratio is
$\Delta K = \Delta I - \Delta E(I - K)\simeq \Delta I-(6/7)\Delta A_I
\simeq 2.85$, where we have adopted $\Delta A_I\sim 0.77$ and assumed
$A_I\sim 7\,A_K$, which is generally valid over the KMT fields.

Given EELT's diffraction-limited $K$-band FWHM of $14\,\mas$,
this contrast ratio should be observable at EELT first light, perhaps
2030, when the source and lens will be separated by about
$\Delta\theta = [77(M/0.98\,M_\odot)^{1/2} + 8]\,\mas$.  If, as seems likely,
the host cannot be directly observed, the proper motion can still be derived
from the difference in positional offsets between two epochs that are
separated by one or two years.


{\subsection{{KMT-2023-BLG-1286}\label{sec:anal-kb231286}}}

Figure~\ref{fig:1286lc} shows an otherwise standard 1L1S light curve
with \citet{pac86} parameters $(t_0,u_0,t_\e)=(118.79,0.13,14.7\,{\rm day})$,
punctuated by a $\Delta t_{\rm bump}=0.1$ day bump, centered at $t_\anom=118.48$.

{\subsubsection{{Heuristic Analysis}\label{sec:heuristic-kb231286}}}

These parameters imply $\tau_\anom=-0.021$, $u_\anom=0.13$, and thus
\begin{equation}
\alpha = 99.2^\circ;
\qquad
s^\dagger_+ = 1.067,
\label{eqn:kb23_1286_heur}
\end{equation}

{\subsubsection{{Light-curve Analysis}
    \label{sec:static-kb231286}}}

The grid search on the $(s,q)$ plane returns two competitive solutions,
plus a third (``resonant'') solution that is disfavored by $\Delta\chi^2>100$,
but which we include for completeness.  The 
refinements of these three solutions, with all parameters set free, are 
shown in
Table~\ref{tab:1286parms}. 
We see that $\alpha=99.1^\circ$
in both competitive solutions, 
while $s^\dagger\equiv \sqrt{s_+ s_-} = 1.070$, which are both
extremely close to the heuristic predictions.

Although the bump is smooth, it is sharply peaked, which indicates that
the source passes close to a cusp but probably not over it.  This
is true of all three models, as seen in the insets to
Figure~\ref{fig:1286lc}.  This narrow peak allows $\rho$ to be
measured, despite the lack of a caustic crossing.  See
Table~\ref{tab:1286parms}.  For comparison, OGLE-2016-BLG-1195 provides
a classic example of this geometry \citep{ob161195a,ob161195b}.
Nevertheless, because the errors in this
measurement are non-Gaussian (in particular, asymmetric), we display
their functional form in Figure~\ref{fig:1286rho}.

In the resonant model, the peak is produced by an approach to an off-axis cusp.
However, as seen in Figure~\ref{fig:1286lc}, this off-axis cusp approach
does not reproduce the small ``dips'' that flank the bump and are
clearly seen in the data.  This is the origin of the
$\Delta\chi^2>100$ that rules out this model.

Because the anomaly is a non-caustic-crossing bump, we also check
for 1L2S models.  Although the best such model reproduces the
main features of the bump, it fails to reproduce the dips that
flank the bump.  This failure is very similar to that of the resonant
model in both morphology (see Figure~\ref{fig:1286lc}) and in
$\Delta\chi^2$ (see Table~\ref{tab:1286parms}).

{\subsubsection{{CMD Analysis}
    \label{sec:cmd-kb231286}}}

There are no major issues in the CMD analysis of KMT-2023-BLG-1286.
We note that by combining the $t_\e$ and $\rho$ measurements from
Table~\ref{tab:1286parms} with the $\theta_*$ determination from
Table~\ref{tab:cmd} yields Einstein-radius and proper-motion estimates of
\begin{equation}
  \theta_\e = 0.365\pm 0.080\,\mas; \qquad
  \mu_\rel =  9.1\pm 2.0\,\masyr
  \label{eqn:mu1286}
\end{equation}

We find that the source is offset from the baseline object by 130 mas,
so the blend is unlikely to be associated with the event.
The calibrated blend magnitude is $I_b=19.37$.  To allow for the mottled
background of bulge fields, we place an upper limit on the lens light
$I_l > 19.1$.

{\subsubsection{{Bayesian Analysis}\label{sec:phys-kb231286}}}

For KMT-2023-BLG-1286, there are three constraints on the
Bayesian analysis.  First, there are the measurements of $t_\e$
from Table~\ref{tab:1286parms}.  Second, there is the measurement of
$\theta_\e$ from Equation~(\ref{eqn:mu1286}).  Finally there is the
upper limit on lens light, $I_l> 19.1$, from Section~\ref{sec:cmd-kb231286}.

The posterior distributions of the lens mass and distance are shown
in Figure~\ref{fig:1286bayes}.  Note, from
Figure~\ref{fig:1286bayes} (see also Section~\ref{sec:phys}),
that the Bayesian posteriors are nearly
identical for the inner and outer solutions, while the resonant solution,
which is ruled out at $\Delta\chi^2>100$, is shown only for completeness.
The host is therefore likely to be a middle M dwarf lying in or near the
Galactic bulge, orbited by a Neptune-class planet.

{\subsection{{KMT-2023-BLG-1746}\label{sec:anal-kb231746}}}

Figure~\ref{fig:1746lc} shows an otherwise standard 1L1S light curve
with \citet{pac86} parameters $(t_0,u_0,t_\e)=(199.81,0.045,113\,{\rm day})$,
punctuated by a $\Delta t_{\rm dip}=1.0$ day dip, centered at $t_\anom=202.5$.

{\subsubsection{{Heuristic Analysis}\label{sec:heuristic-kb231746}}}

These parameters imply $\tau_\anom=0.024$, $u_\anom=0.051$, and thus
\begin{equation}
\alpha = 241.9^\circ;
\qquad
s^\dagger_- = 0.975;
\qquad
q = 7.3\times 10^{-5}.
\label{eqn:kb23_1746_heur}
\end{equation}

{\subsubsection{{Static Analysis}\label{sec:static-kb231746}}}

The grid search on the $(s,q)$ plane returns four solutions,
whose refinements with all parameters set free are 
shown as the ``Standard models'' in 
Tables~\ref{tab:1746parms_inner}--\ref{tab:1746parms_wide}.
This is twice as many solutions as we would
normally expect.  The fundamental reason for this plethora of solutions
is that the dip is traced by only three observations.  As can be seen
in Figure~\ref{fig:1746lc}, this degeneracy could easily have been broken
if there were even one observation at about 201.8 or at about 203.2.
The low effective cadence (about 1 observation per night
per observatory -- see Figure~\ref{fig:1746lc}) over the anomaly derives
from two factors.  First, the event lies in BLG37, where the intrinsic
KMT cadence is $\Gamma=(0.4,0.3,0.3)\,{\rm hr}^{-1}$ for (KMTC,KMTS,KMTA),
respectively.  Second, the anomaly occurred
on Sept 15, when the bulge was observable for only about 4 hours per night.
We will return to the issue of multiple minima below.
For now, we group the four solutions into two pairs, which we call
``inner/outer'' and ``close/wide'', respectively.
We see that $\alpha\simeq 241.7^\circ$ for all four solutions
while $s^\dagger\equiv \sqrt{s_+ s_-} = (0.978,975)$ for `the `inner/outer'' and
``close/wide'' pairs, respectively. Both $\alpha$ and $s^\dagger$ are
very close to the heuristic predictions.  The values of $q$ for the
``close/wide'' solutions are also extremely close to the prediction
of Equation~(\ref{eqn:kb23_1746_heur}), although the ``inner/outer'' pair
has $q$ values that are about 40\% smaller.
From Figure~\ref{fig:1746lc}, we see that the ``inner/outer'' pair has
a set of caustic crossings over the two wings of the planetary caustic,
whereas the ``close/wide'' pair does not.  This is the reason that
each of the light curves (green and cyan) of the first pair has a set of
  two ``bumps'' flanking the dip in Figure~\ref{fig:1746lc}.
One expects (or perhaps, hopes) that
this implies more detailed information about $\rho$.  However, the
issue is complicated by the low cadence, and so we defer consideration
of it to Section~\ref{sec:parallax-kb231746}.

The event is unusually long ($t_\e\sim 113\,$day in the Standard model), so
we expect that parallax and orbital motion may play a significant role.
Hence, we conduct a parallax analysis.

{\subsubsection{{Parallax Analysis}\label{sec:parallax-kb231746}}

As can be seen from Tables~\ref{tab:1746parms_inner}--\ref{tab:1746parms_wide},
there are 16 solutions.  This is twice the eight solutions that
one might naively expect,
namely (inner,outer,close,wide)$\times$($u_0>0$,$u_0<0$).  The reason is
that for each of these eight designations, there are two solutions, one
with $\pi_{\e N}>0$ and the other with $\pi_{\e N}<0$.  We designate these
for example, inner$(+,-)$, which means inner solution with
$u_0>0$ and $\pi_{\e N}<0$.

To untangle this plethora of degenerate solutions, we first ask why
this two-fold north-south degeneracy exists.  From 
Figure~\ref{fig:1746parallax}, we see that in all four cases,
(inner, outer, close, wide), the parallax contours have an
elongated (``one-dimensional'', 1-D) form, and within these 1-D contours, there
are two minima, one near each end of the contour.  The 1-D form is
extremely common in microlens parallax measurements, with the narrow direction
being $\pi_{\e,\parallel}$, that is, the component of $\bpi_\e$ in the
direction of Earth's acceleration (toward the Sun).  See \citet{gmb94}.
In the present case, because
the event is extremely long (so that the direction of the Sun changes
substantially during the event)
and because the post-peak light curve is cut off by the
end of the observing season, 
these contours are not strictly linear.

The two minima
are due to the ``jerk-parallax'' degeneracy \citep{gould04}, so-called
because the jerk (acceleration derivative) of Earth's motion masquerades
as a parallax effect.  We refer the reader to \citet{gould04} for
the mathematical origins of this degeneracy and to \citet{mb03037}
for an explicit calculation
of the vector separation of the two minima on the $\bpi_\e$ plane.
Note in particular that the north-south degeneracy is not symmetric
with respect to the $x$-axis, but to the direction of Earth acceleration.
Thus, in all eight cases, this degeneracy has a major impact on $\bpi_\e$,
not only in its direction, but also its magnitude.  Fortunately, however,
it does not have a major impact on $q$ or $\rho$, which are the parameters
of most immediate physical interest.

We next turn to the $(\pm)$ degeneracy in $u_0$, which is sometimes
called the ``ecliptic'' degeneracy \citep{ob03238} because it is 
exact for events on the ecliptic (other than changing the sign of
$(u_0,\alpha,\pi_{\e N},d\alpha/dt)$).  In the present case, the event lies
$>11^\circ$ from the ecliptic, so this perfect-degeneracy limit does not even
approximately apply.  However, this degeneracy is also very small
for high-magnification events $(u_0\ll 1)$ because, in this case,
$u_0$ does not differ much from $-u_0$.  Hence, we expect that, for example,
inner$(++)$ will be similar to inner$(-+)$, and inner$(+-)$ will be similar to
inner$(--)$.  This expectation is qualitatively 
confirmed, although it completely
fails for the orbital motion parameters, which are poorly measured.

In terms of the parameters that most affect the physical
interpretation ($q$ and $\rho$), there are basically two groups of
solutions, (inner, outer) and (close, wide).
For the first, $q\sim 6\times 10^{-5}$ and $\rho\sim 5\times 10^{-3}$,
while for the second,
$q\sim 1\times 10^{-4}$ and there are only upper limits on $\rho$.

We first consider the (inner, outer) pair.  Very often, when there is
a caustic crossing, $\rho$ is well determined.  However, because of
the low effective cadence, there are models that can ``avoid'' the
caustic at the price of just a few sigma.  Thus, from
Figure~\ref{fig:1746rho}, 
we can see that for both the $u_0>0$ and
$u_0<0$ cases, $\rho$ is well-localized at $\rho\sim 0.0055$
near the $\chi^2$ minima,
but (for all cases except inner(+,+) and inner($-,-$)) suddenly
``blows up'' at about $2.5\,\sigma$, in particular,
quickly permitting values of $\rho$ that are $\sim 3$ times smaller.
While these solutions seem ``obviously disfavored'', albeit at a low
level, we should keep in mind that this corresponds to a $\Delta\chi^2$
difference that is similar to one by which the (inner, outer) models
are favored over the (close, wide) models.

In order to place this issue in context, we 
make use of the measurement, $\theta_*= 0.65\,\muas$, from Table~\ref{tab:cmd}.
Hence, $\rho\sim 0.005$ would imply $\theta_\e\sim 0.12\,\mas$ and
so $\mu_\rel\sim 0.5\,\masyr$.  For typical events, we would regard
such a low proper motion as ``improbable'' and ``suspicious'', but
KMT-2023-BLG-1746 is very long ($t_\e\sim 90\,$day in the parallax models),
and the major cause of extremely long events is low proper motion
\citep{kb162052}.  Moreover, \citet{gould22} showed that planetary events
that enter our sample are strongly biased toward low proper motion relative
to their rate in nature.
Another aspect is that with $\theta_\e\sim 0.12\,\mas$ and
$\pi_\e\geq |\pi_{\e,E}|\ga 0.10$, the lens would have
$\pi_\rel=\theta_\e\pi_\e\ga 0.012$, which would allow it be in the
bulge.  This would naturally account for the direction of $\bpi_\e$
(same as the direction of $\bmu_\rel$) in the negative East direction
($\pi_{\e,E}<0$).  This direction is more difficult (though by no means
impossible) to accommodate for disk lenses.   In this case, 
$M=\theta_\e/\kappa\pi_\e\la 0.15\,M_\sun$, so the lens could be a late
M dwarf in the bulge.

By contrast, the (close, wide) solutions favor much lower
$\rho\la (0.001,0.002)$ at $1\,\sigma$, although they do
accommodate $\rho\sim 0.005$ at $3\,\sigma$.  That is, it is very likely
that they are disk lenses.  Hence, if late-time
imaging (taken when the lens and source had separated) showed
a proper motion $\mu_\rel\sim 0.5\,\masyr$, then the (inner, outer)
solutions would be strongly favored.  Of course, even with the
advent of the
EELT with a
FWHM $\sim 14\,\mas$ in $K$-band, this would require a wait time
of several decades.  On the other hand, if the proper motion
were measured at $\mu_\rel\ga 1.5\,\masyr$, then the (close, wide)
solutions would be strongly favored.

{\subsubsection{{CMD Analysis}
    \label{sec:cmd-kb231746}}}

Recall that for this event, there are 16 separate solutions, four in
each of the four topologies that we have labeled ``inner'', ``outer'',
``close'' and ``wide''.  From the standpoint of the $\theta_\e$
and $\mu_\rel$ measurements, the first two topologies (``inner'' and ``outer'')
are qualitatively different from the latter two because they have
well defined $\rho$ measurements rather than upper limits.  We focus
first on the best solution, inner($- +$), which is described in
Table~\ref{tab:1746parms_inner}.
By combining the $t_\e$ and $\rho$ measurements from
that table with the $\theta_*$ determination from
Table~\ref{tab:cmd}, we obtain the Einstein-radius and proper-motion estimates
for this solution.
\begin{equation}
\theta_\e = 0.127\pm 0.013\,\mas;
\qquad
\mu_\rel = 0.499\pm 0.052\,\masyr,
\qquad (u_0<0)
\label{eqn:1746thetae}
\end{equation}
The values for the other 7 inner+outer solutions are very similar and can
be obtained from Tables~\ref{tab:1746parms_inner} and
\ref{tab:1746parms_outer} by the usual scaling relations.
This value of $\mu_\rel$ would be low for a randomly chosen
microlensing event ($p\sim 0.04\%$), but we should keep in mind three factors.
First, as shown by \citet{gould22}, the distribution of proper motions
for planetary events (with measured proper motions) is weighted toward
low proper motions by a factor $\sim \mu_\rel^{-1}$ relative to the underlying
population of events (that is, $p\rightarrow 0.4\%$).
While \citet{gould22} does not comment on the
reason for this, probably it is that low-proper-motion events are
more likely to enter the sample because lower proper motions imply
more slowly evolving events that, in turn, allow $\rho$ to be 
measured.  Second, \citet{kb162052} have
shown that a major cause of long-$t_\e$ events (in the present case,
$t_\e\sim 90\,$d) is low proper motion.
Finally, half the solutions for this event (that is, all eight of
the close+wide solutions) have only an upper limit on $\rho$ (and so
only a lower limit on $\mu_\rel$).

We calculate the blend light, $I_b$, by subtracting the measured source
flux from the OGLE-III baseline magnitude $I_{\rm base}=18.64$ to
obtain $I_b=19.4$.   Unfortunately, the OGLE-III $V$-band baseline
measurement is not precise enough to obtain a reliable blend color.  To take
account of the mottled background of bulge fields, we place a more
conservative limit on lens light, $I_l > 19.1$.


{\subsubsection{{Bayesian Analysis}\label{sec:phys-kb231746}}}

The Bayesian analysis of KMT-2023-BLG-1746 is broadly similar to that
of KMT-2023-BLG-0164, although in this case there are 16 solutions, rather
than four.  For this event, we also have four constraints,
although they are not exactly the same as for KMT-2023-BLG-0164.
The constraints on $t_\e$,
and $\bpi_\e$ are incorporated in exactly the same way.  We also
handle the constraint on $\rho$ in exactly the same way.  This is essential
for the eight close+wide solutions, for which (as in the case of
KMT-2023-BLG-0164) there are only upper limits on $\rho$.  In principle,
we could treat the four inner solutions as ``measurements'' (as we did
for KMT-2023-BLG-1286), and this would also be basically appropriate for
the four outer solutions.  However, we choose to treat all 16 solutions
in a homogeneous way, and so we derive envelope functions
from Figure~\ref{fig:1746rho}, 
Finally, in this case, the fourth constraint is a limit on lens flux,
$I_l>19.1$, as we discussed Section~\ref{sec:cmd-kb231746}.

The posterior distributions of the lens mass and distance are shown
in Figures~\ref{fig:1746bayes_cw} and \ref{fig:1746bayes_io} for the
eight close/wide and eight inner/outer solutions, respectively.
These posteriors demonstrate, what was already partly clear from
Section~\ref{sec:parallax-kb231746},
that the 16 solutions are broadly divided into
several groups by host mass, planet mass, $\chi^2$, and compatibility
with Galactic-model priors.

We now discuss the origins of these groupings and how the resulting degeneracies
can be partly (or completely) resolved by future AO
imaging.

First, 
as we will tabulate in Section~\ref{sec:phys},
the eight inner/outer solutions are favored by $\Delta\chi^2\sim 5$, while
the eight close/wide solutions are more compatible with the Galactic-model
priors.  The first aspect was already clear from the 
tables in Section~\ref{sec:parallax-kb231746}.
The main reason for the second aspect
is that the inner/outer solutions all have well-localized
$\rho\sim 0.005$
determinations, whereas the close/wide solutions have only upper limits
on $\rho$.  So, at one level, the close/wide solutions appear ``favored''
by the Galactic-model priors simply because they can accommodate a
wider variety of random realizations of these priors.  More specifically,
as already commented upon in 
Section~\ref{sec:parallax-kb231746},
given the
$\theta_*=0.65\,\muas$ measurement from Table~\ref{tab:cmd} and the
$t_\e\sim 90\,$d determinations from Tables~\ref{tab:1746parms_inner} and
\ref{tab:1746parms_outer}, the inner/outer models have
$\theta_\e=0.13\,\mas$ and $\mu_\rel\sim 0.5\,\masyr$.  Such low proper motions
are disfavored by Galactic models.

This same difference also explains why the inner/outer models have
smaller posterior host masses.  That is, with the $\bpi_\e$ relatively
constrained by Figure~\ref{fig:1746parallax},
$M=\kappa\theta_\e/\pi_\e= \kappa\theta_*/\pi_\e\rho\propto \rho^{-1}$.
Thus, a high value of $\rho$ implies a low value of $M$.

Finally, $q$ is about 1.5 times larger for the close/wide models than
the inner/outer models, so $M_{\rm planet}=q M_{\rm host}$ is larger for the
close/wide models according to the product of two independent, larger factors.

Future AO observations (almost certainly requiring EELT or other ELT)
will yield a proper-motion measurement as well
as a host-mass measurements, which will distinguish between the main
degeneracies of the different models.  For example, if the proper motion
is $\mu_\rel\sim 0.5\,\masyr$ this will strongly favor the inner/outer models,
which will in turn resolve the two-fold ambiguity in $q$, whereas a
substantially larger $\mu_\rel$ measurement would strongly favor the
close/wide models.  The same (vector) proper-motion measurement would
distinguish between the $\pi_{\e,N}>0$ and $\pi_{\e,N}<0$ models.  There
would then remain a four-fold degeneracy, for example among the
inner$(-,+)$, inner(+,+), outer$(-,+)$, and outer(+,+) models.  However,
inspection of the four tables in
Section~\ref{sec:parallax-kb231746}
shows that
the differences among these are actually quite minor.

{\subsection{{KMT-2023-BLG-0614}\label{sec:anal-kb230614}}}

Figure~\ref{fig:0614lc} shows an otherwise standard 1L1S light curve
with \citet{pac86} parameters $(t_0,u_0,t_\e)=(62.4,0.3,31\,{\rm day})$,
punctuated by a $\Delta t_{\rm bump}=1.5$ day bump, centered at $t_\anom=61.4$.

{\subsubsection{{Heuristic Analysis}\label{sec:heuristic-kb230614}}}

These parameters imply $\tau_\anom=-0.032$, $u_\anom=0.3$, and thus
\begin{equation}
\alpha = 96^\circ;
\qquad
s^\dagger_+ = 1.16.
\label{eqn:kb23_0614_heur}
\end{equation}

{\subsubsection{{Light Curve Analysis}\label{sec:static-kb230614}}}

The grid search on the $(s,q)$ plane returns two solutions,
whose refinements with all parameters set free are 
shown in Table~\ref{tab:0614parms}.
We see that in the two solutions, $\alpha=(95^\circ,96^\circ)$
while $s^\dagger\equiv \sqrt{s_+ s_-} = 1.18$, which are both
very close to the heuristic predictions.

Next, we note that in both solutions $q\sim 0.01$, which is very
far from the ``low $q$'' regime that defines this paper.  Nevertheless,
the division of Anomaly-Finder candidates by initially-estimated
$q$ is only intended
to facilitate the organization of the light-curve analyses, so we
proceed with a thorough analysis of this event.

The main issue for this event is that the light curve is very well fit
by a 1L2S (binary source) model.  See Figure~\ref{fig:0614lc} and
Table~\ref{tab:0614parms}.  Because the 1L2S model is slightly
favored over the
2L1S models, it is unlikely that the
event would be accepted into a statistical sample unless there were some
other argument that significantly preferred the 2L1S interpretation.

The principal candidates for such arguments would be that the
parameters of the 1L2S solution are physically implausible or that
late-time imaging could yield a $\mu_\rel$ measurement that would
distinguish between the 1L2S and 2L1S solutions.  As we now show,
neither of these arguments is viable.

Regarding the first, in the 1L2S model, the source magnitude is $I_S=20.7$
and the source flux ratio is $q_F= 0.015$ implying that the source
companion has
$I_{S,2} = 25.3$.  The extinction toward this field (from the KMT web site,
ultimately from \citealt{gonzalez12}) is $A_I=2.49$.  Thus, approximately,
$(I_S,I_{S,2})_0 \sim (18.2,22.8)$, corresponding to an early G dwarf and
a middle M dwarf.  M dwarf companions to G dwarfs are very common.
For example, see Table~7 from \citet{dm91}.  If the 1L2S model returned
the normalized source radii of both sources, this might in 
principle
contradict the expected ratio $\rho_2/\rho_1\sim 0.3$.  However, from
Table~\ref{tab:0614parms}, only $\rho_2$ is measured.  

Regarding the second potential argument,
a future $\mu_\rel$ measurement is very unlikely
to distinguish between the models in favor of 2L1S.
The normalized source size is
related to proper motion by $\rho_i = \theta_{*,i}/(\mu_\rel t_{\e,i})$.
From Table~\ref{tab:0614parms}, the values of $t_\e$ are $\sim 1.5$
times larger for 2L1S compared to 1L2S.
Hence, if we adopt $\theta_{*,2,\rm 1L2S}/\theta_{*,\rm 2L1S}\sim 0.35$, then
a given (future) $\mu_\rel$ measurement will predict a value of
$\rho$ that is half as big for the 1L2S model as for the 2L1S model.

From Figure~\ref{fig:0614rho}, we see that if the future $\mu_\rel$
measurement corresponded to any value of $\rho$ for 2L1S that was
within the $2.5\,\sigma$ range (that is, $\rho_{\rm 2L1S}<0.03$), then it
would predict a value in the range $\rho_{\rm 1L2S}<0.015$ for 1L2S, all of
which are acceptable at $\Delta\chi^2<3$.  Hence, the 2L1S solution
could not be confirmed even at marginal confidence.  (Parenthetically,
we note that if 1L2S were correct, and its best fit
$\rho_{\rm 1L2S}\sim 0.025$ were
inferred from a future $\mu_\rel$ measurement, then the corresponding
$\rho_{\rm 2L1S}\sim 0.05$ for the 2L1S case, would be ruled out
at high confidence.  See Figure~\ref{fig:0614rho}.  However,
this does not change the fact that the planetary interpretation
could not be confirmed by such a measurement.)


{\subsubsection{{CMD Analysis}
    \label{sec:cmd-kb230614}}}

There are no OGLE-III catalogs at the position of KMT-2023-BLG-0614,
so we conduct the CMD analysis in the instrumental KMTC pyDIA system.
The KMTC $V$-band data are too noisy to yield a measurement of $V_s$
(and hence of $(V-I)_s$).  As a result, our estimate of $(V-I)_{s,0}$ in
Table~\ref{tab:cmd} is based on the fact that $I_s$ is $4.9\pm 0.3$ mag
below the clump, implying $M_{I,s}\sim 5$, an early K dwarf.

{\subsubsection{{Bayesian Analysis}\label{sec:phys-kb230614}}}

As discussed in
Section~\ref{sec:static-kb230614},
this is not a secure planetary
event due to a competing 1L2S solution.  Therefore, we do not carry
out a Bayesian analysis.

{\subsection{{KMT-2023-BLG-1593}\label{sec:anal-kb231593}}}

Figure~\ref{fig:1593lc} shows an otherwise standard 1L1S light curve
with \citet{pac86} parameters $(t_0,u_0,t_\e)=(136.9,0.25,29\,{\rm day})$,
superposed by a $\Delta t_{\rm bump}=1.4$ day bump, centered at $t_\anom=140.65$.

{\subsubsection{{Heuristic Analysis}\label{sec:heuristic-kb231593}}}

These parameters imply $\tau_\anom=0.129$, $u_\anom=0.281$, and thus
\begin{equation}
\alpha = 62.7^\circ;
\qquad
s^\dagger_+ = 1.15.
\label{eqn:kb23_1593_heur}
\end{equation}

{\subsubsection{{Light Curve Analysis}\label{sec:static-kb231593}}}

The grid search on the $(s,q)$ plane returns two solutions,
whose refinements with all parameters set free are shown
in Table~\ref{tab:1593parms}.
We see that $\alpha=(63^\circ,62^\circ)$
in the  solutions, 
while $s^\dagger\equiv \sqrt{s_+ s_-} = 1.15$, which are both
extremely close to the heuristic predictions.

Next, we note that in both solutions $q\sim 0.003$, which is roughly a factor
20 higher than  the ``low $q$'' regime that defines this paper.  Nevertheless,
as we noted in Section~\ref{sec:static-kb230614}
the division of Anomaly-Finder candidates by $q$ is only intended
to facilitate the organization of the light-curve analyses, so we
proceed with a thorough analysis of this event.

As in the case of KMT-2023-BLG-0614, there is a 1L2S model that is a good
fit to the data.  The argument that this degeneracy is very unlikely
to be resolved is very similar to the one given in
Section~\ref{sec:static-kb230614} for KMT-2023-BLG-0614.  Therefore,
we only sketch this argument, with a focus on making explicit the small
differences relative to the argument just given.

First, from Table~\ref{tab:1593parms}, the 2L1S model is actually preferred
over 1L2S by $\Delta\chi^2=6$.  This is not sufficient to rule out the 1L2S
model by itself, but it does mean that if an additional argument of modest
significance could be found, it might be sufficient to push the
planetary interpretation ``over the threshold''.

However, going through a similar calculation, the extinction is
$A_I=3.01$, and so the dereddened magnitudes 
of the binary source are
$(I_S,I_{S,2})_0 \sim (18.2,23.1)$, that is, again an early G dwarf and middle
M dwarf, which are common.

Again, if there were a future $\mu_\rel$ measurement, the ratio of the
inferred normalized source radii would be
$\rho_{2,\rm 1L2S}/\rho_{\rm 2L1S}\sim 0.5$.  Then, from
Figure~\ref{fig:1593rho}, if a $\mu_\rel$ measurement led to the
conclusion that $\rho_{\rm 2L1S}$ were anywhere within the $2.5\,\sigma$
range (that is, $\rho_{\rm 2L1S}<0.025$), the corresponding value for the 1L2S
model would be $\rho_{\rm 2L1S}<0.0125$), which would be consistent
at the $1\,\sigma$ level.  In brief, it does not appear possible to
rule out (or even argue strongly against) the 1L2S model.

{\subsubsection{{CMD Analysis}
    \label{sec:cmd-kb231593}}}

There are no OGLE-III catalogs at the position of KMT-2023-BLG-1593,
so we conduct the CMD analysis in the instrumental KMTC pyDIA system.
The KMTC $V$-band data are too noisy to yield a measurement of $V_s$
(and hence of $(V-I)_s$).  As a result, our estimate of $(V-I)_{s,0}$ in
Table~\ref{tab:cmd} is based on the fact that $I_s$ is $4.26\pm 0.12$ mag
below the clump, implying $M_{I,s}\sim 4.15$, an early G dwarf.


{\subsubsection{{Bayesian Analysis}\label{sec:phys-kb231593}}}

As discussed in 
Section~\ref{sec:static-kb231593},
this is not a secure planetary
event due to a competing 1L2S solution.  Therefore, we do not carry
out a Bayesian analysis.

\section{{Physical Parameters}
\label{sec:phys}}

In Table~\ref{tab:physall}, we show the posterior estimates and uncertainties
for $M_{\rm host}$, $M_{\rm planet}$, $D_{L}$, and $a_\bot$ for the three
secure planetary systems.  We also highlight specific quotable
results for each case.

\section{{Summary}
\label{sec:conclude}}

We have analyzed all of the 2023 events that were flagged by AnomalyFinder
as having preliminary mass-ratio estimates $q<2\times 10^{-4}$ and were not
already published.  After eliminating false alarms due to data artifacts,
five events remained, of which three proved to be unambiguous low-$q$
planetary systems and the remaining two had ambiguous interpretations as
possible single-lens binary-source (1L2S) events.  All three of the former
indeed have low mass ratios,
KMT-2023-BLG-0164 ($q\sim 1.3\times 10^{-4}$), KMT-2023-BLG-1286
($q\sim 1.9\times 10^{-4}$), and KMT-2023-BLG-1746 ($q\sim 8\times 10^{-5}$).
For the first of these, the planet host is a nearby ($D_L\sim 1.5\,\kpc$) star,
which is either the very bright ($I=16.0$) blend that dominates the
light along the line of sight or is a binary companion to this star.
This ambiguity in host mass can be resolved by future high-resolution
imaging, which will probably require two epochs of EELT observations.
Future high-resolution imaging will also be required to distinguish
between the two groups of solutions for KMT-2023-BLG-1746, which
differ in mass ratio by a factor $\sim 1.5$.

\acknowledgments

This research has made use of the KMTNet system
operated by the Korea Astronomy and Space Science Institute
(KASI) at three host sites of CTIO in Chile, SAAO in South
Africa, and SSO in Australia. Data transfer from the host site to
KASI was supported by the Korea Research Environment
Open NETwork (KREONET). This research was supported by KASI
under the R\&D program (project No. 2025-1-830-05) supervised
by the Ministry of Science and ICT.
W.Z. and H.Y.acknowledge support by the National Natural Science Foundation of China (Grant No. 12133005).
H.Y. acknowledges support by the China Postdoctoral Science Foundation (No. 2024M762938).
W.Zang acknowledges the support from the Harvard-Smithsonian Center for Astrophysics through the CfA Fellowship. 
J.C.Y. acknowledges support from U.S. NSF Grant No. AST-2108414 and U.S. NASA Grant No. 80NSSC25K7146. 
Work by C.H. was supported by the grants of National Research Foundation of Korea (2019R1A2C2085965 and 2020R1A4A2002885). 
J.C.Y. acknowledges support from a Scholarly Studies grant from the Smithsonian Institution.
S.D.\ is supported by the National Natural Science Foundation of China (Grant No. 12133005). This work is supported by the China Manned Space Program with grant no.\ CMS-CSST-2025-A16. S.D.\ acknowledges the New Cornerstone Science Foundation through the XPLORER PRIZE.
Y.H. was supported by the National Natural Science Foundation of China
under grant No. 12422303.
The OGLE project has received funding from the Polish National Science Centre grant OPUS-28 2024/55/B/ST9/00447 to AU.

 \begin{deluxetable}{lrrrrrr}
 \tablecolumns{7} \tablewidth{0pc}
 \tablecaption{\textsc{Event Names, Observational Cadences, Alerts, and Locations}}
 \tablehead{\colhead{Name} & 
\colhead{$\Gamma\,({\rm hr}^{-1})$} &
\colhead{Alert Date} &
\colhead{RA$_{\rm J2000}$} &
\colhead{Dec$_{\rm J2000}$} &
\colhead{$l$} &
\colhead{$b$} }
 \startdata
KMT-2023-BLG-0164 & 2.4 & 27 Mar 2023 & 17:59:11.98 & $-$31:46:14.41 & $-0.95$ & $-3.96$\\ 
OGLE-2023-BLG-0116 & 0.1\\
MOA-2023-BLG-107 & 1.2\\
\hline
KMT-2023-BLG-1286 & 4.0 & 19 Jun 2023 & 18:05:28.46 & $-$27:52:11.32 & $+3.12$ & $-3.23$\\ 
OGLE-2023-BLG-0803 & 1.0\\
\hline
KMT-2023-BLG-1746 & 0.4 & 19 Jul 2023 & 17:48:38.60 & $-$35:17:53.09 & $-5.11$ & $-3.84$\\ 
\hline
KMT-2023-BLG-0614 & 2.0 & 28 Apr 2023 & 17:57:19.04 & $-$27:55:27.19 & $+2.19$ & $-1.69$\\ 
\hline
KMT-2023-BLG-1593 & 4.0 & 10 Jul 2023 & 17:54:34.33 & $-$28:24:47.12 & $+1.46$ & $-1.42$\\ 
\hline
 \enddata
 \tablecomments{The coordinates given here are for the nearest catalog 
   stars (i.e., baseline objects).
}
 \label{tab:names}
 \end{deluxetable}

\begin{deluxetable}{lrrrrr}
\tablecolumns{7} \tablewidth{0pc}
\tablecaption{\textsc{CMD Parameters for Five 2023 Planet Candidates}}
\tablehead{\colhead{Parameter} &
\colhead{KB230164} &
\colhead{KB231286} &
\colhead{KB231746} &
\colhead{KB230614} &
\colhead{KB231593}}
\startdata
$(V-I)_{\rm s}$    & 2.22$\pm$0.02 & 1.61$\pm$0.04   & 1.44$\pm$0.01 & N.A.         & N.A.\\ 
$(V-I)_{\rm cl}$   & 2.50$\pm$0.02 & 1.86$\pm$0.03  & 1.81$\pm$0.03 & 3.28$\pm$0.04 & 3.65$\pm$0.05\\ 
$(V-I)_{\rm cl,0}$  & 1.06           & 1.06         & 1.06           & 1.06          & 1.06 \\
$(V-I)_{\rm s,0}$   & 0.78$\pm0.03$ & 0.81$\pm$0.05 & 0.69$\pm$0.03& 0.90$\pm$0.10& 0.70$\pm$0.10 \\ 
$I_{\rm s}$        & 19.49$\pm$0.03 & 20.38$\pm$0.05 & 19.40$\pm$0.05 & 21.79$\pm$0.31&21.41$\pm$0.11\\ 
$I_{\rm cl}$       & 16.18$\pm$0.04 & 15.30$\pm$0.05 & 15.66$\pm$0.05 & 16.90$\pm$0.05&17.15$\pm$0.05\\ 
$I_{\rm cl,0}$     & 14.50          & 14.35         & 14.62          & 14.37         & 14.38\\ 
$I_{\rm s,0}$      & 17.81$\pm$0.05 & 19.43$\pm$0.07 & 18.36$\pm$0.07 & 19.26$\pm$0.31& 18.64$\pm$0.12\\
$\theta_*$ ($\muas$)& 0.922$\pm$0.054&$0.447\pm$0.036&0.653$\pm$0.045&0.550$\pm$0.113& 0.580$\pm$0.071\\ 
\enddata

\tablecomments{Event names are abbreviations, e.g., KMT-2023-BLG-0164.
$[(V-I),I]_S$ and $[(V-I),I]_{\rm cl}$ 
for KB230164, KB231286, and KB231746 are based on calibrated OGLE-III
photometry, while the other two events have instrumental KMT photometry.
}
\label{tab:cmd}
\end{deluxetable}

\begin{deluxetable}{lcc}
	\tablecolumns{3} \tablewidth{0pc} \tablecaption{\textsc{Standard models for KMT-2023-BLG-0164}} 
	\tablehead{ \colhead{Parameters } & \colhead{${\rm{inner}}$}&	\colhead{${\rm{outer}}$} } \startdata
  $\chi^2/\rm{dof}$                     &6129.514/6080             &6180.430/6080         \\
  $t_0-2460050$                         &1.237 $\pm$ 0.012         &1.231 $\pm$ 0.013    \\
  $u_0$                                 &0.074 $\pm$ 0.002         &0.072 $\pm$ 0.002    \\
  $t_{\rm E}$ $(\rm{days})$             &73.409 $\pm$ 1.473        &75.855 $\pm$ 1.506  \\
  $s$                                   &0.861 $\pm$ 0.004         &0.968 $\pm$ 0.005    \\
  $q$ $(10^{-4})$                       &1.321 $\pm$ 0.153         &0.946 $\pm$ 0.118    \\
  log $q$ (mean)                        &-3.880 $\pm$ 0.050        &-4.024 $\pm$ 0.054   \\
  $\alpha$ $(\rm{rad})$                 &3.575 $\pm$ 0.003         &3.572 $\pm$ 0.003   \\
  $\rho [2\sigma, 3\sigma]$ $(10^{-4})$ &$[<54, <76]$              &$[<62, <81]$  \\
  $I_S$ [OGLE]                          &19.495 $\pm$ 0.026        &19.539 $\pm$ 0.026   \\
  $I_B$ [OGLE]                          &15.885 $\pm$ 0.001        &15.883 $\pm$ 0.001   \\
\enddata
\label{tab:0164parms}
\end{deluxetable}

\begin{deluxetable}{lcccc}
	\tablecolumns{5} \tablewidth{0pc} \tablecaption{\textsc{Parallax + Orbital motion models for KMT-2023-BLG-0164}} 
	\tablehead{ \colhead{Parameters } &\colhead{${\rm{inner(-)}}$} &\colhead{${\rm{outer(-)}}$} &\colhead{${\rm{inner(+)}}$} &\colhead{${\rm{outer(+)}}$}} \startdata
  $\chi^2/\rm{dof}$                      &6075.949/6077              &6075.277/6077              &6078.749/6077              &6078.442/6077\\
  $t_0-2460050$                          &1.211 $\pm$ 0.016          &1.214 $\pm$ 0.016          &1.211 $\pm$ 0.015          &1.212 $\pm$ 0.015  \\
  $u_0$                                  &-0.078 $\pm$ 0.002         &-0.078 $\pm$ 0.002         &0.076 $\pm$ 0.002          &0.076 $\pm$ 0.002  \\
  $t_{\rm E}$ $(\rm{days})$              &69.546 $\pm$ 1.434         &69.554 $\pm$ 1.401         &73.258 $\pm$ 1.593         &73.405 $\pm$ 1.635 \\
  $s$                                    &$0.867_{-0.025}^{+0.020}$  &$1.083_{-0.019}^{+0.037}$  &$0.879_{-0.022}^{+0.017}$  &$1.080_{-0.018}^{+0.027}$\\
  $q$ $(10^{-4})$                        &$1.457_{-0.437}^{+0.646}$  &$0.954_{-0.261}^{+0.842}$  &$1.154_{-0.329}^{+0.470}$  &$0.977_{-0.347}^{+0.540}$\\
  log $q$ (mean)                         &-3.839 $\pm$ 0.150         &-3.974 $\pm$ 0.184         &-3.940 $\pm$ 0.142         &-4.006 $\pm$ 0.178 \\
  $\alpha$ $(\rm{rad})$                  &2.720 $\pm$ 0.065          &$2.647_{-0.037}^{+0.114}$  &$3.611_{-0.062}^{+0.048}$  &$3.635_{-0.077}^{+0.060}$ \\
  $\rho [2\sigma, 3\sigma]$ $(10^{-4})$  &$[<7, <32]$               &$[<1.7, <2.3]$               &$[<5, <48]$               &$[<1.3, <2.0]$\\
  $\pi_{\rm{E},\it{N}}$                  &0.705 $\pm$ 0.094          &0.723 $\pm$ 0.087          &0.507 $\pm$ 0.071          &0.524 $\pm$ 0.064\\
  $\pi_{\rm{E},\it{E}}$                  &0.145 $\pm$ 0.039          &0.155 $\pm$ 0.037          &0.155 $\pm$ 0.039          &0.164 $\pm$ 0.036 \\
  $ds/dt$ $(\rm{yr}^{-1})$               &-0.316 $\pm$ 0.458         &$-3.770_{-0.623}^{+0.387}$ &-0.473 $\pm$ 0.423         &$-3.632_{-0.515}^{+0.392}$  \\
  $d\alpha/dt$ $(\rm{yr}^{-1})$          &0.448 $\pm$ 2.023          &$2.764_{-3.562}^{+1.134}$  &$-0.271_{-1.442}^{+1.925}$ &$-0.958_{-1.837}^{+2.343}$  \\
  $I_S$ [OGLE]                           &19.443 $\pm$ 0.027         &19.444 $\pm$ 0.026         &19.470 $\pm$ 0.028         &19.471 $\pm$ 0.028 \\
  $I_B$ [OGLE]                           &15.886 $\pm$ 0.001         &15.886 $\pm$ 0.001         &15.885 $\pm$ 0.001         &15.885 $\pm$ 0.001 \\
  $\beta$                                &$0.206_{-0.160}^{+0.307}$  &0.431 $\pm$ 0.235          &$0.262_{-0.199}^{+0.302}$  &0.425 $\pm$ 0.236 \\
\enddata
\label{tab:0164par_parms}
\end{deluxetable}

\begin{deluxetable}{lcccc}
	\tablecolumns{5} \tablewidth{0pc}
        \tablecaption{\textsc{Parallax Test for KMT-2023-BLG-0164}}
	\tablehead{ \colhead{Observatory} & \colhead{$\pi_{\e,N}$}&
        \colhead{$\pi_{\e,E}$} &\colhead{Corr.Coef}&\colhead{$\chi^2$}}
        \startdata
KMTC       & $0.391\pm 0.153$ & $0.026\pm 0.078$ & 0.919&  5.70 \\
Complement & $0.384\pm 0.159$ & $0.134\pm 0.076$ & 0.890&  \\
\hline
KMTS       & $0.430\pm 0.221$ & $0.161\pm 0.109$ & 0.890&  2.41 \\
Complement & $0.407\pm 0.127$ & $0.064\pm 0.063$ & 0.910&   \\
\hline
KMTA       & $0.219\pm 0.280$ & $0.059\pm 0.121$ & 0.903&  1.63 \\
Complement & $0.434\pm 0.121$ & $0.083\pm 0.062$ & 0.908&   \\
\hline
OGLE       & $0.652\pm 0.436$ & $0.288\pm 0.252$ & 0.869&  0.76 \\
Complement & $0.404\pm 0.114$ & $0.078\pm 0.056$ & 0.908&   \\
\enddata
\label{tab:0164partest}
\end{deluxetable}

\begin{deluxetable}{lcccc}
\tablecolumns{5} \tablewidth{0pc} \tablecaption{\textsc{Microlens Parameters for KMT-2023-BLG-1286}} \tablehead{\colhead{Parameters} & \colhead{inner} & \colhead{outer} & \colhead{resonant} & \colhead{1L2S}} \startdata
  $\chi^2/\rm{dof}$             &9482.396/9483              &9483.132/9483           &9591.802/9483        &9589.705/9483\\
  $t_0-2460110$                 &8.787 $\pm$ 0.014          &8.787 $\pm$ 0.014       &8.582 $\pm$ 0.018    &8.820 $\pm$ 0.016\\
  $u_0$                         &0.132 $\pm$ 0.006          &0.132 $\pm$ 0.006       &0.111 $\pm$ 0.004    &0.152 $\pm$ 0.007\\
  $t_{\rm E}$ $(\rm{days})$     &14.705 $\pm$ 0.467         &14.677 $\pm$ 0.458      &17.387 $\pm$ 0.511   &13.874 $\pm$ 0.480\\
  $s$                           &1.116 $\pm$ 0.004          &1.025 $\pm$ 0.004       &0.999 $\pm$ 0.001    &...\\
  $q$ $(10^{-4})$               &1.899 $\pm$ 0.177          &1.874 $\pm$ 0.172       &65.867 $\pm$ 4.685   &...\\
  log $q$ (mean)                &-3.722 $\pm$ 0.040         &-3.727 $\pm$ 0.040      &-2.180 $\pm$ 0.032   &...\\
  $\alpha$ $(\rm{rad})$         &1.729 $\pm$ 0.008          &1.729 $\pm$ 0.008       &3.185 $\pm$ 0.017    &...\\
  $\rho$ $(10^{-3})$            &1.225 $\pm$ 0.250          &1.214 $\pm$ 0.269       &1.531 $\pm$ 0.095    &...\\
  $t_{0,2}-2460110$             &...                        &...                     &...                  &8.477 $\pm$ 0.001\\
  $u_{0,2}$ $(10^{-4})$         &...                        &...                     &...                  &$1.123_{-0.722}^{+1.009}$\\
  $\rho_2$ $(10^{-3})$          &...                        &...                     &...                  &1.132 $\pm$ 0.070\\
  $q_F$ $(10^{-3})$             &...                        &...                     &...                  &2.731 $\pm$ 0.130\\
  $I_S$ [OGLE]                  &20.415 $\pm$ 0.049         &20.412 $\pm$ 0.049      &20.660 $\pm$ 0.044   &20.291 $\pm$ 0.056\\
  $I_B$ [OGLE]                  &19.579 $\pm$ 0.021         &19.581 $\pm$ 0.021      &19.493 $\pm$ 0.014   &19.641 $\pm$ 0.029\\
  $I_{S,2}$ [OGLE]              &...                        &...                     &...                  &26.704 $\pm$ 0.065\\
  $t_*$ $(\rm{hours})$          &$0.432_{-0.097}^{+0.079}$  &0.428 $\pm$ 0.094       &0.637 $\pm$ 0.028    &...\\
  $t_{*,2}$ $(\rm{hours})$      &...                        &...                     &...                  &0.377 $\pm$ 0.019\\
  
\enddata
\label{tab:1286parms}
\end{deluxetable}

\begin{deluxetable}{lccccc} \rotate
	\tablecolumns{6} \tablewidth{0pc} \tablecaption{\textsc{Microlens Parameters for KMT-2023-BLG-1746, inner models}} 
	\tablehead{ \colhead{ } & \colhead{	}& \multicolumn{4}{c}{Parallax + orbital motion models} \\
		\cline{3-6} 
		\colhead{Parameters } & \colhead{Standard}&	\colhead{${\rm{inner(+,-)}}$}&\colhead{${\rm{inner(+,+)}}$} &	\colhead{${\rm{inner(-,-)}}$}&\colhead{${\rm{inner(-,+)}}$}\\
		\cline{3-6}
	\colhead{ } & \colhead{}&	\colhead{$u_0>0,\pi_{\rm{E},\it{N}}<0$}&\colhead{$u_0>0,\pi_{\rm{E},\it{N}}>0$} &	   \colhead{$u_0<0,\pi_{\rm{E},\it{N}}<0$}&\colhead{$u_0<0,\pi_{\rm{E},\it{N}}>0$}} \startdata
  $\chi^2/\rm{dof}$             &1193.561/1112               &1109.022/1109              &1109.656/1109              &1109.444/1109              &1109.540/1109\\
  $t_0-2460190$                 &9.808 $\pm$ 0.024           &9.911 $\pm$ 0.025          &9.909 $\pm$ 0.025          &9.910 $\pm$ 0.026          &9.912 $\pm$ 0.025\\
  $u_0$                         &0.044 $\pm$ 0.002           &0.056 $\pm$ 0.003          &0.057 $\pm$ 0.003          &-0.057 $\pm$ 0.003         &-0.058 $\pm$ 0.003\\
  $t_{\rm E}$ $(\rm{days})$     &113.674 $\pm$ 4.515         &92.880 $\pm$ 3.908         &90.026 $\pm$ 4.301         &90.462 $\pm$ 4.163         &89.889 $\pm$ 3.758\\
  $s$                           &$0.968_{-0.006}^{+0.004}$   &$0.960_{-0.009}^{+0.007}$  &$0.961_{-0.009}^{+0.007}$  &$0.961_{-0.009}^{+0.007}$  &0.960 $\pm$ 0.008 \\
  $q$ $(10^{-5})$               &3.981 $\pm$ 0.647           &5.670 $\pm$ 0.799          &5.699 $\pm$ 0.733          &5.900 $\pm$ 0.777          &5.967 $\pm$ 0.865 \\
  log $q$ (mean)                &-4.399 $\pm$ 0.068          &-4.245 $\pm$ 0.065         &-4.246 $\pm$ 0.056         &-4.231 $\pm$ 0.059         &-4.223 $\pm$ 0.067\\
  $\alpha$ $(\rm{rad})$         &4.216 $\pm$ 0.005           &4.244 $\pm$ 0.010          &4.240 $\pm$ 0.014          &2.049 $\pm$ 0.010          &2.042 $\pm$ 0.011\\
  $\rho$ $(10^{-3})$            &4.199 $\pm$ 0.249           &5.136 $\pm$ 0.396          &5.294 $\pm$ 0.424          &5.295 $\pm$ 0.386          &5.278 $\pm$ 0.459\\
  $\pi_{\rm{E},\it{N}}$         &...                         &-0.196 $\pm$ 0.106         &0.086 $\pm$ 0.081          &-0.181 $\pm$ 0.105         &0.096 $\pm$ 0.085\\
  $\pi_{\rm{E},\it{E}}$         &...                         &-0.181 $\pm$ 0.036         &-0.121 $\pm$ 0.018         &-0.167 $\pm$ 0.029         &-0.129 $\pm$ 0.018\\
  $ds/dt$ $(\rm{yr}^{-1})$      &...                         &$-0.603_{-0.961}^{+1.323}$ &$-0.503_{-1.049}^{+1.318}$ &-0.600 $\pm$ 1.172         &-0.656 $\pm$ 1.113\\
  $d\alpha/dt$ $(\rm{yr}^{-1})$ &...                         &0.758 $\pm$ 0.995          &0.034 $\pm$ 1.716          &-0.445 $\pm$ 1.199         &-0.674 $\pm$ 1.345\\
  $I_S$ [KMTC]                  &19.648 $\pm$ 0.048          &19.371 $\pm$ 0.054         &19.356 $\pm$ 0.055         &19.367 $\pm$ 0.053         &19.345 $\pm$ 0.053 \\
  $I_B$ [KMTC]                  &18.521 $\pm$ 0.012          &18.673 $\pm$ 0.023         &18.666 $\pm$ 0.024         &18.673 $\pm$ 0.024         &18.670 $\pm$ 0.024\\
  $t_*$ $(\rm{hours})$          &$11.522_{-0.609}^{+0.414}$  &$11.466_{-0.834}^{+0.656}$ &11.443 $\pm$ 0.689         &11.501 $\pm$ 0.604         &$11.436_{-0.889}^{+0.661}$\\
  $\beta$                                &...                &$0.145_{-0.108}^{+0.142}$  &$0.270_{-0.193}^{+0.291}$  &$0.267_{-0.204}^{+0.286}$  &$0.250_{-0.182}^{+0.300}$\\
\enddata
\label{tab:1746parms_inner}
\end{deluxetable}

\begin{deluxetable}{lccccc} \rotate
	\tablecolumns{6} \tablewidth{0pc} \tablecaption{\textsc{Microlens Parameters for KMT-2023-BLG-1746, outer models}} 
	\tablehead{ \colhead{ } & \colhead{	}& \multicolumn{4}{c}{Parallax + orbital motion models} \\
		\cline{3-6} 
		\colhead{Parameters } & \colhead{Standard}&	\colhead{${\rm{outer(+,-)}}$}&\colhead{${\rm{outer(+,+)}}$} &	\colhead{${\rm{outer(-,-)}}$}&\colhead{${\rm{outer(-,+)}}$}\\
		\cline{3-6}
	\colhead{ } & \colhead{}&	\colhead{$u_0>0,\pi_{\rm{E},\it{N}}<0$}&\colhead{$u_0>0,\pi_{\rm{E},\it{N}}>0$} &	   \colhead{$u_0<0,\pi_{\rm{E},\it{N}}<0$}&\colhead{$u_0<0,\pi_{\rm{E},\it{N}}>0$}} \startdata
  $\chi^2/\rm{dof}$             &1197.211/1112               &1109.570/1109              &1109.811/1109              &1109.627/1109              &1109.757/1109\\
  $t_0-2460190$                 &9.813 $\pm$ 0.023           &9.912 $\pm$ 0.026          &9.911 $\pm$ 0.025          &9.910 $\pm$ 0.026          &9.912 $\pm$ 0.026\\
  $u_0$                         &0.045 $\pm$ 0.002           &0.056 $\pm$ 0.003          &0.058 $\pm$ 0.003          &-0.057 $\pm$ 0.003         &-0.058 $\pm$ 0.003\\
  $t_{\rm E}$ $(\rm{days})$     &113.056 $\pm$ 4.426         &93.184 $\pm$ 4.072         &89.636 $\pm$ 4.276         &90.539 $\pm$ 4.289         &89.821 $\pm$ 3.870\\
  $s$                           &$0.988_{-0.004}^{+0.006}$   &$0.987_{-0.006}^{+0.008}$  &0.985 $\pm$ 0.008          &0.984 $\pm$ 0.007          &0.986 $\pm$ 0.007 \\
  $q$ $(10^{-5})$               &4.591 $\pm$ 0.540           &$5.702_{-0.750}^{+1.116}$  &$6.170_{-0.801}^{+1.076}$  &6.188 $\pm$ 0.805          &5.914 $\pm$ 0.854 \\
  log $q$ (mean)                &-4.338 $\pm$ 0.051          &-4.236 $\pm$ 0.071         &-4.205 $\pm$ 0.067         &-4.210 $\pm$ 0.054         &-4.228 $\pm$ 0.060\\
  $\alpha$ $(\rm{rad})$         &4.218 $\pm$ 0.005           &$4.255_{-0.015}^{+0.010}$  &4.237 $\pm$ 0.014          &2.050 $\pm$ 0.011          &2.034 $\pm$ 0.012\\
  $\rho$ $(10^{-3})$            &4.019 $\pm$ 0.296           &$4.829_{-0.623}^{+0.424}$  &$5.192_{-0.577}^{+0.443}$  &5.179 $\pm$ 0.445          &$5.093_{-0.511}^{+0.407}$\\
  $\pi_{\rm{E},\it{N}}$         &...                         &-0.186 $\pm$ 0.107         &0.083 $\pm$ 0.080          &-0.176 $\pm$ 0.104         &0.101 $\pm$ 0.087\\
  $\pi_{\rm{E},\it{E}}$         &...                         &-0.177 $\pm$ 0.036         &-0.123 $\pm$ 0.019         &-0.166 $\pm$ 0.029         &-0.129 $\pm$ 0.018\\
  $ds/dt$ $(\rm{yr}^{-1})$      &...                         &-0.462 $\pm$ 1.111         &-0.318 $\pm$ 1.281         &-0.217 $\pm$ 1.231         &$-0.549_{-0.954}^{+1.219}$\\
  $d\alpha/dt$ $(\rm{yr}^{-1})$ &...                         &$-0.766_{-1.075}^{+1.794}$ &$0.731_{-2.041}^{+1.565}$  &-0.703 $\pm$ 1.279         &0.439 $\pm$ 1.513\\
  $I_S$ [KMTC]                  &19.641 $\pm$ 0.047          &19.375 $\pm$ 0.056         &19.350 $\pm$ 0.055         &19.367 $\pm$ 0.054         &19.344 $\pm$ 0.055 \\
  $I_B$ [KMTC]                  &18.524 $\pm$ 0.012          &18.672 $\pm$ 0.024         &18.669 $\pm$ 0.024         &18.673 $\pm$ 0.024         &18.671 $\pm$ 0.026\\
  $t_*$ $(\rm{hours})$          &$10.951_{-0.765}^{+0.565}$  &$10.895_{-1.446}^{+0.757}$ &$11.233_{-1.179}^{+0.733}$ &$11.290_{-0.918}^{+0.670}$ &$11.013_{-1.009}^{+0.704}$\\
  $\beta$                                &...                &0.392 $\pm$ 0.261          &0.346 $\pm$ 0.248          &$0.286_{-0.208}^{+0.306}$  &$0.260_{-0.173}^{+0.276}$\\
\enddata
\label{tab:1746parms_outer}
\end{deluxetable}

\begin{deluxetable}{lccccc} \rotate
	\tablecolumns{6} \tablewidth{0pc} \tablecaption{\textsc{Microlens Parameters for KMT-2023-BLG-1746, close models}} 
	\tablehead{ \colhead{} & \colhead{}& \multicolumn{4}{c}{Parallax + orbital motion models} \\
		\cline{3-6} 
		\colhead{Parameters} & \colhead{Standard}&	\colhead{${\rm{close(+,-)}}$}&\colhead{${\rm{close(+,+)}}$} &	\colhead{${\rm{close(-,-)}}$}&\colhead{${\rm{close(-,+)}}$}\\
		\cline{3-6}
	\colhead{} & \colhead{}&	\colhead{$u_0>0,\pi_{\rm{E},\it{N}}<0$}&\colhead{$u_0>0,\pi_{\rm{E},\it{N}}>0$} &	   \colhead{$u_0<0,\pi_{\rm{E},\it{N}}<0$}&\colhead{$u_0<0,\pi_{\rm{E},\it{N}}>0$}} \startdata
  $\chi^2/\rm{dof}$             &1205.003/1112               &1115.207/1109              &1115.668/1109              &1114.831/1109              &1115.499/1109\\
  $t_0-2460190$                 &9.813 $\pm$ 0.023           &9.916 $\pm$ 0.026          &9.915 $\pm$ 0.026          &9.916 $\pm$ 0.026          &9.915 $\pm$ 0.026\\
  $u_0$                         &0.044 $\pm$ 0.002           &0.057 $\pm$ 0.003          &0.059 $\pm$ 0.003          &-0.058 $\pm$ 0.003         &-0.058 $\pm$ 0.003\\
  $t_{\rm E}$ $(\rm{days})$     &113.582 $\pm$ 4.491         &91.796 $\pm$ 4.235         &87.937 $\pm$ 4.093         &87.975 $\pm$ 4.335         &89.729 $\pm$ 3.700\\
  $s$                           &$0.929_{-0.004}^{+0.006}$   &0.927 $\pm$ 0.008          &0.926 $\pm$ 0.008          &0.927 $\pm$ 0.008          &0.929 $\pm$ 0.008 \\
  $q$ $(10^{-5})$               &7.487 $\pm$ 0.870           &9.380 $\pm$ 1.157          &10.059 $\pm$ 1.234         &10.164 $\pm$ 1.390         &9.545 $\pm$ 1.173 \\
  log $q$ (mean)                &-4.128 $\pm$ 0.049          &-4.028 $\pm$ 0.052         &-3.999 $\pm$ 0.053         &-3.994 $\pm$ 0.057         &-4.020 $\pm$ 0.051\\
  $\alpha$ $(\rm{rad})$         &4.219 $\pm$ 0.005           &4.252 $\pm$ 0.010          &4.242 $\pm$ 0.009          &2.046 $\pm$ 0.010          &2.039 $\pm$ 0.009\\
  $\rho$ $(10^{-3})$            &$<3.8$                      &$<5.2$                     &$<6.3$                     &$<6.2$                     &$<6.4$\\
  $\pi_{\rm{E},\it{N}}$         &...                         &-0.234 $\pm$ 0.107         &0.117 $\pm$ 0.086          &-0.247 $\pm$ 0.095         &0.113 $\pm$ 0.089\\
  $\pi_{\rm{E},\it{E}}$         &...                         &-0.195 $\pm$ 0.039         &-0.124 $\pm$ 0.019         &-0.188 $\pm$ 0.031         &-0.129 $\pm$ 0.018\\
  $ds/dt$ $(\rm{yr}^{-1})$      &...                         &$-0.551_{-0.646}^{+0.859}$ &$-0.473_{-0.636}^{+0.834}$ &$-0.705_{-0.639}^{+0.913}$ &$-0.694_{-0.583}^{+0.742}$\\
  $d\alpha/dt$ $(\rm{yr}^{-1})$ &...                         &0.053 $\pm$ 1.047          &0.138 $\pm$ 0.957          &$-0.315_{-0.891}^{+1.169}$ &-0.435 $\pm$ 0.819\\
  $I_S$ [KMTC]                  &19.647 $\pm$ 0.048          &19.355 $\pm$ 0.059         &19.329 $\pm$ 0.054         &19.338 $\pm$ 0.056         &19.342 $\pm$ 0.052 \\
  $I_B$ [KMTC]                  &18.522 $\pm$ 0.012          &18.678 $\pm$ 0.025         &18.675 $\pm$ 0.025         &18.681 $\pm$ 0.025         &18.670 $\pm$ 0.024\\
  $\beta$                                 &...                &0.356 $\pm$ 0.251          &0.354 $\pm$ 0.251          &0.410 $\pm$ 0.250          &0.357 $\pm$ 0.248\\
\enddata
\label{tab:1746parms_close}
\end{deluxetable}

\begin{deluxetable}{lccccc} \rotate
	\tablecolumns{6} \tablewidth{0pc} \tablecaption{\textsc{Microlens Parameters for KMT-2023-BLG-1746, wide models}} 
	\tablehead{ \colhead{} & \colhead{}& \multicolumn{4}{c}{Parallax + orbital motion models} \\
		\cline{3-6} 
		\colhead{Parameters} & \colhead{Standard}&	\colhead{${\rm{wide(+,-)}}$}&\colhead{${\rm{wide(+,+)}}$} &	\colhead{${\rm{wide(-,-)}}$}&\colhead{${\rm{wide(-,+)}}$}\\
		\cline{3-6}
	\colhead{} & \colhead{}&	\colhead{$u_0>0,\pi_{\rm{E},\it{N}}<0$}&\colhead{$u_0>0,\pi_{\rm{E},\it{N}}>0$} &	   \colhead{$u_0<0,\pi_{\rm{E},\it{N}}<0$}&\colhead{$u_0<0,\pi_{\rm{E},\it{N}}>0$}} \startdata
  $\chi^2/\rm{dof}$             &1204.215/1112               &1113.598/1109              &1114.355/1109              &1114.012/1109              &1114.488/1109\\
  $t_0-2460190$                 &9.814 $\pm$ 0.023           &9.917 $\pm$ 0.026          &9.917 $\pm$ 0.025          &9.915 $\pm$ 0.025          &9.916 $\pm$ 0.026\\
  $u_0$                         &0.045 $\pm$ 0.002           &0.058 $\pm$ 0.003          &0.060 $\pm$ 0.003          &-0.057 $\pm$ 0.003         &-0.059 $\pm$ 0.003\\
  $t_{\rm E}$ $(\rm{days})$     &113.322 $\pm$ 4.494         &91.578 $\pm$ 4.171         &86.535 $\pm$ 3.960         &89.488 $\pm$ 4.271         &88.877 $\pm$ 3.845\\
  $s$                           &$1.023_{-0.007}^{+0.005}$   &1.011 $\pm$ 0.008          &1.010 $\pm$ 0.008          &$1.016_{-0.010}^{+0.008}$  &1.012 $\pm$ 0.008 \\
  $q$ $(10^{-5})$               &7.285 $\pm$ 0.923           &9.251 $\pm$ 1.134          &9.936 $\pm$ 1.241          &9.823 $\pm$ 1.415          &9.558 $\pm$ 1.236 \\
  log $q$ (mean)                &-4.139 $\pm$ 0.055          &-4.034 $\pm$ 0.051         &-4.005 $\pm$ 0.053         &-4.010 $\pm$ 0.061         &-4.021 $\pm$ 0.055\\
  $\alpha$ $(\rm{rad})$         &4.219 $\pm$ 0.005           &4.254 $\pm$ 0.009          &4.247 $\pm$ 0.008          &2.046 $\pm$ 0.009          &2.035 $\pm$ 0.009\\
  $\rho$ $(10^{-3})$            &$<3.8$                      &$<5.0$                     &$<5.6$                     &$<6.1$                     &$<5.9$\\
  $\pi_{\rm{E},\it{N}}$         &...                         &-0.239 $\pm$ 0.105         &0.142 $\pm$ 0.082          &-0.212 $\pm$ 0.107         &0.120 $\pm$ 0.089\\
  $\pi_{\rm{E},\it{E}}$         &...                         &-0.198 $\pm$ 0.039         &-0.124 $\pm$ 0.019         &-0.177 $\pm$ 0.031         &-0.132 $\pm$ 0.018\\
  $ds/dt$ $(\rm{yr}^{-1})$      &...                         &$0.634_{-0.838}^{+0.594}$  &$0.578_{-0.723}^{+0.581}$  &-0.105 $\pm$ 0.840         &$0.266_{-0.892}^{+0.700}$\\
  $d\alpha/dt$ $(\rm{yr}^{-1})$ &...                         &-0.192 $\pm$ 0.789         &-0.549 $\pm$ 0.643         &-0.363 $\pm$ 0.840         &-0.056 $\pm$ 0.858\\
  $I_S$ [KMTC]                  &19.644 $\pm$ 0.048          &19.351 $\pm$ 0.059         &19.312 $\pm$ 0.053         &19.356 $\pm$ 0.054         &19.330 $\pm$ 0.055 \\
  $I_B$ [KMTC]                  &18.523 $\pm$ 0.012          &18.680 $\pm$ 0.025         &18.679 $\pm$ 0.025         &18.676 $\pm$ 0.024         &18.676 $\pm$ 0.025\\
  $\beta$                                &...                &0.357 $\pm$ 0.241          &0.386 $\pm$ 0.255          &0.328 $\pm$ 0.252          &0.356 $\pm$ 0.252\\
\enddata 
\label{tab:1746parms_wide}
\end{deluxetable}

\begin{deluxetable}{lccc}
\tablecolumns{4} \tablewidth{0pc} \tablecaption{\textsc{Microlens Parameters for KMT-2023-BLG-0614}} \tablehead{\colhead{Parameters} & \colhead{inner} & \colhead{outer} & \colhead{1L2S}} \startdata
  $\chi^2/\rm{dof}$             &6654.314/6654              &6653.790/6654               &6652.735/6654\\
  $t_0-2460060$                 &2.346 $\pm$ 0.241          &2.433 $\pm$ 0.249           &2.670 $\pm$ 0.335\\
  $u_0$                         &0.287 $\pm$ 0.067          &$0.305_{-0.069}^{+0.088}$   &0.678 $\pm$ 0.187        \\
  $t_{\rm E}$ $(\rm{days})$     &31.823 $\pm$ 5.121         &30.635 $\pm$ 5.279          &$20.437_{-3.026}^{+4.578}$\\
  $s$                           &1.470 $\pm$ 0.051          &0.947 $\pm$ 0.066           &...\\
  $q$ $(10^{-2})$               &$1.347_{-0.296}^{+0.373}$  &$1.175_{-0.268}^{+0.331}$   &...        \\
  log $q$ (mean)                &-1.872 $\pm$ 0.106         &-1.932 $\pm$ 0.111          &...\\
  $\alpha$ $(\rm{rad})$         &1.662 $\pm$ 0.026          &1.670 $\pm$ 0.027           &...\\
 $\rho$ $(10^{-2})$            & $<3.0$                       & $<4.0$            &...\\
  $t_{0,2}-2460060$             &...                        &...                         &1.497 $\pm$ 0.057          \\
  $u_{0,2}$                     &...                        &...                         &0.015 $\pm$ 0.008          \\
  $\rho_2$ $(10^{-2})$          &...                        &...                         &  $<4.5$   \\
  $q_F$ $(10^{-2})$             &...                        &...                         &$1.478_{-0.390}^{+0.535}$          \\
  $I_S$ [KMTC]                  &21.855 $\pm$ 0.311         &21.788 $\pm$ 0.346          &20.718 $\pm$ 0.512\\
  $I_B$ [KMTC]                  &$18.210_{-0.009}^{+0.012}$ &$18.212_{-0.010}^{+0.015}$  &$18.281_{-0.045}^{+0.064}$       \\
  $I_{S,2}$ [KMTC]              &...                        &...                         &25.340 $\pm$ 0.302         \\
  
\enddata
\label{tab:0614parms}
\end{deluxetable}

\begin{deluxetable}{lccc}
\tablecolumns{4} \tablewidth{0pc} \tablecaption{\textsc{Microlens Parameters for KMT-2023-BLG-1593}} \tablehead{\colhead{Parameters} & \colhead{inner} & \colhead{outer} & \colhead{1L2S}} \startdata
  $\chi^2/\rm{dof}$             &11950.198/11947            &11946.860/11947             &11952.946/11947\\
  $t_0-2460130$                 &6.903 $\pm$ 0.082          &6.834 $\pm$ 0.084           &6.522 $\pm$ 0.105\\
  $u_0$                         &0.256 $\pm$ 0.023          &0.242 $\pm$ 0.021           &0.260 $\pm$ 0.024\\
  $t_{\rm E}$ $(\rm{days})$     &28.703 $\pm$ 1.814         &29.591 $\pm$ 1.811          &29.161 $\pm$ 1.939\\
  $s$                           &1.371 $\pm$ 0.021          &0.966 $\pm$ 0.021           &...\\
  $q$ $(10^{-3})$               &3.307 $\pm$ 0.502          &3.215 $\pm$ 0.492           &...\\
  log $q$ (mean)                &-2.481 $\pm$ 0.066         &-2.494 $\pm$ 0.066          &...\\
  $\alpha$ $(\rm{rad})$         &1.105 $\pm$ 0.015          &1.087 $\pm$ 0.015           &...\\
  $\rho$ $(10^{-2})$            &$<2.6$  &$<2.5$   &...\\
  $t_{0,2}-2460130$             &...                        &...                         &10.577 $\pm$ 0.041\\
  $u_{0,2}$                     &...                        &...                         &$0.013_{-0.005}^{+0.003}$\\
  $\rho_2$ $(10^{-2})$          &...                        &...                         &$<2.6 $ \\
  $q_F$ $(10^{-2})$             &...                        &...                         &1.162 $\pm$ 0.233\\
  $I_S$ [KMTC]                  &21.213 $\pm$ 0.117         &21.285 $\pm$ 0.111          &21.251 $\pm$ 0.123\\
  $I_B$ [KMTC]                  &$21.812_{-0.169}^{+0.233}$ &$21.698_{-0.137}^{+0.180}$  &$21.749_{-0.163}^{+0.221}$\\
  $I_{S,2}$ [KMTC]              &...                        &...                         &26.102 $\pm$ 0.202\\
  
\enddata
\label{tab:1593parms}
\end{deluxetable}

\begin{deluxetable}{lccccccc}
\tablecolumns{8} 
\tablewidth{0pc}\tablecaption{\textsc{Physical properties}} 
\tablehead{\colhead{Event} & \multicolumn{4}{c}{Physical Properties} & \colhead{} &
\multicolumn{2}{c}{Relative Weights}\\
\cline{7-8} \colhead{Models}&\colhead{$M_{\rm host}$ $[M_\sun]$} &\colhead{$M_{\rm planet}$ $[M_\earth]$} 
&\colhead{$D_{\rm L}$ [kpc]} &\colhead{$a_\bot$ [au]} &  &
\colhead{Gal.Mod.} & \colhead{$\chi^2$}} \startdata
 KB230164  \\
 inner$(-)$ &$0.29_{-0.11}^{+0.14}$  &$13.98_{-5.21}^{+6.88}$  &$0.97_{-0.21}^{+0.37}$  &$1.19_{-0.18}^{+0.31}$   &&0.391 &0.715\\
 inner$(+)$ &0.44 $\pm$ 0.22         &16.75 $\pm$ 8.55         &$1.16_{-0.33}^{+0.64}$  &1.57 $\pm$ 0.27          &&1.000 &0.176\\
 outer$(-)$ &$0.39_{-0.11}^{+0.06}$  &$12.52_{-3.43}^{+1.95}$  &$0.87_{-0.23}^{+0.11}$  &$1.61_{-0.26}^{+0.16}$   &&0.000 &1.000\\
 outer$(+)$ &0.79 $\pm$ 0.05         &25.75 $\pm$ 1.57         &$0.66_{-0.01}^{+0.08}$  &2.12 $\pm$ 0.08          &&0.000 &0.206\\
 Adopted     &$0.32_{-0.13}^{+0.21}$  &$14.95_{-6.01}^{+8.21}$  &$1.04_{-0.27}^{+0.51}$  &$1.33_{-0.25}^{+0.39}$  && &\\[1ex]
         
 \cline{1-8}\\[-1ex]
  KB231286   \\
  inner     &$0.31_{-0.18}^{+0.29}$  &$19.89_{-11.07}^{+18.53}$     &$6.15_{-1.35}^{+0.99}$  &2.09 $\pm$ 0.61     &&0.998 &1.000\\
  outer     &$0.31_{-0.18}^{+0.29}$  &$19.58_{-10.90}^{+18.30}$     &$6.15_{-1.35}^{+0.99}$  &1.92 $\pm$ 0.56     &&1.000 &0.692\\
  Resonant  &$0.34_{-0.19}^{+0.30}$  &$750.74_{-418.27}^{+659.90}$  &$6.16_{-1.38}^{+0.99}$  &1.97 $\pm$ 0.58     &&0.814 &0.000\\
  Adopted   &$0.31_{-0.18}^{+0.29}$  &$19.78_{-11.01}^{+18.48}$     &$6.15_{-1.35}^{+0.99}$  &2.02 $\pm$ 0.60     && &\\[1ex]
 \cline{1-8}\\[-1ex]
 
 KB231746    \\  
 inner$(+,-)$ &$0.11_{-0.04}^{+0.11}$  &$2.00_{-0.78}^{+2.14}$   &7.46 $\pm$ 0.95         &$1.04_{-0.18}^{+0.75}$ &&0.004 &1.000\\
 inner$(+,+)$ &0.11 $\pm$ 0.03         &2.05 $\pm$ 0.48          &8.16 $\pm$ 0.81         &1.00 $\pm$ 0.14        &&0.010 &0.728\\
 inner$(-,-)$ &$0.08_{-0.02}^{+0.03}$  &$1.65_{-0.47}^{+0.67}$   &7.80 $\pm$ 0.84         &0.93 $\pm$ 0.13        &&0.004 &0.810\\
 inner$(-,+)$ &$0.12_{-0.03}^{+0.09}$  &$2.31_{-0.62}^{+1.73}$   &7.83 $\pm$ 0.92         &$1.05_{-0.17}^{+0.62}$ &&0.010 &0.772\\
 
 outer$(+,-)$ &$0.14_{-0.07}^{+0.18}$  &$2.73_{-1.31}^{+3.43}$   &$7.22_{-1.27}^{+0.99}$  &$1.34_{-0.39}^{+0.96}$ &&0.009 &0.760\\
 outer$(+,+)$ &$0.13_{-0.04}^{+0.08}$  &$2.63_{-0.74}^{+1.67}$   &7.83 $\pm$ 0.89         &$1.13_{-0.20}^{+0.58}$ &&0.015 &0.674\\
 outer$(-,-)$ &$0.10_{-0.03}^{+0.05}$  &$1.95_{-0.65}^{+1.10}$   &7.67 $\pm$ 0.88         &$1.01_{-0.14}^{+0.35}$ &&0.004 &0.739\\
 outer$(-,+)$ &$0.11_{-0.03}^{+0.04}$  &$2.20_{-0.54}^{+0.81}$   &7.99 $\pm$ 0.85         &$1.07_{-0.16}^{+0.27}$ &&0.010 &0.693\\
 
  Adopted     &$0.11_{-0.03}^{+0.08}$   &$2.23_{-0.68}^{+1.52}$   &7.78 $\pm$ 0.94         &$1.06_{-0.17}^{+0.55}$ && &\\[1ex]
  \cline{2-8}\\[-1ex]
 
 close$(+,-)$ &0.40 $\pm$ 0.20         &12.39 $\pm$ 6.10         &$4.36_{-1.75}^{+2.17}$  &2.32 $\pm$ 0.64        &&0.182 &0.045\\
 close$(+,+)$ &0.46 $\pm$ 0.19         &15.51 $\pm$ 6.48         &$4.80_{-1.28}^{+1.78}$  &2.64 $\pm$ 0.61        &&1.000 &0.036\\
 close$(-,-)$ &0.37 $\pm$ 0.18         &12.34 $\pm$ 6.12         &$3.77_{-1.40}^{+2.30}$  &2.20 $\pm$ 0.57        &&0.167 &0.055\\
 close$(-,+)$ &0.46 $\pm$ 0.19         &14.69 $\pm$ 6.18         &$4.81_{-1.30}^{+1.78}$  &2.64 $\pm$ 0.62        &&0.896 &0.039\\
 
  wide$(+,-)$ &0.40 $\pm$ 0.19         &12.29 $\pm$ 5.90         &$4.11_{-1.61}^{+2.25}$  &2.53 $\pm$ 0.68        &&0.152 &0.102\\
  wide$(+,+)$ &0.45 $\pm$ 0.19         &15.02 $\pm$ 6.13         &$4.50_{-1.20}^{+1.78}$  &2.83 $\pm$ 0.62        &&0.956 &0.070\\
  wide$(-,-)$ &0.39 $\pm$ 0.20         &12.59 $\pm$ 6.53         &4.81 $\pm$ 1.92         &2.55 $\pm$ 0.72        &&0.270 &0.083\\
  wide$(-,+)$ &0.45 $\pm$ 0.19         &14.39 $\pm$ 6.03         &$4.72_{-1.27}^{+1.77}$  &2.84 $\pm$ 0.65        &&0.943 &0.065\\

  Adopted     &0.44 $\pm$ 0.19         &14.34 $\pm$ 6.30         &$4.63_{-1.37}^{+1.85}$  &2.70 $\pm$ 0.66        && &\\[1ex]
  \cline{2-8}\\[-1ex]
  Adopted (Total) &0.39 $\pm$ 0.23    &12.72 $\pm$ 7.92          &$5.03_{-1.63}^{+2.30}$  &$2.54_{-1.05}^{+0.72}$ && &\\[1ex]

 \enddata
 \label{tab:physall}
\end{deluxetable}

\clearpage

\begin{figure}
\plotone{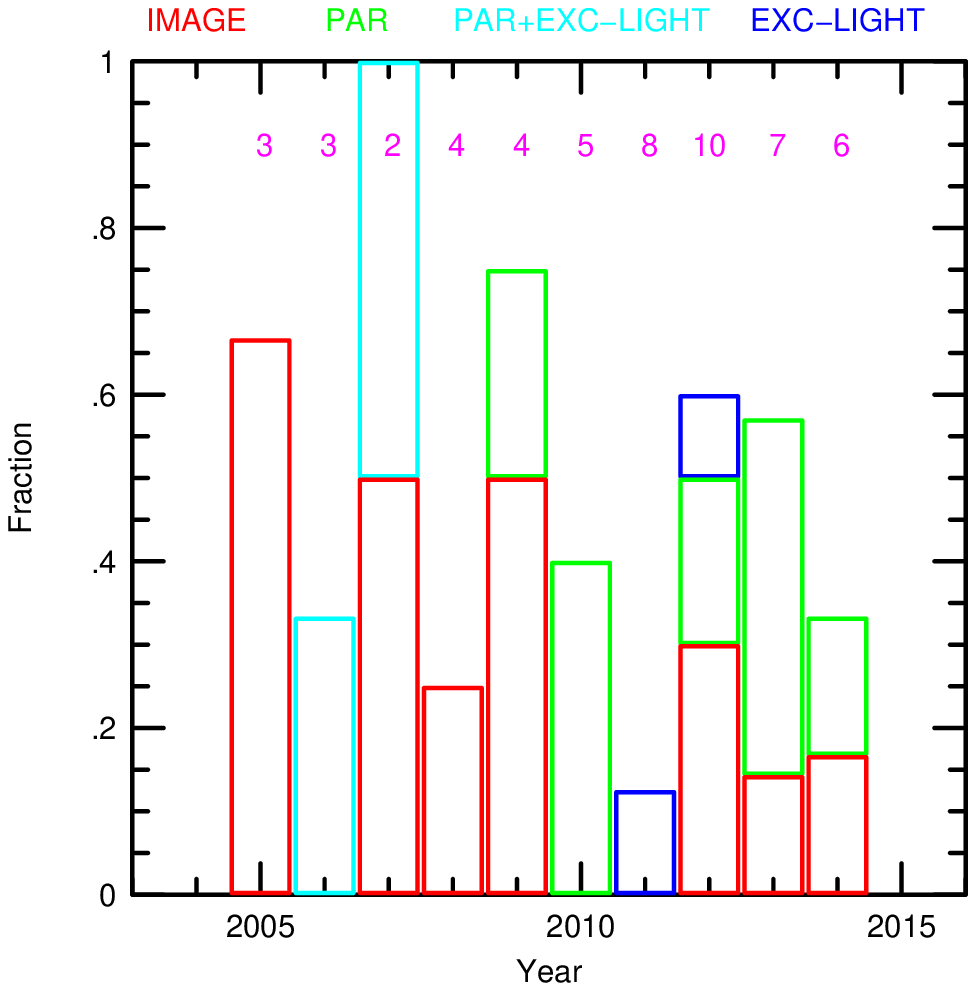}
\caption{Mass-Measurements of 24 microlensing hosts for events occurring
  2005-2014.  The magenta numbers show the total number of planetary events
  for each year.  The colors give the method of mass measurement, as
  indicated just above the image.
   Specifically:
    (1) Late-time high-resolution imaging \citep{ob05169bat,ob05169ben,gould22}
    (``IMAGE'', red, 11);
    (2) Combining light-curve measurements of the microlens parallax ($\pi_\e$)
    and Einstein radius ($\theta_\e$) to yield $M=\theta_\e/\kappa\pi_\e$
    \citep{gould92} (``PAR'', green, 9);
    (3) Combining light curve measurements of $\pi_\e$ with excess light,
    relative to the source flux derived from the light curve \citep{ob06109b},
    (``PAR+EXC-LIGHT'', cyan, 2);
    (4) Mass derived from excess-light alone \citep{mb11293b}
    (``EXC-LIGHT'', blue, 2).
}
\label{fig:event-hist}
\end{figure}

\begin{figure}
\plotone{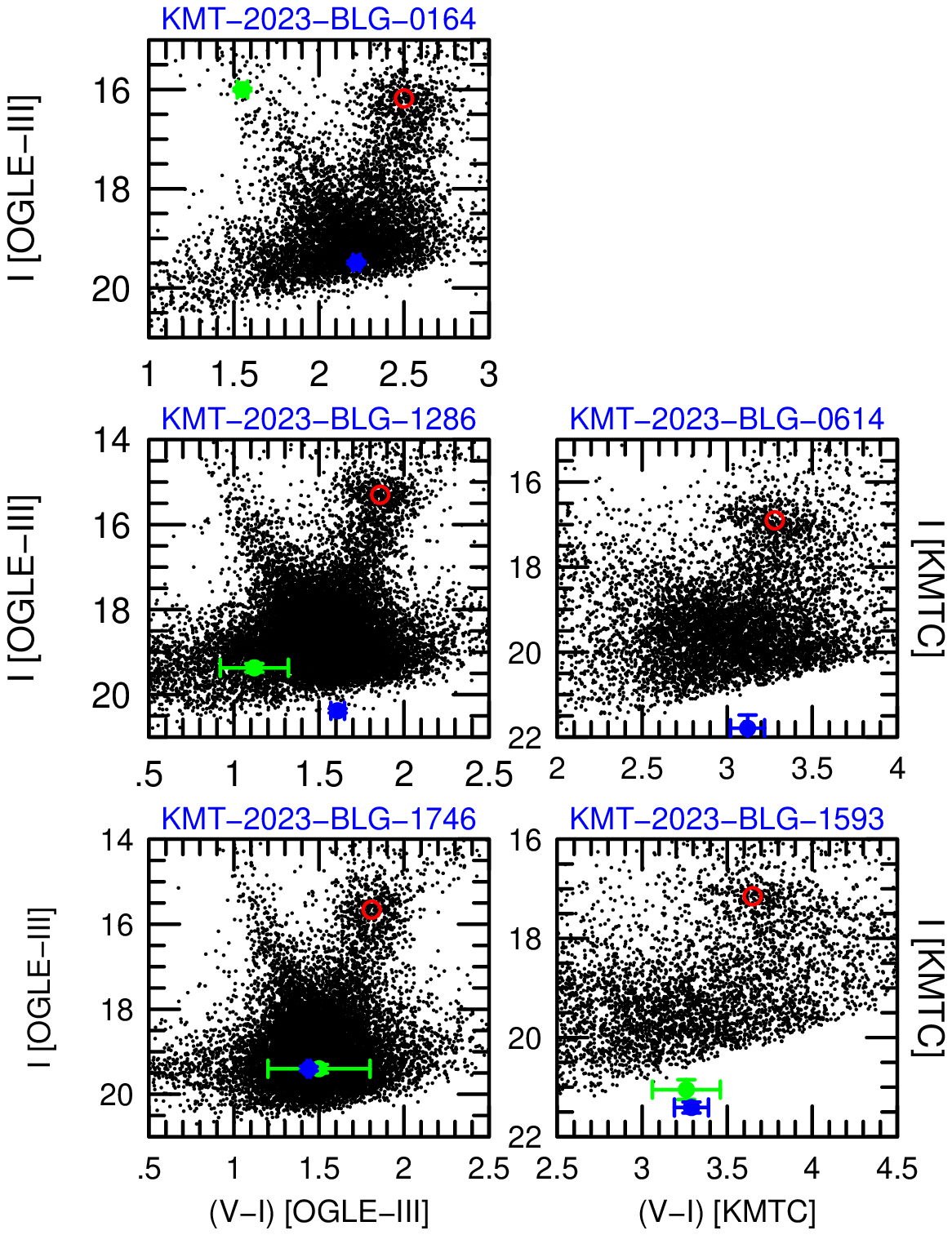}
\caption{Color-magnitude diagrams (CMDs) for the five events analyzed in
  this paper.  The positions of the source (blue), clump-giant centroid (red),
  and blend (green) are marked.
}
\label{fig:allcmd}
\end{figure}

\begin{figure}
\plotone{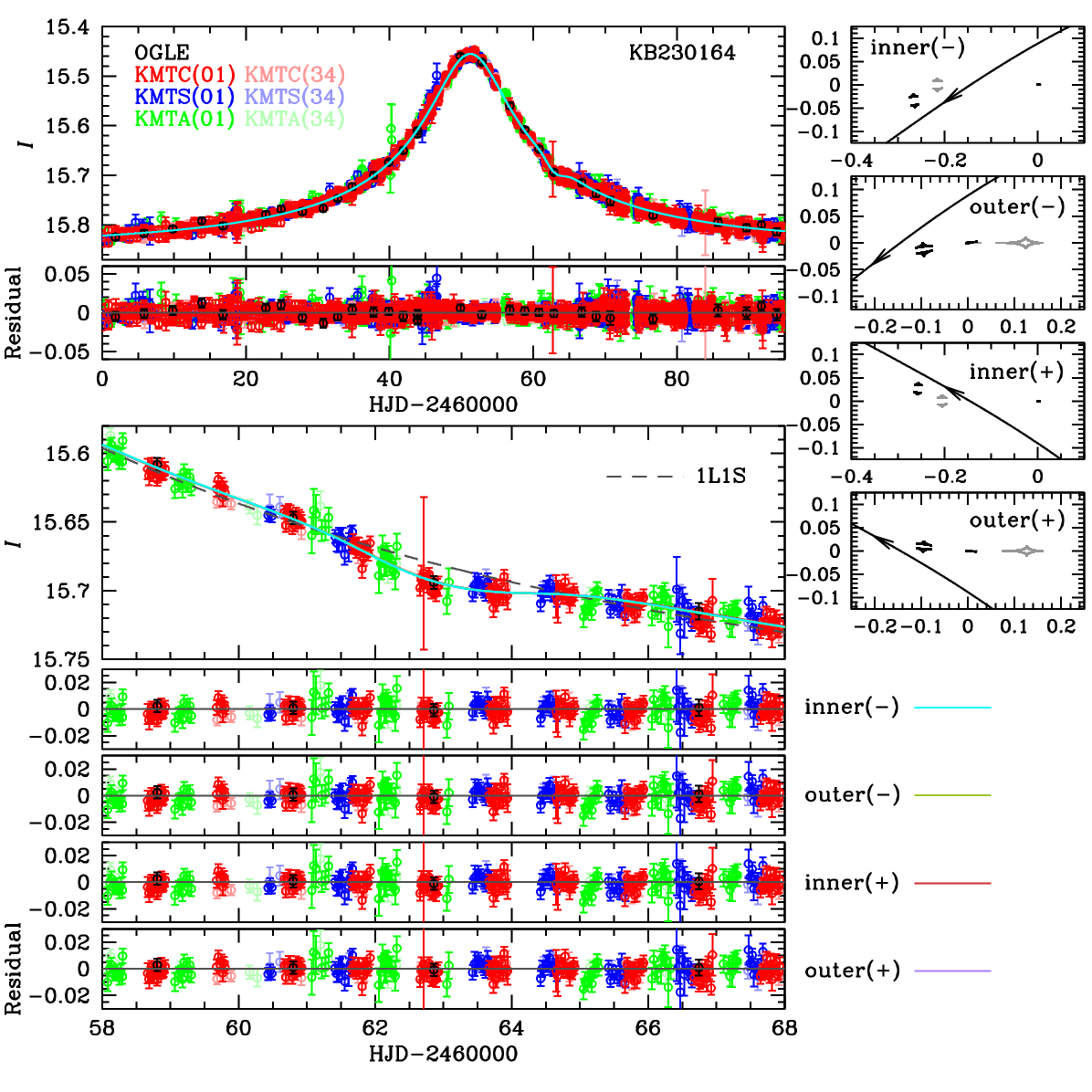}
\caption{Light curve and models for KMT-2023-BLG-0164.  For the
  four caustic-geometry insets at the right, the light-colored caustics
  represent the configurations at $t=t_0$, while the dark-colored caustics
  represent $t=t_{\rm anom}$.
}
\label{fig:0164lc}
\end{figure}

\begin{figure}
\plotone{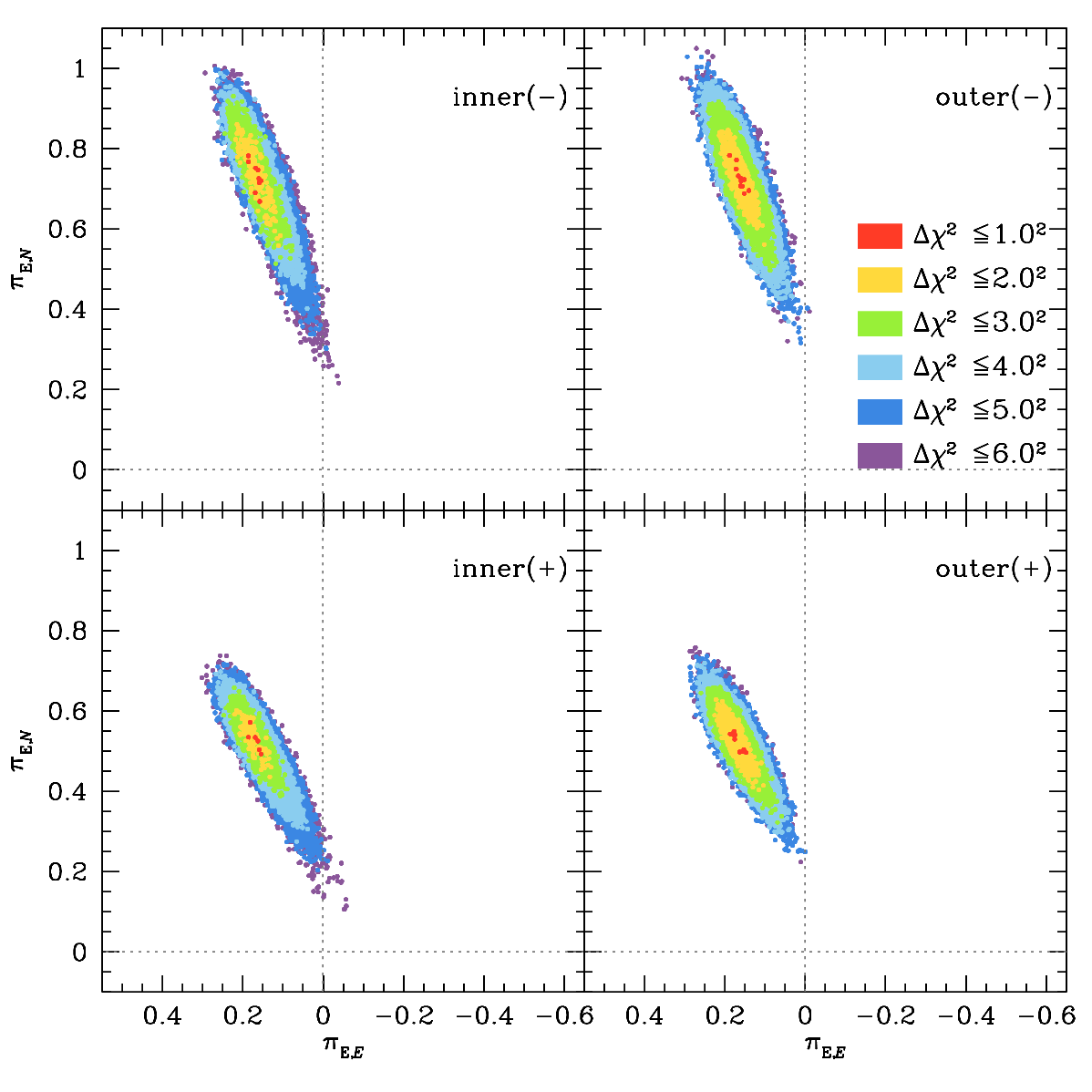}
\caption{Parallax contours for the four solutions of
  KMT-2023-BLG-0164.
  Colors indicate values of $\Delta\chi^2<1,4,9,16,25,36$.
}
\label{fig:0164parallax}
\end{figure}

\begin{figure}
\plotone{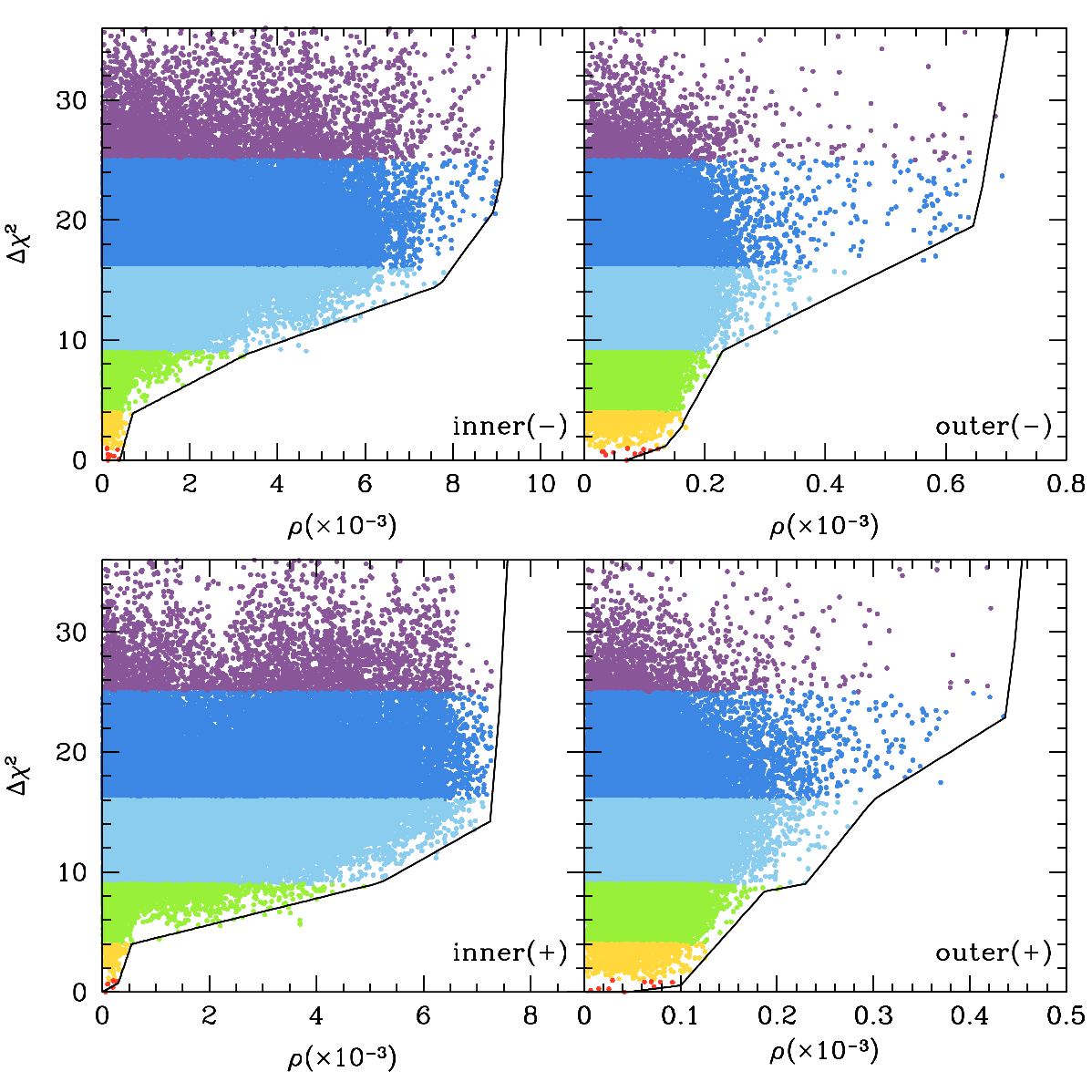}
\caption{Scatter plots of the normalized source radius, $\rho$, against
  the values of $\Delta\chi^2$ relative to the best fit for
  each of the four solutions of KMT-2023-BLG-0164.  Colors indicate
  values of $\Delta\chi^2<1,4,9,16,25,36$.  Also shown is the adopted
  envelope function, which is used in Section~\ref{sec:phys}.
}
\label{fig:0164rho}
\end{figure}

\begin{figure}
\plotone{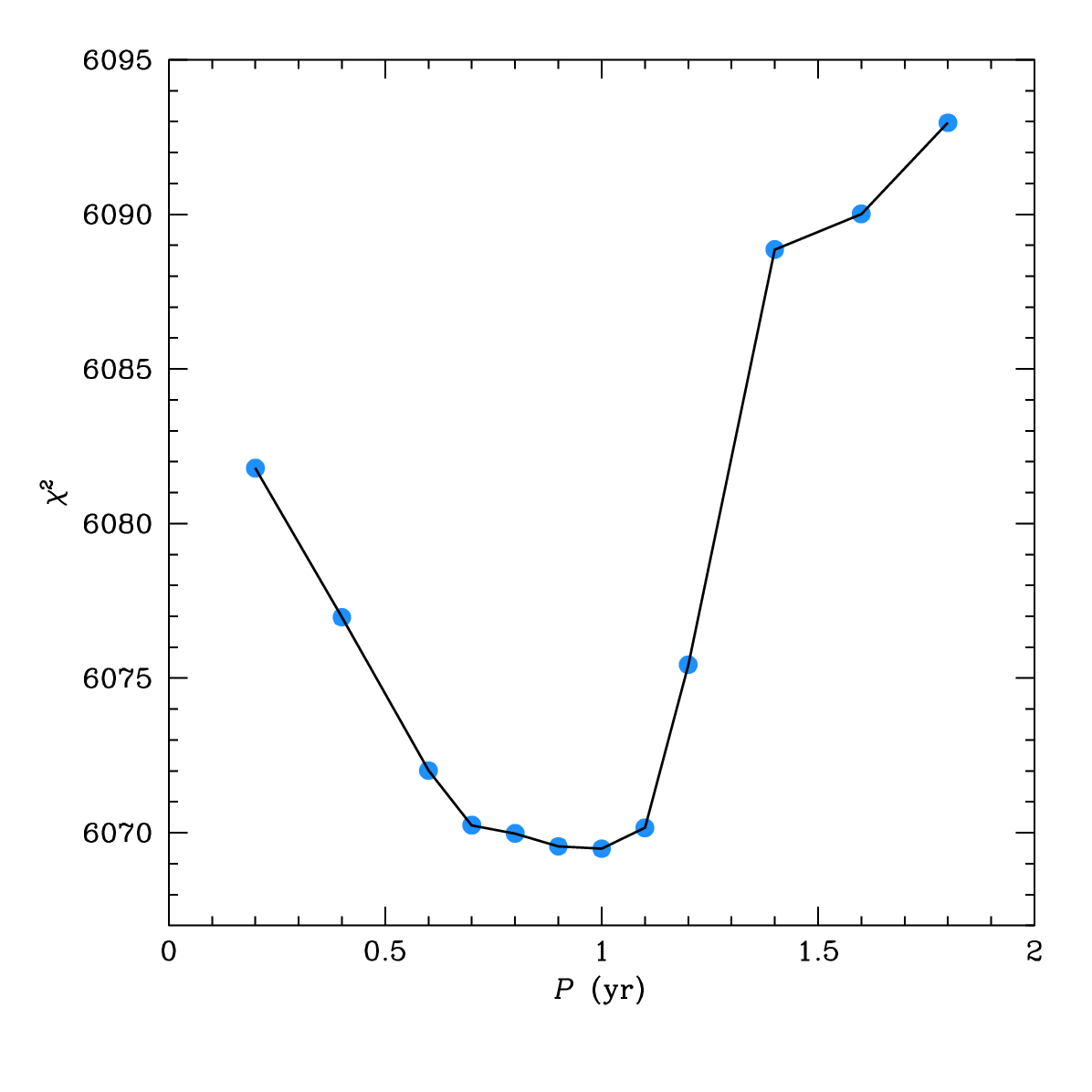}
\caption{Evolution of $\chi^2$ with assumed period $P$ for xallarap models
  of KMT-2023-BLG-0164 that allow for free fits to the orientation 
  parameters $(\alpha,\delta)$.  The xallarap period $P=1\,$yr is the best fit,
  as one would expect if the actual origin of the light-curve distortion
  is the parallax effect induced by Earth's annual orbital motion.
}
\label{fig:0164xalper}
\end{figure}

\begin{figure}
\plotone{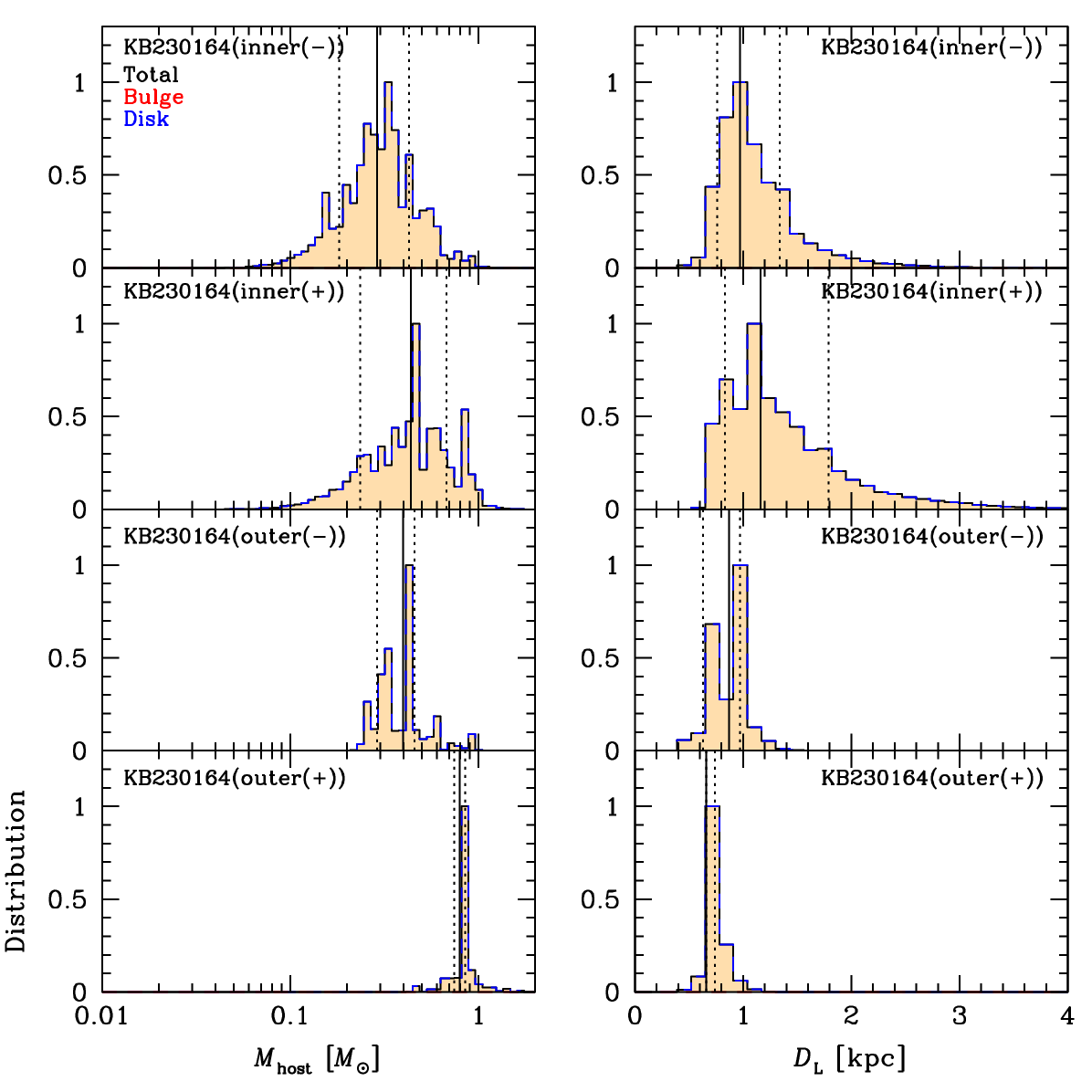}
\caption{Posterior lens mass and distance
  distributions for the four solutions for KMT-2023-BLG-0164.
}
\label{fig:0164bayes}
\end{figure}

\begin{figure}
\plotone{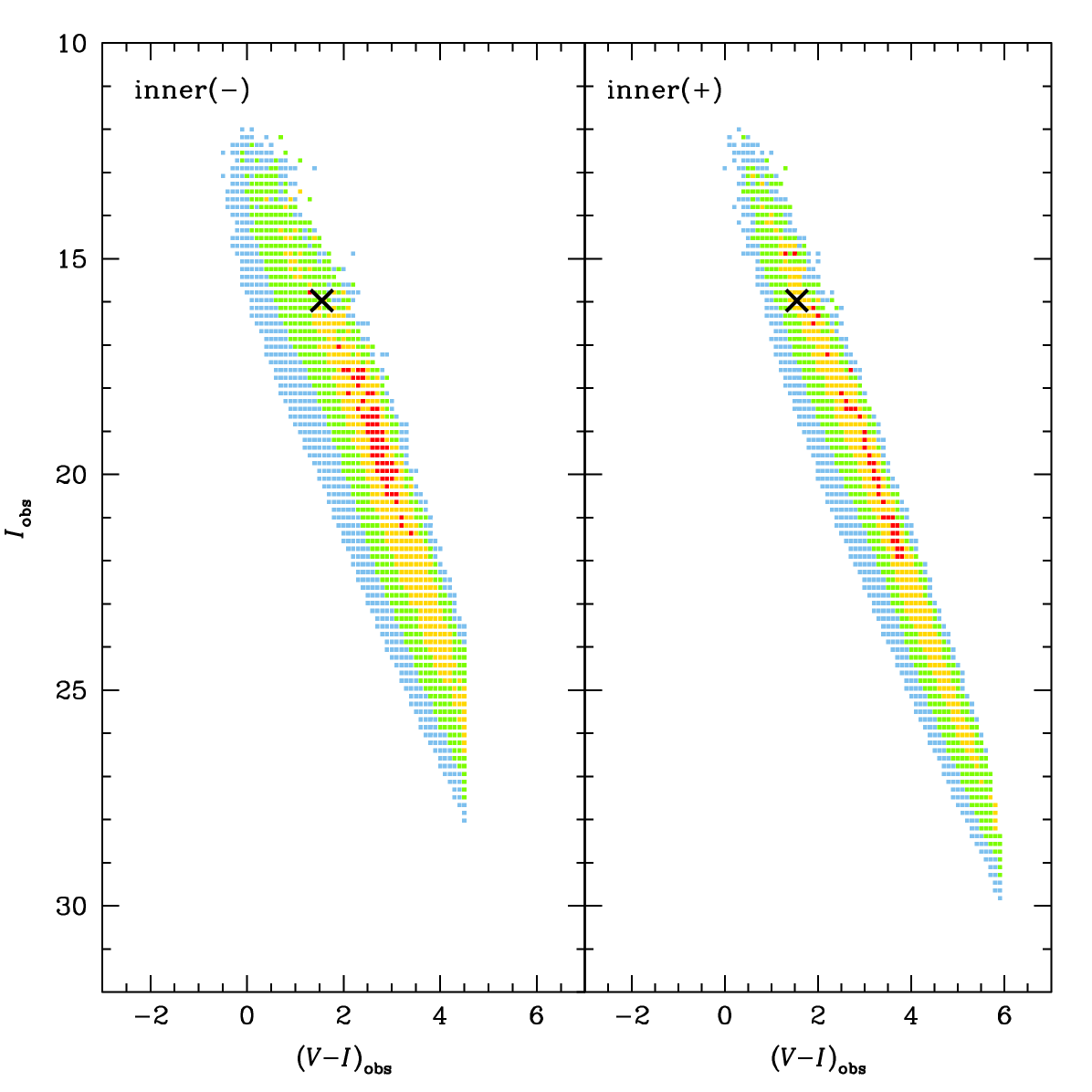}
\caption{Two-dimensional color-magnitude posterior distribution
  for the two ``inner'' solutions for KMT-2023-BLG-0164.
  The ``X'' marks the position of the blend.
}
\label{fig:0164bayescmd}
\end{figure}

\begin{figure}
  \plotone{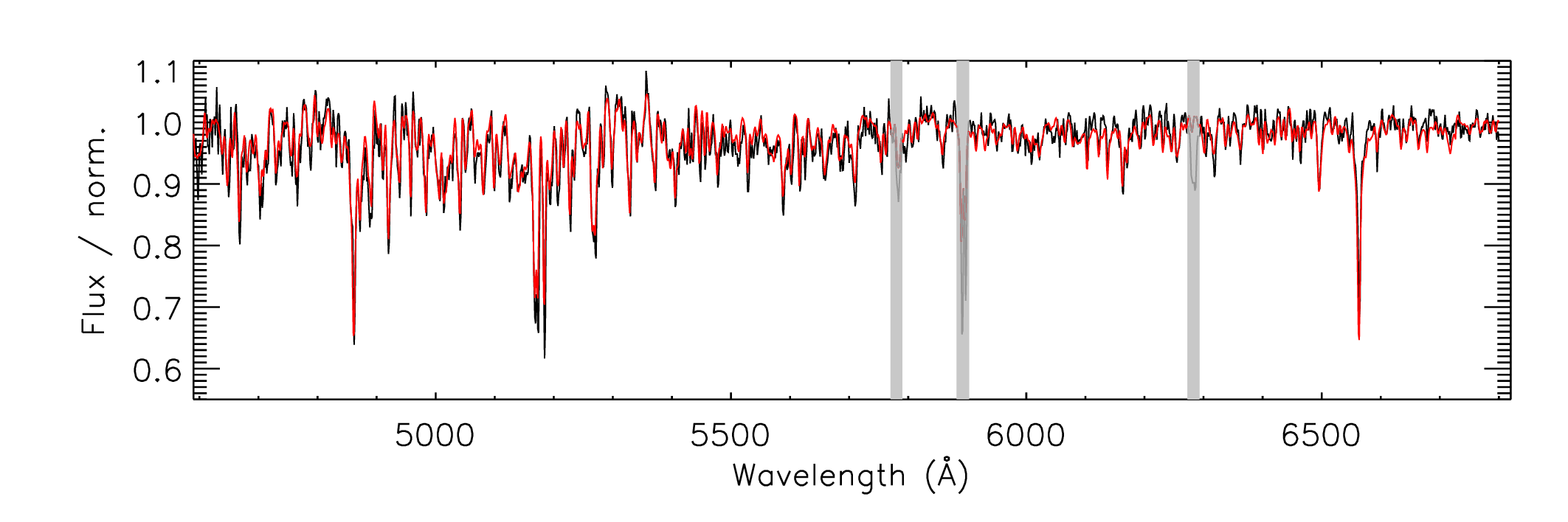}
  \caption{The spectrum of KMT-2023-BLG-0164 with the best-matched
    template. The black line shows the observed spectrum of the target
    obtained with IMACS, while the red line represents one of the
    best-matched MILES template spectrum (HD186427; $T_{\rm eff} =
    5762$\,K, $\log\,g = 4.43$, and [Fe/H] = 0.07), convolved to a
    resolution of $R = 1000$. All spectra are
    continuum-normalized. The three shaded regions indicate the masked
    wavelength ranges affected by interstellar absorption features,
    including the Na\,\textsc{i}\,D doublet and the diffuse
    interstellar bands (DIBs) at $\lambda\lambda$\,5780 and 6283\,\AA.
  }
  \label{fig:results_spmatch}
\end{figure}

\begin{figure}
\plotone{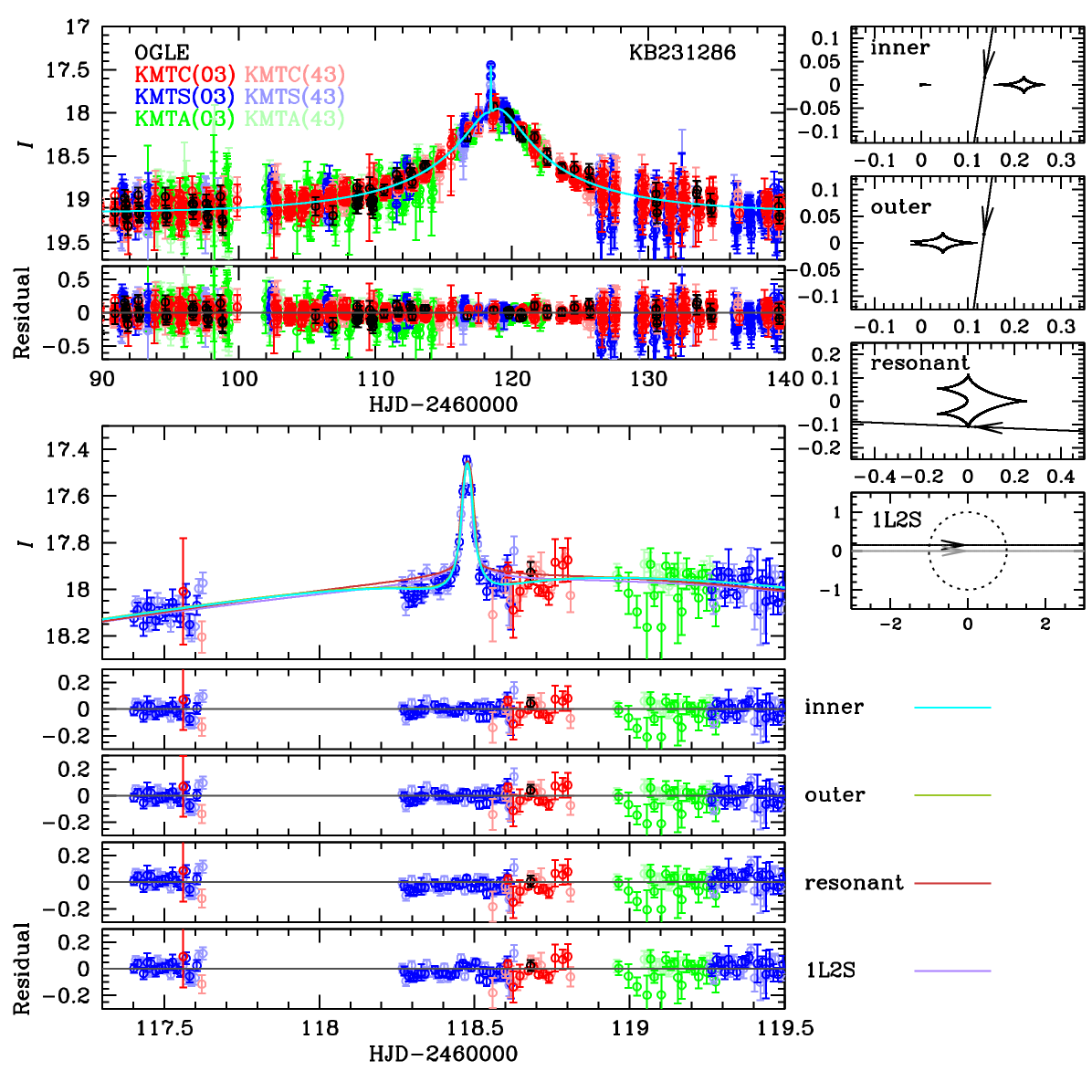}
  \caption{Light curve and models for KMT-2023-BLG-1286.
}
\label{fig:1286lc}
\end{figure}

\begin{figure}
\plotone{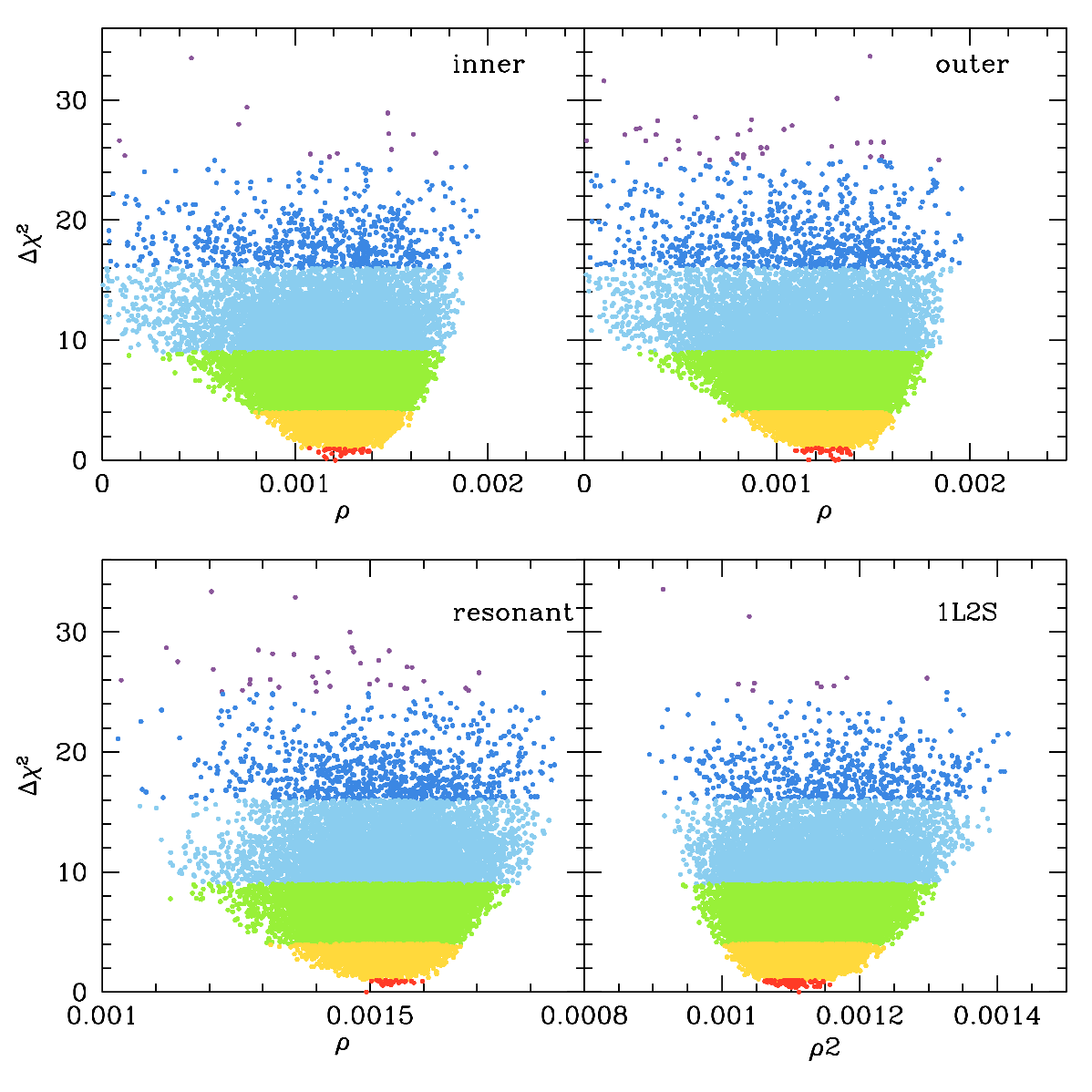}
\caption{Scatter plots of the normalized source radius, $\rho$, against
  the values of $\Delta\chi^2$ relative to the best fit
  for each of the four tabulated solutions of KMT-2023-BLG-1286.
  Colors indicate values of $\Delta\chi^2<1,4,9,16,25,36$.
}
\label{fig:1286rho}
\end{figure}

\begin{figure}
\plotone{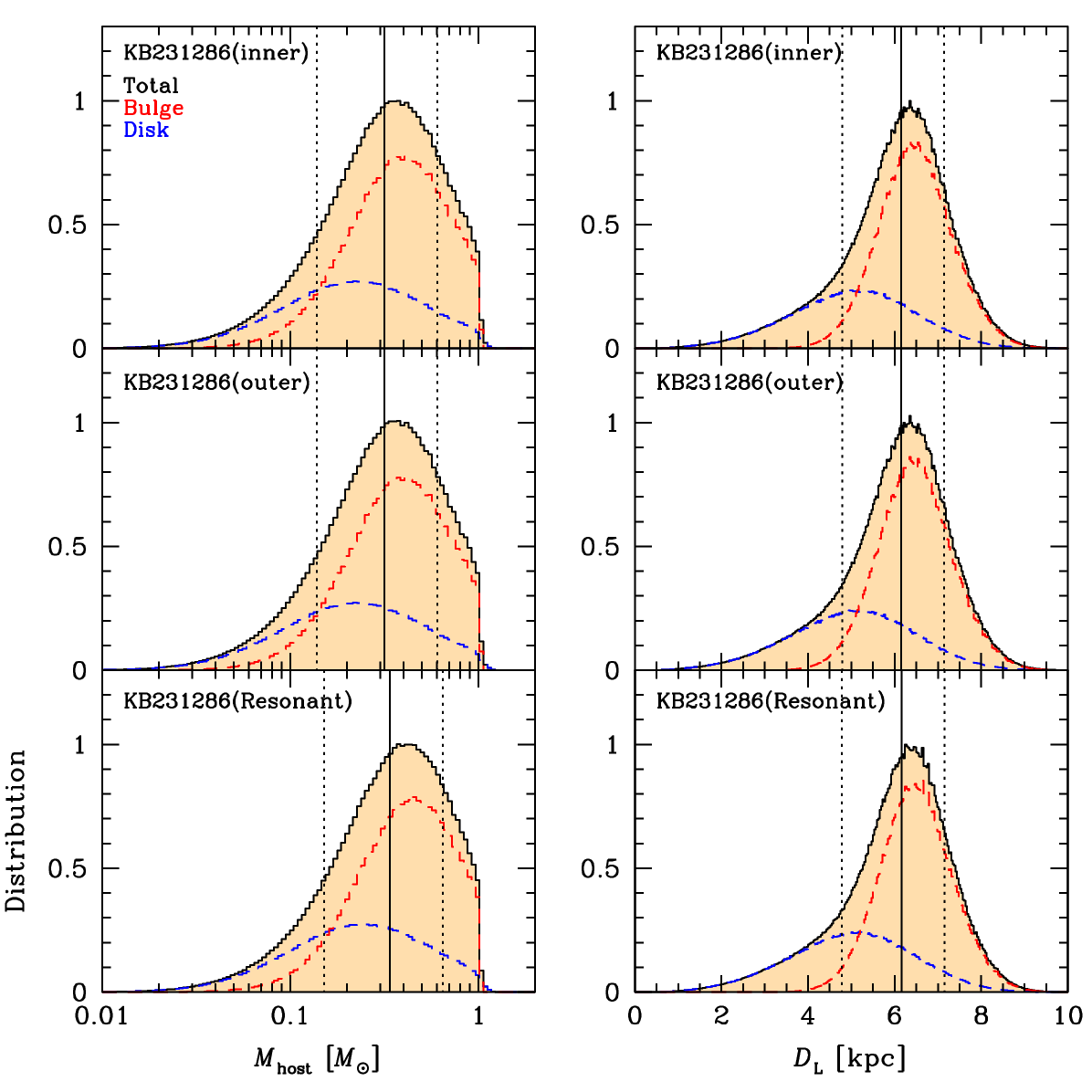}
\caption{Posterior lens mass and distance
  distributions for the three solutions for KMT-2023-BLG-1286.
}
\label{fig:1286bayes}
\end{figure}

\begin{figure}
\plotone{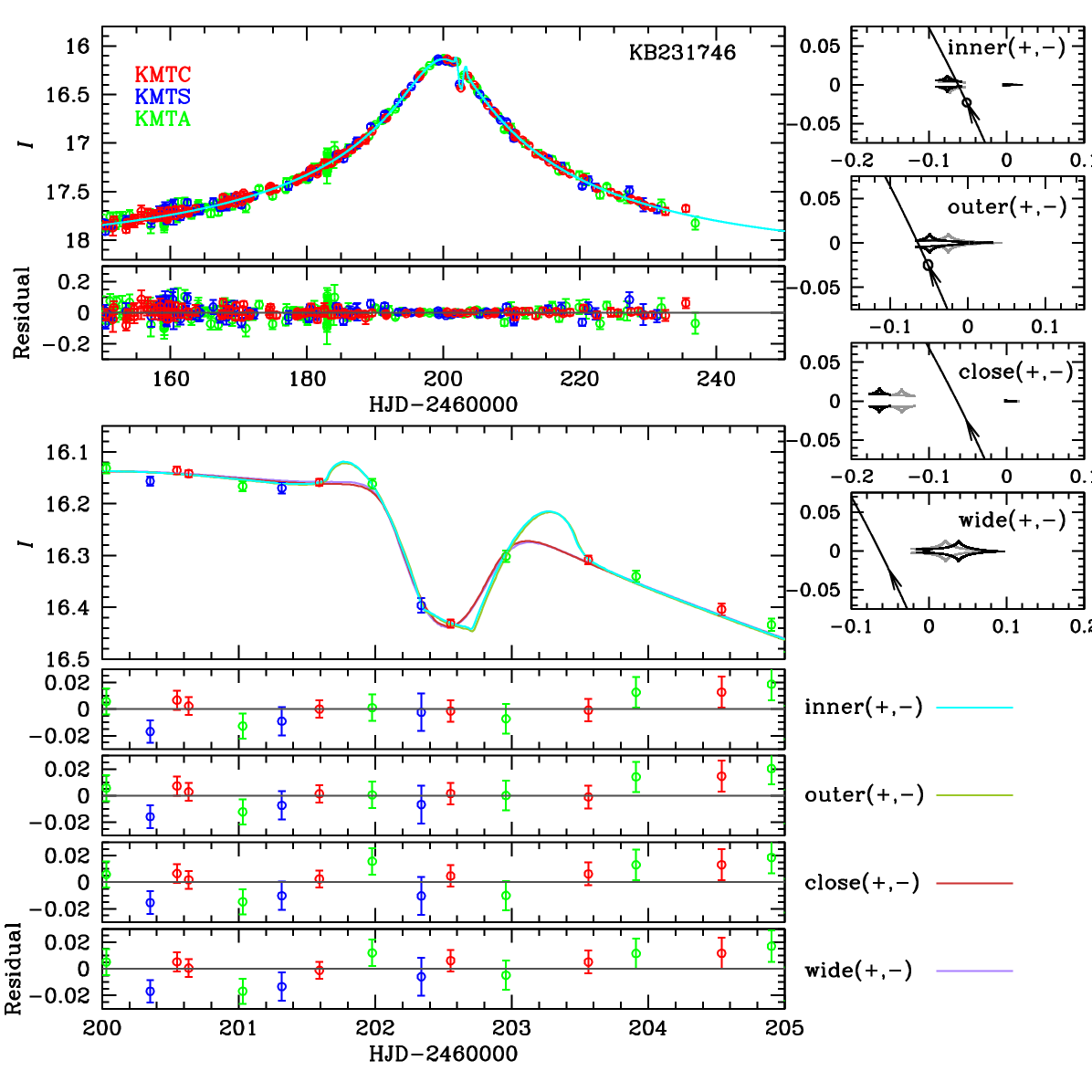}
\caption{Light curve and models for the quartet of four $(+,-)$ solutions  of
  KMT-2023-BLG-1746 with $u_0>0$ and $\pi_{\e N}<0$.  The other three
  quartets look very similar and therefore are not shown explicitly.
  The insets to the right show the caustic geometries.  Note that in
  the (inner, outer) pair of solutions, the source crosses the caustic, but
  in the (close,wide) pair, it does not. In these four panels,
  the light-colored caustics
  represent the configurations at $t=t_0$, while the dark-colored caustics
  represent $t=t_{\rm anom}$.
}
\label{fig:1746lc}
\end{figure}

\begin{figure}
\plotone{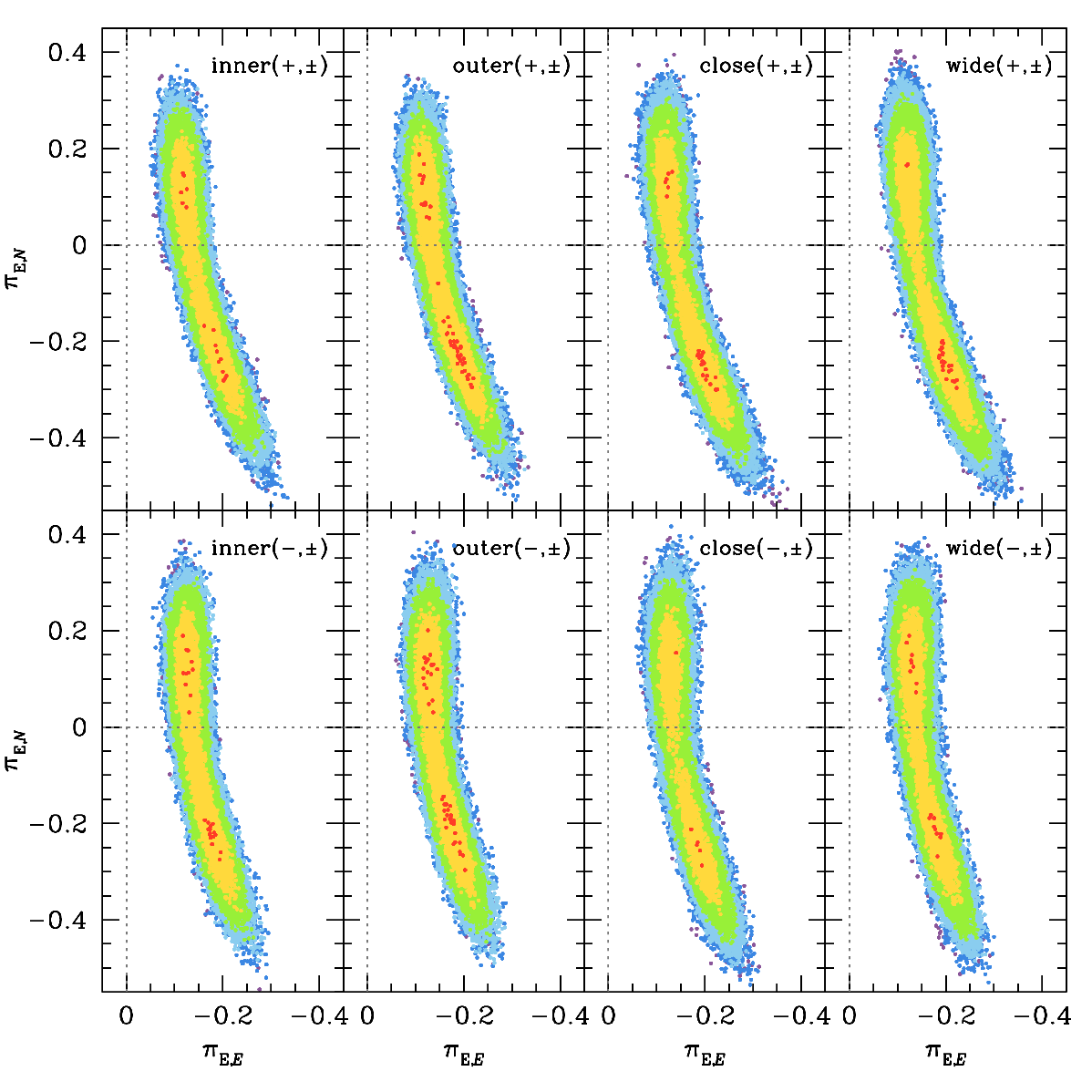}
\caption{Parallax contours for the eight pairs of solutions of
  KMT-2023-BLG-1746.  Colors indicate values of $\Delta\chi^2<1,4,9,16,25,36$.
}
\label{fig:1746parallax}
\end{figure}

\begin{figure}
\plotone{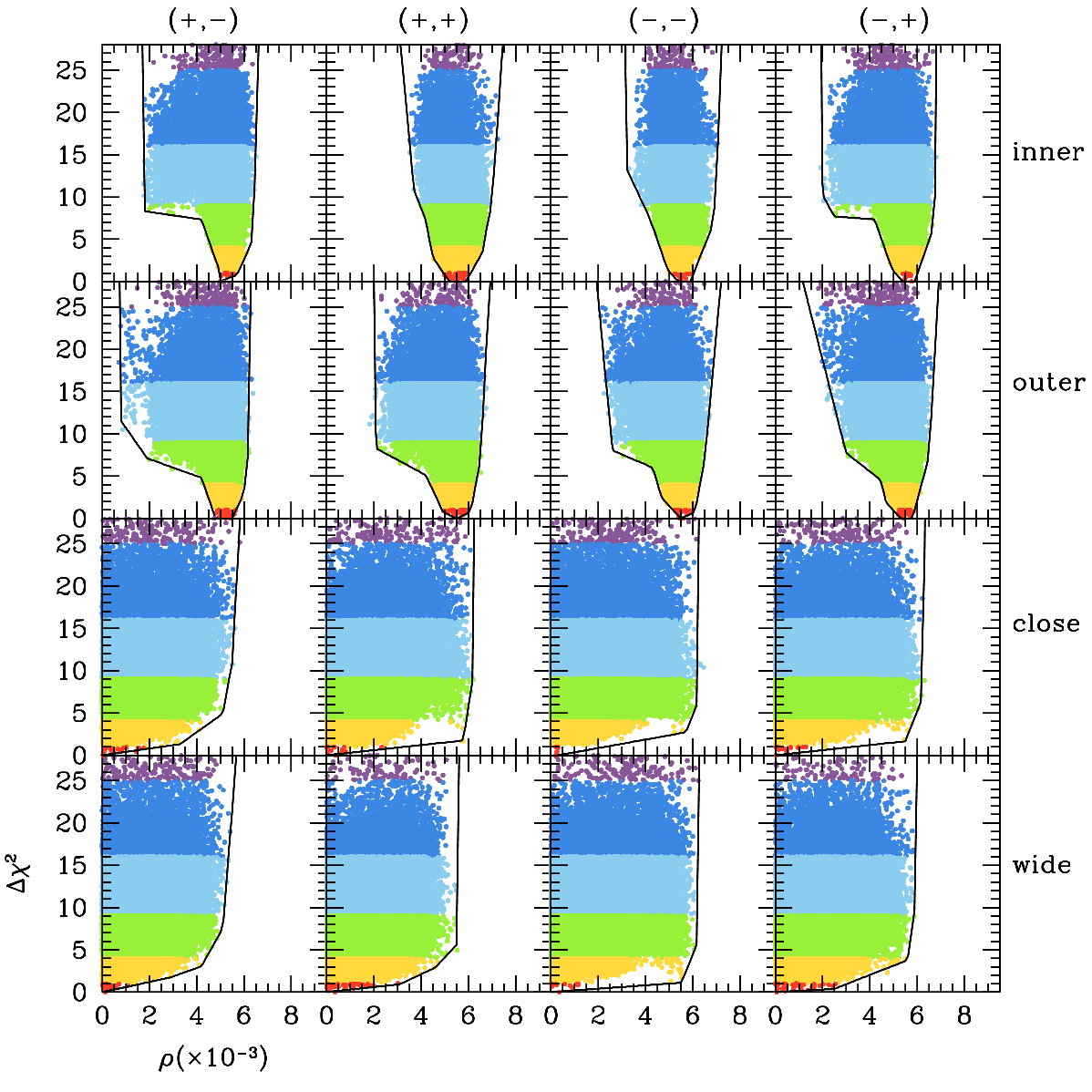}
\caption{Scatter plots, for the 16 solutions of KMT-2023-BLG-1746, 
  of the normalized source radius, $\rho$, against
  the values of $\Delta\chi^2$ relative to the best fit.  Colors indicate
  values of $\Delta\chi^2<1,4,9,16,25,36$.  Also shown is the adopted
  envelope function, which is used in Section~\ref{sec:phys}.
}
\label{fig:1746rho}
\end{figure}

\begin{figure}
\plotone{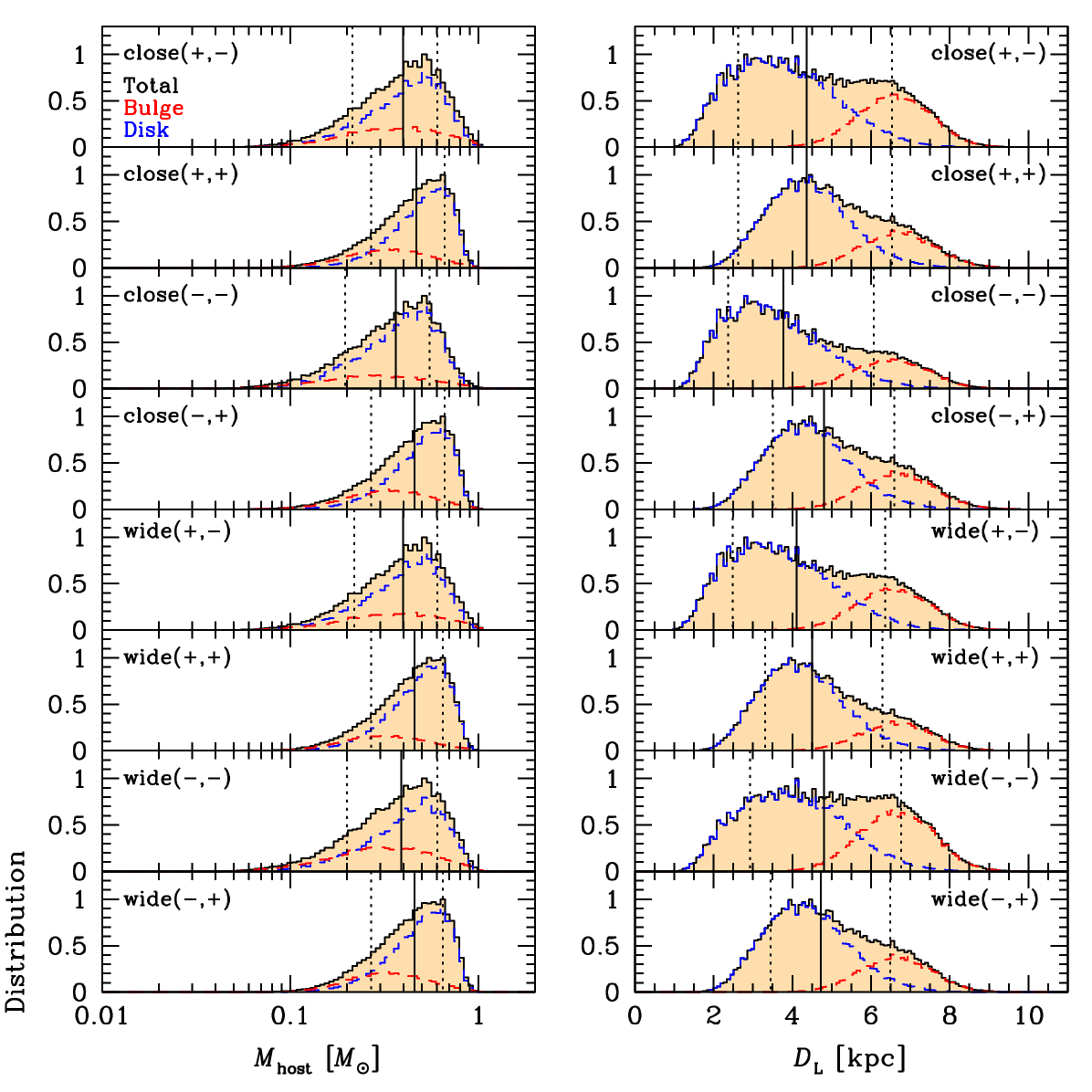}
\caption{Posterior lens mass and distance
  distributions for the eight close/wide solutions for KMT-2023-BLG-1746.
}
\label{fig:1746bayes_cw}
\end{figure}

\begin{figure}
\plotone{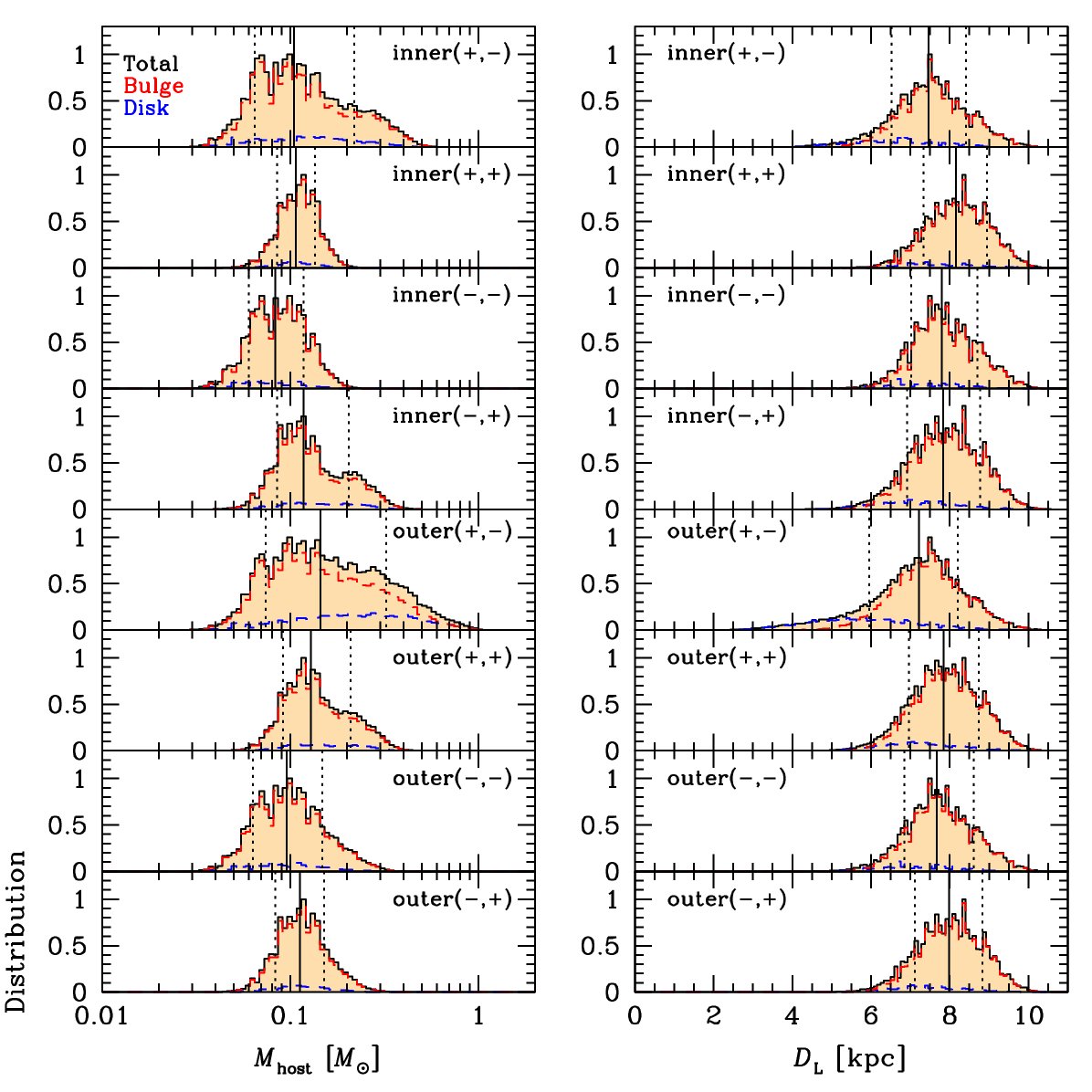}
\caption{Posterior lens mass and distance
  distributions for the eight inner/outer solutions for KMT-2023-BLG-1746.
}
\label{fig:1746bayes_io}
\end{figure}

\begin{figure}
\plotone{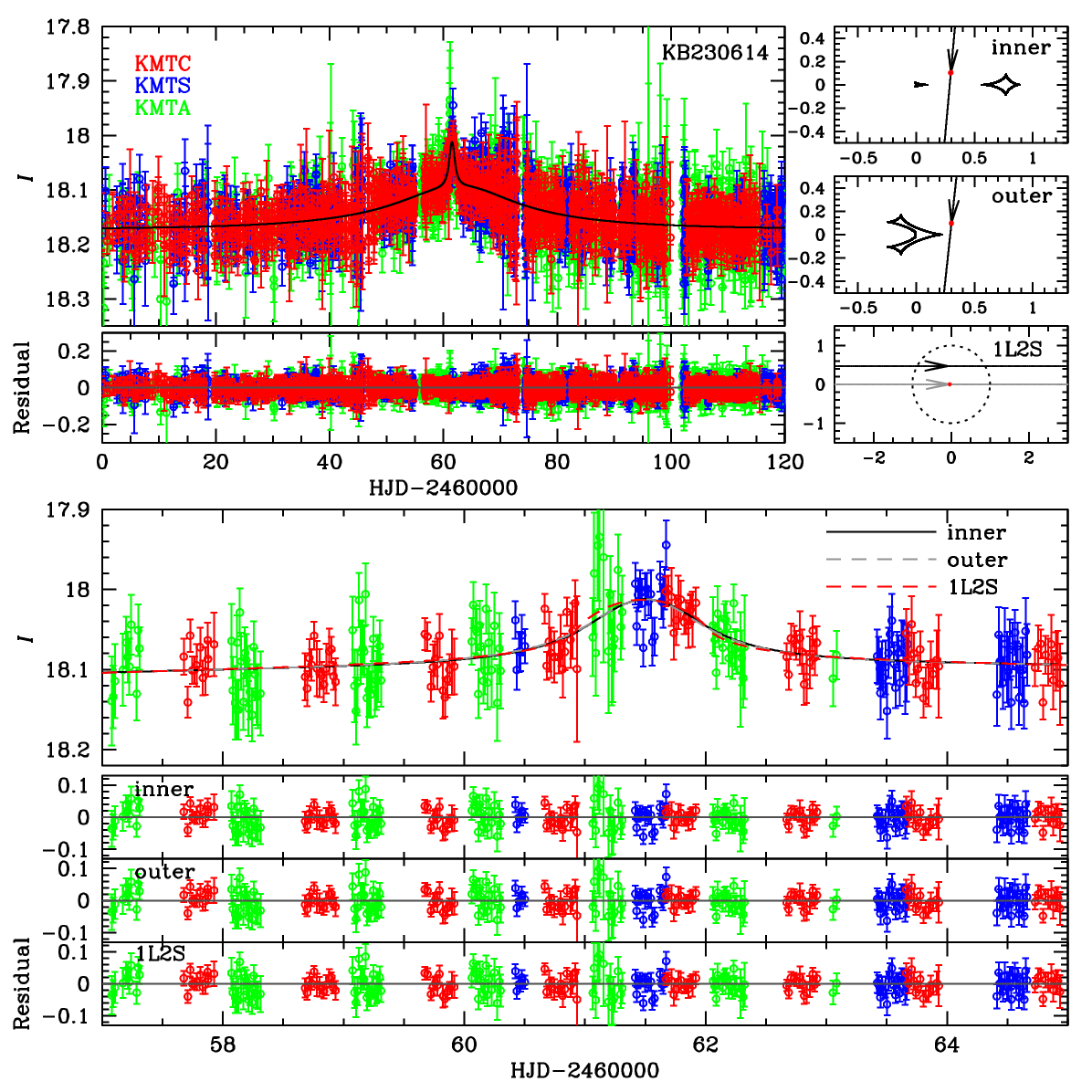}
  \caption{Light curve and models for KMT-2023-BLG-0614.
}
\label{fig:0614lc}
\end{figure}

\begin{figure}
\plotone{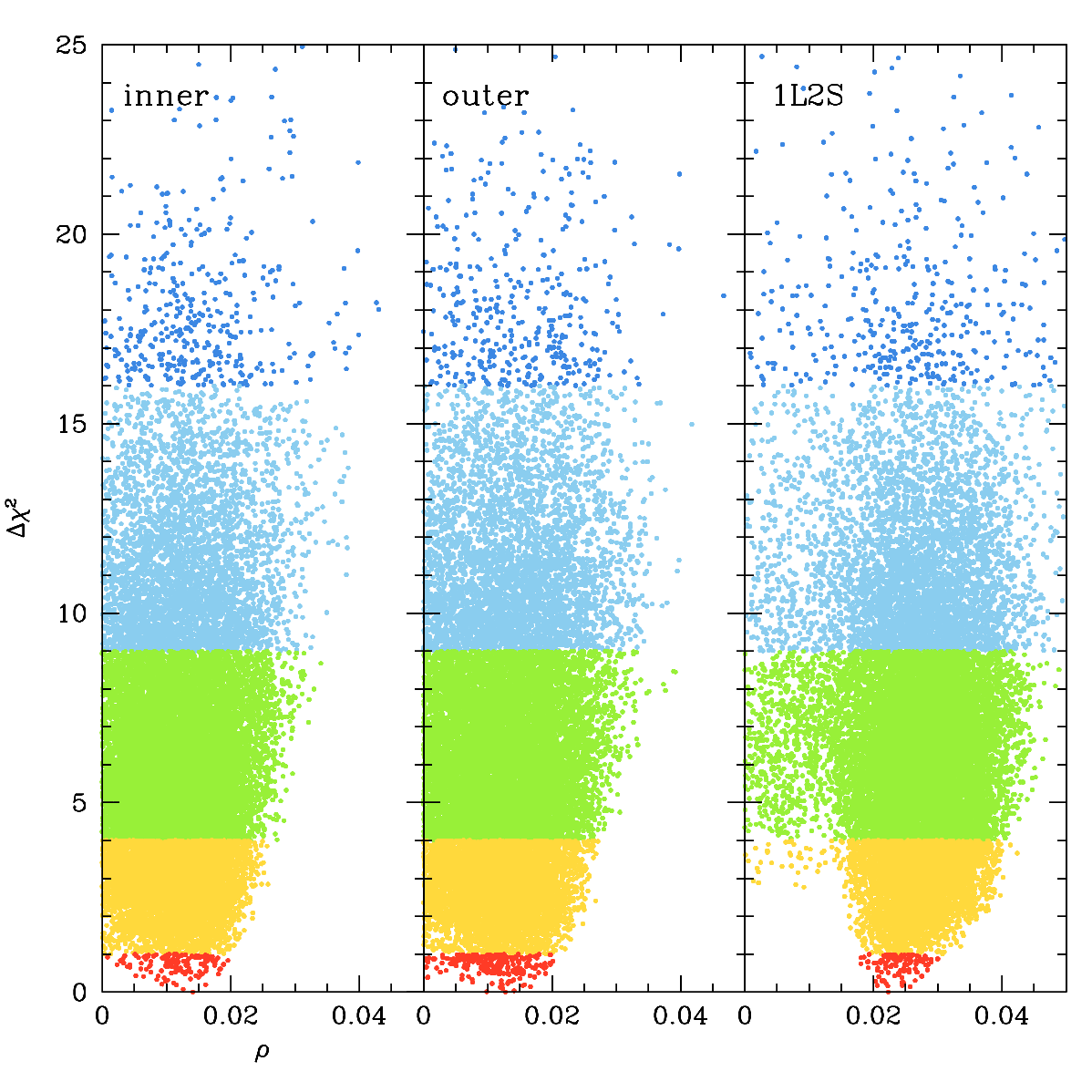}
\caption{Scatter plot of the normalized source radius, $\rho$, against
  the values of $\Delta\chi^2$ relative to the best fit
  for the three solutions of KMT-2023-BLG-0614.  Colors indicate
  values of $\Delta\chi^2<1,4,9,16,25,36$.
}
\label{fig:0614rho}
\end{figure}

\begin{figure}
\plotone{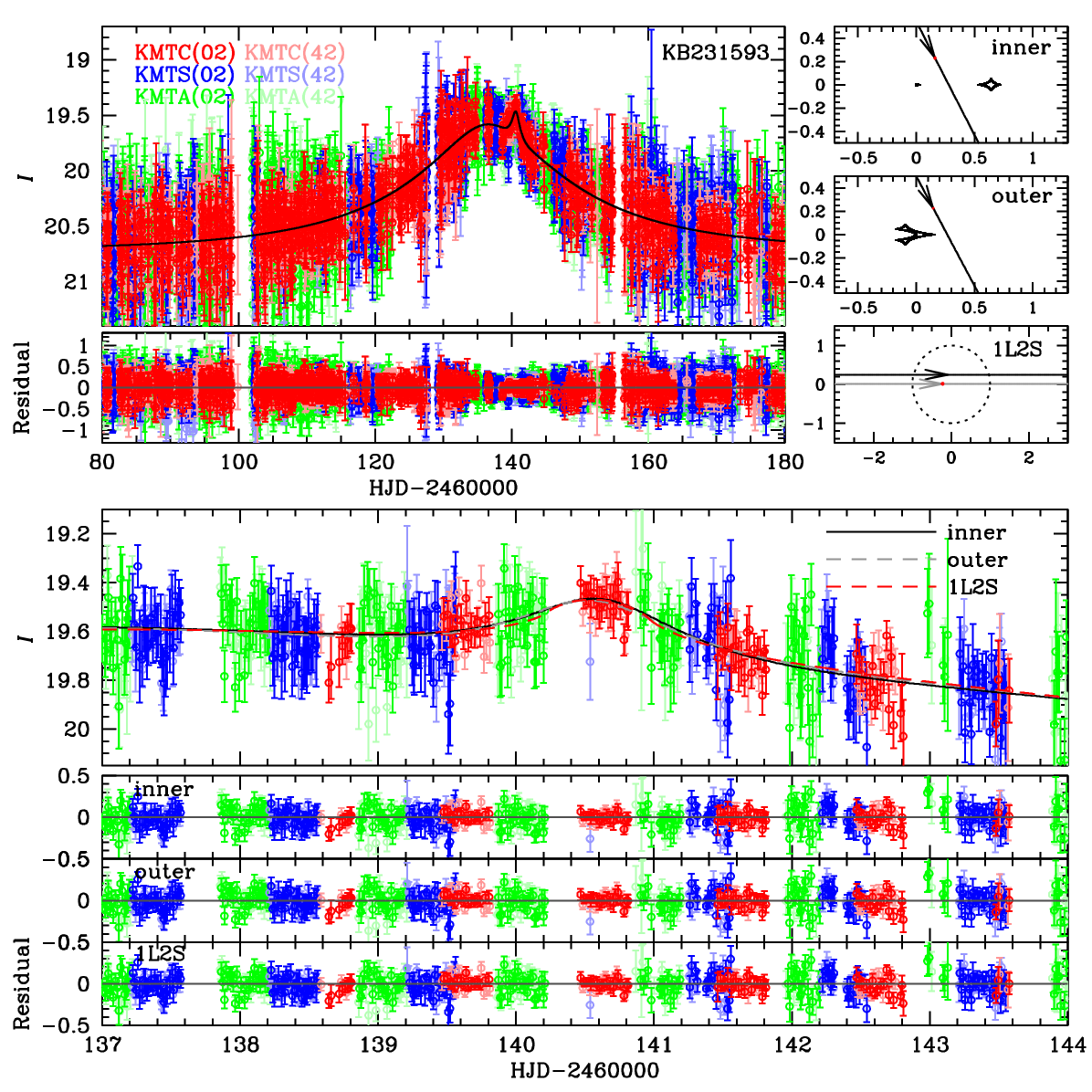}
  \caption{Light curve and models for KMT-2023-BLG-1593.
}
\label{fig:1593lc}
\end{figure}

\begin{figure}
\plotone{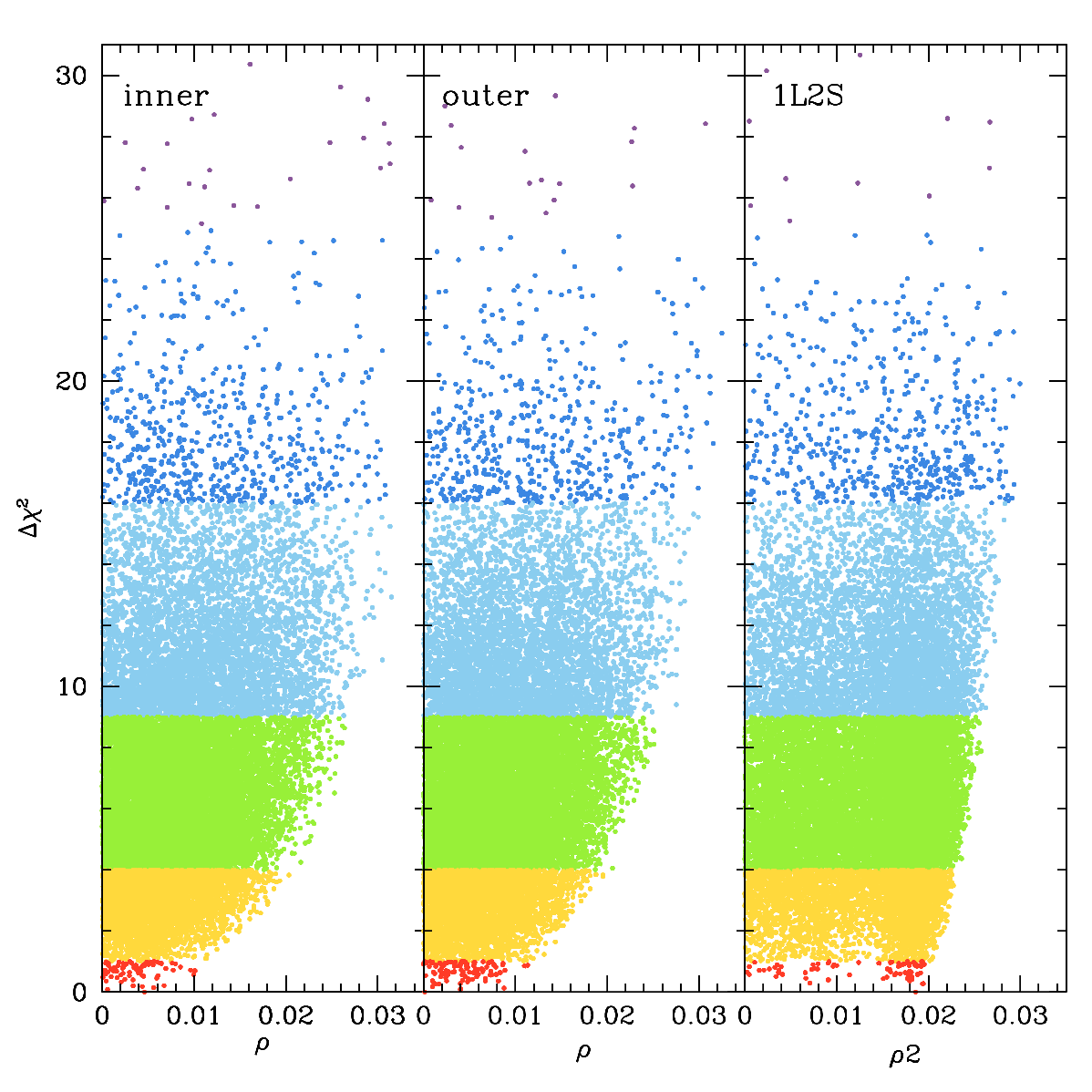}
\caption{Scatter plot of the normalized source radius, $\rho$, against
  the values of $\Delta\chi^2$ relative to the best fit
  for the three solutions of KMT-2023-BLG-1593.  Colors indicate
  values of $\Delta\chi^2<1,4,9,16,25,36$.
}
\label{fig:1593rho}
\end{figure}



\begin{thebibliography}{99}



\bibitem[Alard \& Lupton(1998)]{alard98} Alard, C. \& Lupton, R.H.,1998, \apj, 503, 325

\bibitem[Albrow(2017)]{pydia}Albrow, M.D. Michaeldalbrow/Pydia: InitialRelease On Github., vv1.0.0, Zenodo



\bibitem[Albrow et al.(2009)]{albrow09}Albrow, M.\ D., Horne, K., Bramich, D.\ M., et al.\ 2009, \mnras, 397, 2099





\bibitem[An et al.(2002)]{eb2k5}An, J.H., Albrow, M.D., Beaulieu, J.-P. et al.\ 2002, \apj, 572, 521






\bibitem[Batista et al.(2014)]{mb11293b} Batista, V., Beaulieu, J.-P., Gould, A., et al. 2014, \apj, 780, 54

\bibitem[Batista et al.(2015)]{ob05169bat} Batista, V., Beaulieu, J.-P., Bennett, D.P., et al. 2015, \apj, 808, 170





\bibitem[Bennett et al.(2010)]{ob06109b}Bennett, D.P., Rhie, S.H., Nikolaev, S. et al. 2010, \apj, 713, 837




\bibitem[Bennett et al.(2015)]{ob05169ben} Bennett, D.P., Bhattacharya, A., Anderson, J., et al. 2015, \apj, 808, 169


\bibitem[Bennett et al.(2020)]{ob05071c}Bennett, D. P., Bhattacharya, A., Beaulieu, J. P., et al. 2020, \aj, 159, 68

\bibitem[Bennett et al.(2024a)]{mb08379b}Bennett, D. P., Bhattacharya, A., Beaulieu, J. P., et al. 2024a, \aj, 168, 15

\bibitem[Bennett et al.(2024b)]{ob120563b}Bennett, D. P., Bhattacharya, A., Beaulieu, J. P., et al. 2024b, arXiv:2412.03651



\bibitem[Bensby et al.(2013)]{bensby13} Bensby, T. Yee, J.C., Feltzing, S.\ et al.\ 2013, \aap, 549, A147


\bibitem[Bessell \& Brett(1988)]{bb88} Bessell, M.S., \& Brett, J.M.\ 1988, \pasp, 100, 1134

\bibitem[Bhattacharya et al.(2019)]{ob120950b} Bhattacharya, A., Beaulieu, J.-P., Bennett, D.P., et al. 2019, \aj, 156, 289

\bibitem[Bhattacharya et al.(2021)]{mb09400b} Bhattacharya, A., Bennett, D.P., Beaulieu, J.-P., et al. 2021, \aj, 162, 60



\bibitem[Bond et al.(2017)]{ob161195a} Bond, I.A., Bennett, D.P., Sumi, T. et al.\ 2017, \mnras, 469, 2434




\bibitem[Bressan et al.(2012)]{2012MNRAS.427..127B} Bressan, A., Marigo, P., Girardi, L., et al.\ 2012, \mnras, 427, 127














\bibitem[Dong et al.(2009)]{ob05071b} Dong, S., Gould, A., Udalski, A., et al. 2009, \apj, 695, 970




\bibitem[Dressler et al.(2011)]{Dressler11} Dressler, A., Bigelow, B., Hare, T., et al.\ 2011, \pasp, 123, 901, 288 

\bibitem[Duquennoy \& Mayor(1991)]{dm91}Duquennoy, A., \& Mayor, M. 1991, \aap, 248, 485

  



\bibitem[Gaia Collaboration et al.(2023)]{gaia}Gaia Collaboration,Vallenari, A., Brown, A.G.A.,  et al., 2023, \aap, 674, A1

\bibitem[Gaudi(1998)]{gaudi98} Gaudi, B.S.\ 1998, \apj, 506, 533










\bibitem[Gonzalez et al.(2012)]{gonzalez12} Gonzalez, O.~A., Rejkuba, M., Zoccali, M., et al.\ 2012, \aap, 543, A13

\bibitem[Gould(1992)]{gould92} Gould, A. 1992, \apj, 392, 442






\bibitem[Gould(2000)]{gould00} Gould, A. 2000, \apj, 542, 785

\bibitem[Gould(2004)]{gould04} Gould, A. 2004, \apjl, 606, 319




\bibitem[Gould(2022)]{gould22} Gould, A. 2022, arXiv:2209.12501








\bibitem[Gould et al.(1994)]{gmb94} Gould, A., Miralda-Escud\'e, J. \&  Bahcall, J.N. 1994, \apj, 423, L105



















\bibitem[Griest \& Hu(1992)]{griest-hu} Griest, K.\ \& Hu, W.\ 1992, \apj, 397, 362

  



\bibitem[Han \& Gould(1997)]{han97} Han, C. \& Gould, A.\ 1997, \apj, 480, 196







\bibitem[Han et al.(2018)]{kb162052} Han, C., Jung, Y.K., Shvarzvald, Y.\ 2018, \aj, 865, 14










\bibitem[Han et al.(2024a)]{kb230416}Han, C., Udalski, A., Lee, C.-U., et al. 2024a, \aap, 683A, 187 

\bibitem[Han et al.(2024b)]{kb231866}Han, C., Bond, I.A., Udalski, A., et al. 2024b, \aap, 687A, 241 

\bibitem[Han et al.(2025a)]{kb231896}Han, C., Bond, I.A., Jung, Y.K., et al. 2025a, \aap, 694A, 90 

\bibitem[Han et al.(2025b)]{kb232209}Han, C., Albrow, M.D., Lee, C.-U., et al. 2025b, \aj, 169, 288








\bibitem[Huang et al.(2022)]{2022ApJ...925..164H} Huang, Y., Beers, T.~C., Wolf, C., et al.\ 2022, \apj, 925, 164





\bibitem[Hwang et al.(2022)]{kb190253} Hwang, K.-H., Zang, W., Gould, A., et al. 2022, \aj, 163, 43 


\bibitem[Jiang et al.(2004)]{ob03238} Jiang, G., DePoy, D.L., Gal-Yam, A., et al. 2004, \apj, 617, 307
  







\bibitem[Jung et al.(2021)]{ob180567}Jung, Y.\ K., Han, C., Udalski, A.,. et al. 2021, \aj, 161, 293 

\bibitem[Jung et al.(2022)]{2018subprime}Jung, Y.\ K., Zang, W., Han, C., et al. 2022, \aj, 164, 262

\bibitem[Jung et al.(2023)]{2019subprime}Jung, Y.\ K., Zang, W., Wang., H., et al. 2023, \aj, 165, 226




\bibitem[Kervella et al.(2004a)]{kervella04a} Kervella, P., Th{\'e}venin, F., Di Folco, E., \& S{\'e}gransan, D.\ 2004a, \aap, 426, 297

\bibitem[Kervella et al.(2004b)]{kervella04b} Kervella, P., Bersier, D., Mourard, D., et al.\ 2004b, \aap, 428, 587  
  
\bibitem[Kim et al.(2016)]{kmtnet} Kim, S.-L., Lee, C.-U., Park, B.-G., et al.  2016, JKAS, 49, 37 


\bibitem[Kim et al.(2018)]{alertfinder} Kim,  H.-W., Hwang, K.-H., Shvartzvald, et al. 2018, arXiv:1806.07545










\bibitem[Marigo et al.(2017)]{2017ApJ...835...77M} Marigo, P., Girardi, L., Bressan, A., et al.\ 2017, \apj, 835, 77

  











\bibitem[Nataf et al.(2013)]{nataf13} Nataf, D.M., Gould, A., Fouqu\'e, P. et al. 2013, \apj, 769, 88


\bibitem[Paczy\'nski(1995)]{pac95} Paczy\'nski, B.\ 1995, Acta Astron., 45, 345

\bibitem[Paczy\'nski(1986)]{pac86} Paczy\'nski, B.\ 1986, \apj, 304, 1


\bibitem[Park et al.(2004)]{mb03037}Park, B.-G., DePoy, D.L.., Gaudi, B.S.,  et al.\ 2004, \apj, 609, 166






\bibitem[Poindexter et al.(2005)]{poindexter05} Poindexter, S., Afonso, C.,  Bennett, D.P., et al.\ 2005, \apj, 633, 914

  







\bibitem[Rektsini et al.(2024)]{ob130132b} Rektsini, N.E., Batista, V., Ranc, C., et al. 2024, \aj, 167, 145

\bibitem[Rektsini et al.(2025)]{ob141760b} Rektsini, N.E., Ranc, C., Koshimoto, N. et al. 2025, arXiv:2502.12942

  





\bibitem[Ryu et al.(2022)]{kb211391} Ryu, Y.-H., Jung, Y.K., Yang, H., et al. 2022, \aj, 164, 180 
  
\bibitem[Ryu et al.(2023)]{kb210712} Ryu, Y.-H., Shin, I.-G., Yang, H., et al. 2023, \aj, 165, 83 %

\bibitem[Sánchez-Blázquez et al.(2006)]{2006MNRAS.371..703S} Sánchez-Blázquez, P., Peletier, R.~F., Jiménez-Vicente, J., et al.\ 2006, \mnras, 371, 703

  





















\bibitem[Shvartzvald et al.(2017)]{ob161195b} Shvartzvald, Y., Yee, J.C., Calchi Novati, S.\ et al.\ 2017, \apjl, 840, L3






\bibitem[Smith et al.(2003)Smith, Mao \& Paczy\'nski]{smp03} Smith, M., Mao, S., \& Paczy\'nski, B., 2003, \mnras, 339, 925










\bibitem[Szyma\'nski et al.(2011)]{oiiicat}Szyma\'nski, M.K., Udalski, A., Soszy\'nski, I., et al. 2011, Acta Astron., 61, 83

\bibitem[Terry et al.(2021)]{mb09319b} Terry, S.K., Bhattacharya, A., Bennett, D.P., et al. 2021, \aj, 161, 54
  
\bibitem[Terry et al.(2024)]{mb07192b} Terry, S.K., Beaulieu, J.-P., Bennett, D.P., et al. 2024, \aj, 168, 72
  

\bibitem[Tomaney \& Crotts(1996)]{tomaney96} Tomaney, A.B. \& Crotts, A.P.S. 1996, \au, 112, 2872














\bibitem[Vandorou et al.(2020)]{mb13220b} Vandorou, A., Bennett, D.P., Beaulieu, J.-P., et al. 2020, \aj, 160, 121




\bibitem[Yang et al.(2024)]{yang24} Yang, H., Yee, J.C., Hwang, K.-H.,, et al.\ 2024, \mnras, 528, 11

\bibitem[Yee et al.(2012)]{mb11293} Yee, J.C., Shvartzvald, Y., Gal-Yam, A.\ et al.\ 2012, \apj, 755, 102





\bibitem[Yoo et al.(2004)]{ob03262} Yoo, J., DePoy, D.L., Gal-Yam, A.\ et al.\ 2004, \apj, 603, 139



\bibitem[Zang et al.(2021)]{ob191053} Zang, W., Hwang, K.-H., Udalski, A., et al. 2021, \aj, 162, 163 

\bibitem[Zang et al.(2022)]{af2} Zang, W., Yang, H., Han, c., et al.\ 2022, \mnras, 515, 928 

\bibitem[Zang et al.(2025)]{massrat} Zang, W., Jung, Y.K., et al.\ 2024, Science, 




\bibitem[Zhang et al.(2021)]{zhang22} Zhang, K., Gaudi, B.S., Bloom, J.S, 2021, arXiv:2111.13696










\end{thebibliography}
\end{document}